\documentstyle[preprint,tighten,eqsecnum,aps]{revtex}

\begin{document}

\draft
\title{ {\bf Towards softer scales in hot QCD} }
\author{F. Guerin }
\address{  Institut Non Lineaire de Nice, 1361 route des Lucioles,  06560 Valbonne, France }

\maketitle

\begin{abstract}

An effective theory for the soft field modes in hot QCD has been obtained recently by integrating out the field modes of momenta of order $T$ and $g T$. The mean hard particle   distribution obeys a transport equation with a collision term. The scale that comes out for the soft field is $g^2 T \ln 1/g $ .  \\
 $n$-gluon soft amplitudes are shown to be  a sum of tree-diagrams and to obey tree-like Ward identities. These effective amplitudes are used to compute
 the one-soft-loop contribution to the gluon self-energy when the loop momentum is of order $ g^2 T \ln 1/g$. It allows to identify a new collision term which, when inserted into the transport equation, takes into account collisions with gluon exchange $\sim g^2T\ln 1/g$.  The infrared behaviour of the collision term exhibits a significant change.

\end{abstract}


 \section{Introduction}
 \setcounter{equation}{0}
 
The framework of this paper is pure gauge theory in thermal equilibrium. 
with $g<<1$ \\
It has been realized that the near-equilibrium dynamics of high temperature QCD
 involves a hierarchy of length scales \\
scale $(T)^{-1}$ inverse of the typical momentum of the plasma particles \\
scale $(g T)^{-1}$ electric screening length and first scale of collective excitations \\
scale $(g^2 T \ln 1/g)^{-1}$  damping length of colour excitations \\
scale ($g^2 T)^{-1} $ non perturbative magnetic fluctuations.

Effective theories can be constructed at a given scale that integrate out 
the smaller scales. The prototype is the hard thermal loop ( HTL) effective 
theory that integrate out the scale $(T)^{-1}$  \cite{Pi1,FTay,MLB}. 
To integrate out the hard modes $p\sim T$ means to calculate loop diagrams 
with external momenta $<<T$ and internal momenta of order $T$. It generates 
effective propagators and effective vertices for the field modes with 
momenta $<<T$. To leading order, only one-loop diagrams survive and the result 
is the hard thermal loop $n$-point functions. Remarkably, the generating 
functional of the hard thermal loops is gauge invariant.

B\"{o}deker's break-through  \cite{Bod1}  was to realize that one may 
integrate out the scale $(g T)^{-1}$. The summation now involves an infinite 
set of multiloop diagrams with loop momenta $ g T$ and effective HTL 
propagators and effective vertices.

One way to construct effective theories for the soft field modes of the 
plasma is to use classical transport equations. The soft field modes behave 
classically, this is due to the fact that the momentum scales of interest are 
small compared to the temperature, so that the modes have large occupation
numbers. The resulting set of equations that characterize the effective theory has now
 been obtained through different approaches \cite{Bod1,Bod2,AY2,BI2,Lit}. 
 This work will follow the approach of Blaizot and Iancu \cite{BI3,BI1}
 to the construction of the effective theories, we summarize it here. \\
 
 At high temperature, the non-abelian pure gauge theory describes a weakly
 coupled plasma whose constituents, gluons have a typical momenta $k\sim T$ and a
 typical distance $(T)^{-1}$. These plasma particles may take part in
 long-wavelength collective excitations, which are fluctuations in the average
 density of the plasma particles. These excitations may be induced by a weak 
 external disturbance. The collective excitations at scale $x$ are described 
 by a field $W(x,{\bf v})$, a colour matrix in the adjoint representation 
 $W(x,{\bf v})=W^a(x,{\bf v})\ T^a$, where ${\bf v}$ is the direction of 
 propagation of the excitation. Transport equations describe the space-time 
 evolution of those long-wavelenth excitations. \\
 The soft fields that one is interested in, are represented by mean fields. In
 the transport equation, they couple to the collective excitations via a
 mean-field term ${\bf v}.{\bf E}^a(x)$ and via the covariant derivative. The
 soft fields obey Maxwell equations, where the collective excitations act as an
 induced source
 \begin{equation}
 D^{\nu}\ F_{\mu\nu}(x) = j_{\mu}^{ext}(x) + j_{\mu}^{ind}(x)
 \label{RA1}
 \end{equation}
 where the induced current is the response of the plasma to the initial
 perturbation $j_{\mu}^{ext}(x)$
 \begin{equation}
 j_{\mu}^{a \ ind}(x) =m_D^2\int {d\Omega_{\bf v}\over 4\pi}\ v_{\mu} \ 
 W^a(x,{\bf v})
 \label{RA2}
 \end{equation}
 with
 \begin{equation}
 m_D^2={g^2NT^2\over 3}
 \label{RA3}
 \end{equation}
 On the space-time scale $(gT)^{-1}$, the transport equation is
 \begin{equation}
 v.D_x^{ab} \ W^b(x, {\bf v}) = {\bf v}.{\bf E}^a(x)
 \label{RA4}
 \end{equation}
 with $v_{\mu}(v_0=1,{\bf v})$. The effective theory for the soft field modes is
 obtained as follows. By solving the transport equation via a Green function,
 one can express the collective excitation $W(x,{\bf v})$ as a functional of
 the soft fields. Then the induced current, Eq.(\ref{RA2}), acts as a generating
 functional for the $n$-point amplitudes of the soft fields. The resulting
 amplitudes are the HTL amplitudes.

When one is interested in collective excitations involving colour fluctuations
on larger wavelength, one integrates out both scales $(T)^{-1}$ and $(gT)^{-1}$.
It turns out that the resulting set of equations is similar. The new element is
the inclusion of the effect of collisions and they are dominated by small angle
scattering. Indeed a scattering process which hardly changes the momentum of a
hard particle, can change its colour charge. The transport equation at larger
wavelength $x>> (gT)^{-1}$ is
\begin{equation}
 [v.D_x^{ab}  + \hat{C}] \ W^b(x, {\bf v}) = {\bf v}.{\bf E}^a(x)
 \label{RA5}
 \end{equation}
 The collision term $\hat{C} =C({\bf v},{\bf v'})$ is non local in ${\bf v}$,
 but local in $x$ and blind to colour. The gluons $p\sim T$ take part in the
 collective motion, the gluons $p\sim gT$ are exchanged in the collision
 process. \\
 The soft fields obey the Maxwell equations (\ref{RA1}) and the effective theory
 is obtained from the induced current (\ref{RA2}). The resulting new feature is 
 the effect of dissipation as a result of collisions. A new scale appears, the
 colour relaxation time $1/\gamma$ with $\gamma \sim g^2T\ln 1/g$. 

This purely dissipative description is appropriate to the study of the
 relaxation of a weak initial disturbance at scale $x$. However colour
 excitations may also be generated by thermal fluctuations in the plasma. The
 fluctuation-dissipation theorem relates the two phenomena. In B\"odeker's
 alternative construction of the effective theory (without an external source)
 \cite{Bod1}, the collision term in Eq.(\ref{RA5}) is accompanied by a Gaussian
 white noise term, which arises from the thermal fluctuations of initial
 conditions of the field with momentum $\sim gT$. The noise term keeps the soft
 modes in thermal equilibrium, it injects energy which compensates for the loss
 of energy at scale $x$ from the damping term.

 One quantity that turns out to be particularly sensitive to the change of 
 scale  is the polarization tensor in the magnetic sector. Precisely that 
 quantity enters the collision term of the transport equation in a strategic 
 way. Indeed the interaction rate between two hard particles is dominated by 
 soft momentum transfer $q \sim  gT$ where the squared modulus of the resummed 
 gluon propagator enters 
$$ I    \simeq m_D^2 \int_{q_0 <<q}  { dq_0   d_3 q \over{ q^4 + ({ \mathrm{Im}} 
\Pi^t)^2} }   {1\over q^2}   $$
where $ {1 /  q^2} $ comes from energy-momentum conservation at each vertex 
 \\
When the scale $T$ has been integrated out, $\Pi^t$ has the well known HTL 
expression  $ {\mathrm{Im}} \Pi^t =- \pi m_D^2 \  q_0  / q$ for  $q_0 << q$   
so that $ I  \sim \int dq / q   $ leading to the $\ln 1/g$ factor in the scale 
of the collision term. That expression for $\Pi^t$ turns out to be valid at 
the scale $g T $ only. Indeed, when the scale $g T$ is integrated out the 
resulting $ \Pi^t$ is \cite{BI1,AY1}   
$ {\mathrm{Im}} \Pi^t  =-m_D^2  \ q_0 / {3 \gamma}$ for  $q_0 << q << \gamma $ 
 so that the integral $I$ now behaves as $ I \simeq \int d q / q^2 $. 
 Again this $\Pi^t$ is valid at the scale $g^2 T \ln 1/g$. If one wants to 
 reach the scale $g^2 T$, one has to integrate out the scale $g^2 T \ln 1/g$ in
 order to see how it affects the physics at the scale $g^2 T$.

Arnold and Yaffe \cite{AY1} have recently attacked the problem of taking
 the scale $g^2 T \ln 1/g$ into account. As they are interested in the colour conductivity, 
 they concentrate on static quantities. They make use of effective theories and 
 have computed the one-loop 
 contribution to $\Pi^l(q_0=0, q)$ in an effective theory.

In this work, the one-loop contribution to the polarization tensor 
$\Pi^{\mu \nu} (q_0, q)$ will be computed for loop momenta 
$\sim g^2 T \ln 1/g$ and external momenta $\sim g^2 T$ assuming $\ln 1/g >>1$. 
It is only the first term of an infinite series of multiloop diagrams. 
However, following a remark made by Blaizot and Iancu  \cite{BI1} for the case 
of momenta $\sim gT$, it will be possible to identify in the one-loop term, 
the new collision term that should be added in the transport equation in order 
to sum the infinite series. This is only
a first step towards integrating out the scale $g^2 T \ln 1/g$. We will not
address the more ambitious question whether integrating out this scale  amounts
to take into account collisions with exchange gluon $\sim g^2 T \ln 1/g$.

 The one-soft-loop contribution to the polarization tensor $\Pi^{\mu\nu}(q_0,q)$
 comes from a well known pair of diagrams \cite{Pi1,MLB} whose contribution 
 adds up to
 give a transverse $\Pi^{\mu\nu}$. One diagram possesses two effective
 three-gluon vertices and two effective propagators, the other one has an
 effective four-gluon vertex and an effective propagator. For the case of a loop
 momentum $\sim g^2 T\ln 1/g$, up to now explicit expressions have only been
 written for the polarization tensor and for the linearized form of the
 transport equation (\ref{RA5}) \cite{AY2,BI1}. \\
 One part of this work will be devoted to write down the explicit form of the
 needed vertices. In the following, HTL amplitudes refer to the effective 
 amplitudes when the scale $(T)^{-1}$ is integrated out, soft amplitudes refer
 to the case when the scales $(T)^{-1}$ and $(gT)^{-1}$  are integrated out.
 The remarkable similarities between the HTL amplitudes and the soft amplitudes
 will be a recurrent theme in this work: \\
 The $n$-point amplitudes are obtained from the induced current through the same
 steps, they obey the same Ward identities. In the one-soft-loop diagram, this
 similarity will show up again and it will lead to the existence of the new
 collision operator $\hat{C'}$ which exhibits features similar to the operator
 $\hat{C}$. There are also interesting differences, the soft amplitudes contain
 damping in a way that does not conflict with gauge symmetry.
 \\ The transport equation (\ref{RA4}) has been shown to be gauge invariant, the
 transport equation (\ref{RA5}) has been established in the Coulomb gauge,
 however the collision operator is expected to be gauge-independent \cite{BI1}.
 
 Section 2 is devoted to the polarization tensor $\Pi^{\mu\nu}$. An explicit
 Lorentz-covariant form is obtained for the case when the scale $(gT)^{-1}$ is
 integrated out. $\Pi^{\mu\nu}$ appears under a form that allows to interpolate
 between the scale $gT$ and the scale $g^2T\ln 1/g$. Then it is detailed how the one-loop contribution to the gluon 
 self-energy with loop  momenta $g^2 T \ln 1/g$ allows to identify the new 
 collision operator, under which assumptions.  In Section 3, explicit forms of
 the effective three-gluon and four-gluon vertices are obtained. The response
 approach makes use of a Retarded Green function, it leads to the use of the
 Retarded/Advanced formalism.
 The $n$-point Retarded soft amplitudes are obtained from the induced current
 and are shown to obey tree-like Ward identities in any leg. The reason is, they
 are a sum of tree diagrams, the propagator along the tree is the (linearized) 
 one of $W^b(x, {\bf v})$, i.e. the soft fields induce the long-wavelength
 collective excitations. A needed generalization is then made for the four-gluon
 vertex. In Section 4 the one-soft-loop self-energy diagrams are evaluated in
 the Retarded/Advanced formalism. The  diagram 
 with 3-gluon effective vertices is evaluated first. 
 Loop momenta $g T$ and $g^2 T \ln 1/g $ are successively considered.
 For the case $g^2T \ln1/g$ one part of the new collision term is identified, 
 its physical interpretation, its scale are examined.
 Then, the one-loop self-energy diagram with a 4-gluon vertex is treated 
 in a similar way. The total one-loop $\Pi^{\mu \nu}$ is transverse, 
 the total collision operator has a zero mode. 
 The similarities between the two collision operators appear. In Section 5, some
 properties and consequences of the new collision operator $\hat{C'}$ are 
 studied, in particular its infrared behaviour.  Conclusions are in Section 6. \\
In an Appendix, the explicit expression of $\Pi(q_0, q)$ at the scale 
$g^2 T \ln 1/g$ is written down and its analytical properties in the 
complex $q_0$ plane are studied. Another Appendix contains explicit expressions
for the eigenvalues of the operator $\hat{C'}$ for the pure magnetic sector's case.

\section{The Polarization tensor}
\label{sec2}
Our  notations are $K (k_0, {\mathbf{k}}), \ K^2=k_0^2-k^2$
\subsection{The collision-resummed $W$  propagator}
\label{sec2.1}
The transport equation for the collective $W$ field at a scale $x>>(gT)^{-1}$
is \cite{Bod1,AY1,BI1}
\begin{equation}
(v.D_x+\hat{C})W(x,{\mathbf {v}}) ={\mathbf{v}.\mathbf{E}} (x)
\end{equation}
where $D_x$ is the covariant derivative and $\hat{C}$ the collision operator. 
The linearized equation is in Fourier space
\begin{equation}
(-iv.K+\hat{C}) \ W(K,{\mathbf{v}}) = {\mathbf{v}.\mathbf{E}} (K)
\label{A2}
\end{equation}
where $v.K = v_0 k_0 -  \mathbf{v}.\mathbf{k}$   and  $v (v_0=1,\mathbf{v})$ 
with ${\mathbf{v}}^2 =1$ , and $\hat{C}$ is an operator in $\mathbf{v}$ space
\begin{equation}
\hat{C} \  W  =\int { d{\Omega}_{\mathbf{v'}} \over{4\pi}} \ C({\mathbf{v}, \mathbf{v'}}) W(K,{\mathbf{v'}})
\end{equation}
$ C(\mathbf{v}, \mathbf{v'}) $ is a real function, symmetric in 
$\mathbf{v}$ and $\mathbf{v'}$
\begin{equation}
C({\mathbf{v}, \mathbf{v'}}) = \gamma \  {\delta}_{S_2} ({\mathbf{v} - \mathbf{v'}}) - m^2_D {g^2 N T\over 2} \Phi({\mathbf{v} . \mathbf{v'}})
\label{A9}
\end{equation} 
\begin{equation}
\int { d{\Omega}_{\mathbf{v'}} \over{4\pi}} C({\mathbf{v}, \mathbf{v'}}) = 0
\label{A10}
\end{equation}
and the scale of $\hat{C}$ is set by
\begin{equation}
 \gamma \sim g^2 T \ln  1/g 
\end{equation}
The explicit form of $ \Phi(\mathbf{v} . \mathbf{v'})$ is in Sec.\ref{sec4.2}.
The eigenvalues of the $\hat{C}$ operator are all positive, as the first term in
(\ref{A9}) dominates over the second one, except for the eigenvalue zero  of 
$\hat{C}$, i.e. Eq.(\ref{A10}).

The linearized $W$ propagator is the Green function associated 
with Eq.(\ref{A2}) i.e. the inverse operator  \cite{AY1}
\begin{equation}
\hat{G}_R(K, v) =(v.K + i  \hat{C} )^{-1}
\label{A4}
\end{equation}
When $\hat{C}$ is replaced by $\epsilon>0$, $G_R$ is the retarded Green function associated with  the drift equation. Note that this operator in $\mathbf{v}$ space should properly be written
\begin{equation}
\hat{G}_R(K, v) =(v.K \ { \mathcal{I}}+ i v_0  \hat{C} )^{-1}
\end{equation}
where $\mathcal{I}$ is the identity operator in $\mathbf{v}$ space, so that
\begin{equation}
\hat{G}_R(K,v_0,{ \mathbf{v}}) = - \hat{G}_R(K, -v_0, -{\mathbf{v}})
\end{equation}
The advanced propagator is
\begin{equation}
\hat{G}_A (K, v ) = (v.K - i \hat{C} )^{-1}
\label{A7}
\end{equation}
and one has
\begin{equation}
\hat{G}_A (K, v) = {\hat{G}_R}(K, v)^ *  \  \  \  \  \ \  \  
\hat{G}_A (K, v) = - \hat{G}_R ( - K, v)
\end{equation}

\noindent \textit{Operations in $\mathbf{v}$ space} \\
We adopt the notations of Arnold  and Yaffe  \cite{AY1}
The measure is  $\int { d{\Omega}_{\mathbf{v'}}  / (4\pi)}$ \\
$< \ \ \ \ \ >_{v}$ denotes averaging over the direction $\mathbf{v}$
\begin{eqnarray}
<\delta_{S_2}({\mathbf{v} - \mathbf{v'}})>_{v'} = 1 = < {\mathcal{I}} >_{v'} \\
\hat{C} \  W  = < C({\mathbf{v},\mathbf{v'}}) W({\mathbf{v'}}) >_{v'} \\
< v_i  \hat{C} v_j > = < v_i  C({\mathbf{v},\mathbf{v'}}) {v'}_j >_{v,v'}
\end{eqnarray}
so that $>$ represent any function independent of $\mathbf{v}$ . As stressed in \cite{AY1} an important property of $\hat{C}$ is
\begin{equation}
\hat{C} > = 0   \ \ \      { \mathrm{or}}   \ \ \ <  \hat{C} = 0
\end{equation}
from  (\ref{A10}),  with the useful consequence 
\begin{equation}
(v.K + i \hat{C})^{-1} v.K > =   (v.K + i \hat{C})^{-1} (v.K + i \hat{C})  > = \mathcal{I} > =  \ >
\label{A12}
\end{equation}
and a similar one for $(v.K - i \hat{C})^{-1}$

\subsection{The collision-resummed  polarisation tensor}
\label{sec2.2}

 Blaizot  and Iancu \cite{BI1}  have extracted from the linearized induced 
 current, the form of the polarization tensor that takes into account the 
 full effect of collisions . Defining 
\begin{equation}
W(K,{\mathbf{v}}) = i \ W^i(K,{\mathbf{v}}) E^i(K)
\end{equation}
the new functions $W^i(K,\mathbf{v})$ satisfy, from  (\ref{A2})
\begin{equation}
(v.K + i  \hat{C})\  W^i(K,{\mathbf{v}}) = v^i
\label{B2}
\end{equation}
and they obtained (see Eqs.(4.6-4.7) in \cite{BI1})
\begin{eqnarray}
\Pi^{\mu i}  & = & q_0 m_D^2  \int { d{\Omega}_{\mathbf{v}} \over{4\pi}} 
\ v^{\mu}  \ W^i(Q,\mathbf{v}) \label{B3b}  \\
\Pi^{\mu 0} & = & q^i m_D^2  \int { d{\Omega}_{\mathbf{v}} \over{4\pi}} \ 
 v^{\mu} \   W^i(Q,\mathbf{v})
\label{B3}
\end{eqnarray}
with  $m_D^2 = g^2 N T^2 /3 $. The solution of (\ref{B2}) may be written in terms of the retarded inverse operator defined in Eq. (\ref{A4})
\begin{equation}
W_R^i (K,{\mathbf{v}}) = (v.K + i \hat{C} )^{-1} v^i
\label{B4}
\end{equation}
Eq.(\ref{B3}) become
\begin{equation}
\Pi_R^{\mu i} (Q) = m_D^2 q_0  < v^{\mu} (v.Q + i \hat{C})^{-1} v^i >_{v,v'}
\end{equation}
\begin{equation}
\Pi_R^{\mu o} (Q) = m_D^2 < v^{\mu}  (v.Q + i \hat{C})^{-1} {\mathbf{v}.\mathbf{q}} >_{v,v'}
\label{B'5}
\end{equation}
With $  {\mathbf{v}.\mathbf{q}} = q_0 - v.Q $ and  Eq.(\ref{A12}), (\ref{B'5})  
becomes
\begin{equation} 
\Pi^{\mu 0} =  m_D^2 [ q_0 < v^{\mu} (v.Q + i \hat{C})^{-1} >_{v,v'} - < v^{\mu} {\mathcal {I}}>_{v,v'} ]
\end{equation}
so that $\Pi^{\mu \nu}$ may be written in the compact form
\begin{equation}
\Pi_R^{\mu \nu}(Q) = m_D^2 [\  q_0 < v^{\mu} (v.Q + i \hat{C})^{-1}  v^{\nu}>_{v,v'} -  g^{\mu 0} g^{\nu 0} \  ]
\label{B7}
\end{equation}
(\ref{B7})  explicits many properties of $\Pi^{\mu \nu}$. Indeed
$ \Pi^{\mu \nu}(Q) = \Pi^{\nu \mu}(Q) $ since $\hat{C}$ is symetric in $\mathbf{v}$ space, and one readily verifies that $ Q_{\mu} \Pi^{\mu \nu} = Q_{\nu}\Pi^{\mu \nu} = 0 $
since, again with Eq.(\ref{A12})
\begin{equation}
Q_{\mu}\Pi^{\mu \nu} = m_D^2 [ q_0 <v^{\nu} >_{v.v'} - q_0 g^{\nu 0} ] = 0
\end{equation}
Moreover, for momenta $ q_0, q >> g^2 \ln 1/g $ , one may replace the 
operator $\hat{C}$ by $\epsilon>0$, the $W$ propagator is now diagonal in 
$\mathbf{v}$ space and one recovers the well-known HTL form for $\Pi^{\mu \nu}$
\begin{equation}
\Pi_R^{\mu \nu}(Q) = m_D^2 [  \ q_0 < v^{\mu} (v.Q + i \epsilon)^{-1}  v^{\nu}>_{v,v'} -  g^{\mu 0} g^{\nu 0} \  ]
\label{B9}
\end{equation}
(\ref{B7}) interpolates between the collisionless scale $(T)^{-1}$ where 
the mean field $W$ propagates on a straight line and the scale $(gT)^{-1}$ where collisions cause fluctuations of the direction $\mathbf{v}$ of the W field.

 Another aspect of (\ref{B7}) follows. An interpretation in terms of tree 
 diagrams may be attached to the first term, i.e. to the Landau damping term. 
 The propagator along the tree is the collision-resummed $W$ propagator. 
 It carries the momentum of the incoming gluon, $v^{\mu}$ is the vertex for a 
 gluon  (polarisation $\mu$) attached to the $W$ line. 
 The W propagator is blind to colour. The retarded prescription of the 
 incoming gluon determines the retarded prescription for the $W$ propagator.
 In Sec.\ref{sec3.2} the expression (\ref{B7}) will be recovered in the
 framework of the functional derivatives of the induced current.

In Appendix \ref{secA} , an explicit form of $\Pi^{\mu \nu}$ is given, 
as a continuous fraction, and its analytic properties in the complex $q_0$ 
plane are examined.

\subsection{Collision operator and  one-loop soft exchange in $\Pi^{ji}$}

In his quest for an effective theory for the soft field modes $Q\sim g^2T$ ,   
B\"{o}deker's first step was to estimate the contribution from the one-soft-loop diagrams 
 to the polarization tensor $ \Pi^{\mu \nu}(Q) $ for loop mpmentum $K\sim gT$ 
 and $Q\sim g^2T$. \cite{Bod1,Bod2,Bod3}  Those one-loop diagrams involve 
 hard thermal loop effective vertices and propagators. 

Then higher loop diagrams with  momentum $K\sim gT$ were resummed via a 
transport equation, so that both scales $T$ and $gT$ were integrated out  
\cite{Bod1,Bod2}. This transport equation is in terms of the field 
$W(K,\mathbf{v})$ that describes the collective excitations of the hard
particles.  Fluctuations in $\mathbf{v}$ are caused by collisions and taken 
into account via the operator $\hat{C}({\mathbf{v}, \mathbf{v}'})$. 
The linearized transport equation may be written (see Eq.(\ref{B2}))
\begin{equation}
(v.K + i \hat{C}) \  W^i(K,{\mathbf{v}}) =v^i
\label{D1}
\end{equation}
while the integration over the scale $T$ only, lead to the linearized transport equation \cite{BI3}
\begin{equation}
v.K \ \ W^i = v^i
\end{equation}
The complete solution to the transport equation (\ref{D1}) involves the inverse operator $(v.K+i\hat{C})^{-1}$
\begin{equation}
W_R^i(K,v)=(v.K+i\hat{C})^{-1} v^i
\end{equation}
As recently stressed by Blaizot and Iancu \cite{BI1}, alternatively one may 
solve the transport equation (\ref{D1}) by iteration. One obtains a series 
expansion in powers of $\hat{C}$ whose first terms reproduce the expansion in 
the number of loops with momentum $K\sim gT$. Indeed, writing Eq.(\ref{D1})
\begin{equation}
v.K \ W^i = v^i - i \hat{C}\  W^i
\end{equation}
one obtains
\begin{equation}
W_R^{i(0)} ={v^i\over{v.K+i\epsilon}}   \ \ \  , \ \ \ \epsilon>0
\end{equation}
\begin{equation}
W_R^{i(1)}={1\over{v.K+i\epsilon}} (-i)\hat{C} {v^i\over{v.K+i\epsilon}}=
 {1\over{v.K+i\epsilon}}<(-i) C({\mathbf{v}},{\mathbf{v}'}){v'^i\over{v'.K+i\epsilon}}>_{v'}
\end{equation}
Since the polarization tensor is simply related to the $W^i$ field 
(see Eq. (\ref{B3b}))
\begin{equation}
\Pi^{j i}(Q) = q_0 m_D^2<v^j \ W^i(Q,{\mathbf{v}})>_v 
\end{equation}
one obtains
\begin{equation}
\Pi^{j i  (0)}(Q) =q_0 m_D^2 <v^j {1\over{v.Q+i\epsilon}}v^i>_v
\end{equation}
\begin{eqnarray}
\Pi^{j i (1)}(Q) &  = &  q_0 m_D^2  <{v^j \over{v.Q+i\epsilon}} \  (-i)\hat{C} \  {v^j \over{v.Q+i\epsilon}}>_{v,v'} \nonumber \\
 &  = &  q_0 m_D^2  <{v^j \over{v.Q+i\epsilon}} (-i) 
 C({\mathbf{v}},{\mathbf{v}}') {v'^j \over{v'.Q+i\epsilon}}>_{v,v'}
\label{D9}
\end{eqnarray}
$\Pi^{j i  (0)}$ is the hard thermal loop contribution to the polarization 
tensor, $\Pi^{j i  (1)}$ is the one-loop soft momentum contribution  
for $K\sim gT$  with $C(\mathbf{v},\mathbf{v}')$ as in (\ref{A9}) 
\cite{BI1,Bod3}.  The conclusion is, one can identify the collision operator $\hat{C}$ if $\Pi^{j i  (1)}$ is known and appears under the form (\ref{D9}).

We shall follow this strategy. We want to integrate out the momenta 
$K \sim g^2T \ln 1/g$ to see how they affect the dynamics of the soft fields 
$Q \sim g^2 T$. We shall assume $\ln 1/g >>1$ so that the scales are well 
separated.  If one assumes that the summation over the scale $g^2 T \ln 1/g$ 
can be made via a transport equation  which involves a collision operator 
$\hat{C}'$, i.e.
\begin{equation}
(v.K + i \hat{C} + i \hat{C}') W^i(K, {\mathbf{v}}) = v^i
\label{D10}
\end{equation}
the full solution will be in terms of an inverse operator
\begin{equation}
W^i = (v.K + i \hat{C} + i \hat{C}')^{-1} v^i
\label{D11}
\end{equation}
and the iterative solution will be
\begin{equation} 
W^{i (0)} = (v.K + i \hat{C} )^{-1} v^i
\end{equation}
\begin{equation}
W^{i (1)} = (v.K + i \hat{C} )^{-1}  (-i) \hat{C}' (v.K + i \hat{C} )^{-1} v^i 
\end{equation}
so that
\begin{equation}
{\Pi'}^{j i  (0)}(Q)= q_0 m_D^2 <v^j (v.Q +i \hat{C})^{-1} v^i>_{v,v'}
\end{equation}
\begin{equation}
{\Pi'}^{j i  (1)}(Q)= q_0 m_D^2 <v^j (v.Q +i \hat{C})^{-1} (-i)\hat{C}' 
(v.Q +i \hat{C})^{-1} v^i>_{all\  v}
\label{D15}
\end{equation}
 ${\Pi'}^{j i  (0)}$ is the form found in Eq. (\ref{B7}) for the polarization 
 tensor, when the scales  $T$ and $gT$ have been integrated out. 
 ${\Pi'}^{j i  (1)}$ will correspond to the one-loop soft momentum exchange 
 $K\sim g^2T \ln 1/g$, where the one-loop diagrams involve effective vertices 
 and resummed propagators with the scales $T$ and $gT$ integrated out.

The effective vertices will be  constructed in Sec.\ref{sec3}. The one-loop 
soft momentum exchange $K\sim g^2T \ln 1/g$ contribution to the polarization 
tensor  ${\Pi'}^{j i  (1)}$ will be computed in Sec.\ref{sec4}, it will indeed 
show up as in (\ref{D15}). This fact will allow to identify the operator 
$\hat{C}'$.

The expression for  $C(\mathbf{v},\mathbf{v}') $ involves by construction an 
integration over the momentum $K\sim gT$. So does  $C'(\mathbf{v},\mathbf{v}')$
 with momentum $K \sim g^2T \ln 1/g$. There wont be any over-counting in 
 $ \hat{C} + \hat{C}' $ in Eqs.(\ref{D10},\ref{D11}) if one limits the integration range over the space momentum $k$
\begin{eqnarray}
\lefteqn{\mu_2 < k < \mu_1\  \ {\mathrm{for}}\ \  \hat{C}  \ \ \ \  \ \ \   \mu_3 < k < \mu_2 \ \ {\mathrm{for}}\ \  \hat{C}'} \label{D14}\\  &  &
  gT<\mu_1 < T   \ \ \ , \ \ \   g^2T \ln 1/g  < \mu_2 < gT    \ \ \ , \ \ \   g^2 T < \mu_3 < g^2  T\ln 1/g  \nonumber
\end{eqnarray}

\section{The effectives vertices}
\label{sec3}
In the response approach, a weak external perturbation is applied to the plasma
at some scale $x$ with an adiabatic switching-on at time $-\infty$ . The
response of the plasma is an induced current. The $n$-point soft amplitudes are
obtained as functional derivatives of this current with respect to the mean
field. The response approach involves the Retarded 2-point Green function (the
appropriate one to study the response of the plasma to a perturbation which
vanishes at time $-\infty$) and it gives out the $n$-point Retarded amplitudes
$V( X_1 ; X_2 ....X_n )$ where $X_1^0$ is the largest time 
(see Sec. \ref{sec3.2}).
In momentum space, the 2-point Retarded function is the analytical continuation
into the upper complex $p_0$ plane of the Imaginary Time (IT) Green function,
i.e. one approaches the real energy axis from above $G(p_0+i\epsilon) = G_{ret}$
. The $n$-point Retarded amplitude is the analytical continuation $p_i^0 + i
\epsilon_i$ of the IT amplitude such that $\epsilon_1 > 0$ and all other 
$\epsilon_i < 0$. (The constraint $\sum_i p_i^0 = 0 $ is imposed on any
analytical continuation into the complex energy variables $p_i^0$, in particular
 $\sum_{i=1}^n \epsilon_i =0$ ) \cite{Ev} . \\
 Since the known amplitudes will be the Retarded ones, it is natural to use the
 Retarded/Advanced formalism in computing the two one-soft-loop diagrams that
 contribute to $\Pi_{\mu \nu}$. Are needed the propagator, the 3-point and
 4-point effective vertices. We summarize first the features of this formalism
 as they will be needed later on, in particular for the 4-point vertex.
 
 \subsection{A summary of the R/A formalism}
 \label{sec3.1}
 It is a Real Time formalism where the propagators are the Retarded and
 Advanced propagators of the zero temperature Field Theory. General properties
 have been established \cite{Au1,Au2,vW}. An external leg $l$ may be of type $R$
 or of type $A$, it incoming energy is $p_l^0 +i\epsilon_l , \epsilon_l >0$ type
 $R$, $\epsilon_l < 0$ type $A$. The $n$-point amplitudes are defined with all
 momenta incoming  $$\sum_{l=1}^n p_l^0 = 0  \ \ \ \ {\mathrm{and}}\ \  \sum_{l=1}^n
 \epsilon_l =0 $$
 Then the only non-zero three-point functions have two legs of type $R$ and one
 leg of type $A$, or two legs of type $A$ and one leg of type $R$.
  The general properties are, for bosons
\begin{equation}
V (P_{1 R}, P_{2 A}, P_{3 A} )  =  V ^* (P_{1 A}, P_{2 R}, P_{3 R})
\label{C1}
\end{equation}
and for massless bosonic fields in the QED/QCD case  \cite{Au2}
\begin{equation}
V ( (-P_1)_A, (-P_2)_R, (-P_3)_R)  =   - V ( P_{1 R}, P_{2 A}, P_{3 A})
\label{C2}
\end{equation}
With the same notation, the two-point function, should be written
\begin{equation}
\Pi_R(P)  \equiv \Pi(P_R) = \Pi ( P_R, (-P)_A)
\end{equation}
and the relations analogous to (\ref{C1},\ref{C2}) are
\begin{eqnarray}
\Pi( P_A) &=& \Pi^* (P_R)
\label{C4} \\
\Pi((-P)_R) &=& \Pi ( P_A)
\label{C5}
\end{eqnarray}
Our results will obey those general properties. See (\ref{B7}) for 
$\Pi^{\mu \nu}$.  \\
The $n$-point Retarded functions that appear in the response approach are
$V(P_{1R}, P_{2A}, \ldots P_{nA})$ \cite{Ev}, they are related by complex 
conjugation to the Advanced ones
\begin{equation}
V(P_{1R}, P_{2A},  \ldots P_{nA}) = V^* (P_{1A}, P_{2R}, \ldots P_{nR})
\end{equation}
When constructing a diagram, because any $n$-point vertex is defined with all
 incoming momenta, a propagator joins an $R$ leg to an $A$ leg since 
 a $P_R(p_0+i\epsilon,\bf{p})$ incoming a vertex comes from another vertex 
 with incoming  momentum $(-p_0-i\epsilon,-\bf{p})$ i.e.$ (-P)_A$. 
 One can define an $\epsilon$-flow along any tree diagram. This
flow is incoming the $R$ legs and outgoing the $A$ legs, it obeys the rules 
of an electric current since there is no source or sink at a vertex 
(where $\sum_i\epsilon_i=0$).

In this R/A formalism, the thermal Bose weight $n(p_0)$ that are associated 
with the loop momenta in the Imaginary Time formalism, are carried by the vertex 
in a very specific way, they are attached to the vertices that possess two 
(or more) legs of type $A$
\begin{equation}
\Gamma(P_{1R}, P_{2R}, P_{3A}) = V (P_{1R}, P_{2R}, P_{3A})
\label{RC6}
\end{equation}
\begin{equation}
\Gamma(P_{1R}, P_{2A}, P_{3A}) =  - V (P_{1R}, P_{2A}, P_{3A}) \ \  N(P_2, P_3)
\label{RC7}
\end{equation}
with
\begin{equation}
N(P_2,P_3) = n(p_2^0)+n(p_3^0) + 1 = {n(p_2^0) n(p_3^0) \over {n(p_2^0 + p_3^0)}}
\label{RC8}
\end{equation}
\begin{equation}
N(P, -P) = n(p^0) + n(-p^0) + 1 = 0
\label{RC9}
\end{equation}
For three legs of type $A$
\begin{equation}
\Gamma (P_{1R}, P_{2A}, P_{3A}, P_{4A}) = V (P_{1R}, P_{2A}, P_{3A}, P_{4A})\ \
{\cal{N}}(P_2, P_3, P_4)
\label{RC10}
\end{equation}
with
\begin{equation}
{\cal{N}}(P_2, P_3, P_4) = {n(p_2^0) n(p_3^0) n(p_4^0)
\over {n(p_2^0 + p_3^0+p_4^0})} 
= N(P_2,P_3) N(P_2 + P_3 , P_4)
\label{RC11}
\end{equation}
In order to build one of the pair of one-loop diagrams contributing to the
self-energy, a 4-point amplitude with two legs of type R and two legs of type A 
is needed. Indeed a loop has to be made by joining an R leg to an A leg after
the insertion of a propagator. That amplitude possesses a new feature, the
prescription $p_i^0 + i\epsilon_i$ on the external legs does not constrain all
the energy variables of the amplitude \cite{Ev,FG1}. For the case
$\epsilon_1>0, \epsilon_2>0, \epsilon_3<0, \epsilon_4<0$, 
 $(\epsilon_1 +\epsilon_2 + \epsilon_3 + \epsilon_4 =0)$, the energy variable
 $p_1^0+p_3^0 = -(p_2^0+p_4^0) $ may have $ \epsilon_1 + \epsilon_3 >0 $ or $<0$.
 The same for the variable $p_1^0 + p_4^0$. As a result there are, for this
 case,
 four distinct analytical continuations of the Imaginary time amplitude, known
 as Generalized-Retarded amplitudes and they are linked by one relation (see
 references in \cite{FG1})
 \begin{eqnarray}
 \lefteqn{V^{(4)}(\epsilon_1+\epsilon_3>0, \epsilon_1+\epsilon_4>0) +
 V^{(4)}(\epsilon_1+\epsilon_3<0, \epsilon_1+\epsilon_4<0)}\nonumber \\
 & & =
 V^{(4)}(\epsilon_1+\epsilon_3>0, \epsilon_1+\epsilon_4<0)+
 V^{(4)}(\epsilon_1+\epsilon_3<0, \epsilon_1+\epsilon_4>0) 
 \end{eqnarray}
 The 4-point amplitude RRAA that occurs in the R/A formalism is a specific
 combination of those analytic continuations \cite{FG1,Au2}. The rules are
 easily stated when the amplitude is a sum of tree diagrams, the only case to be
 encountered in this work. One splits the tree diagrams into three groups, which
 depend respectively on $P_1+P_2, P_1+P_3, P_1+P_4$ and
 \begin {eqnarray}
 \lefteqn{- \Gamma(P_{1R}, P_{2R}, P_{3A}, P_{4A}) = F_1((P_1+P_2)_R) 
 N(P_3, P_4)}
 \nonumber\\
 & &+ F_2((P_1+P_3)_R) N(P_3, P_2+P_4) +F_2((P_1+P_3)_A) N(P_4, P_1+P_3)
 \nonumber\\
 & &+ F_3((P_1+P_4)_R) N(P_4, P_2+P_3) +F_3((P_1+P_4)_A) N(P_3, P_1+P_4)
 \label{RC12}
 \end{eqnarray}
 From Eqs. (\ref{RC8}, \ref{RC9})
 \begin{equation}
 N(P_3, P_2+P_4) + N(P_4, P_1+P_3) = N(P_3, P_4) = N(P_4, P_2+P_3)
+ N(P_3, P_1+P_4) 
\label{RC13}
\end{equation}
Those weights are compatible with the complex conjugate relation \cite{vW}
\begin{equation}
{\Gamma(P_{1R}, P_{2R}, P_{3A}, P_{4A}) \over{\Gamma^*(P_{1A}, P_{2A}, P_{3R},
P_{4R})}} = { N(P_3, P_4) \over{ N(P_1, P_2)}}
\label{RC14}
\end{equation}
indeed
\begin{equation}
{N(P_3, P_4)\over{N(P_1, P_2)}} = -{n(p_3^0) n(p_4^0) \over {n(-p_1^0)
n(-p_2^0)}} = {N(P_3, P_2+P_4)\over{N(P_2, P_1+ P_3)}} = 
{N(P_4, P_1 +P_3)\over{N(P_1, P_2+P_4)}} = \cdots
\label{RC15b}
\end{equation}
as may be shown by writing $N(P_i, P_j) = -N(-P_i, -P_j)$ in each denominator.

\subsection{Soft Amplitudes from the induced current}
\label{sec3.2}

The soft amplitudes will be obtained as functional derivative of the induced
current with respect to the mean field in a way that closely parallels the case
of the HTL amplitudes (see Sec.5 of \cite{BI3}). One set of Ward Identities
satisfied by the amplitudes will follow from the conservation law obeyed by the
induced current.

It is convenient to introduce the Retarded Green function associated with the
transport equation (\ref{RA5}) for the $W(X, {\bf v})$ field.
\begin{eqnarray}
i\ (v.D_x + \hat{C}) \ G_{ret}( X, Y ;{\bf v}, {\bf v'}) & = &
 \delta^{(4)}(X-Y) \ \delta_{S_2} ({\bf v} - {\bf v'})  \label{RC15}\\
 G_{ret} ( X, Y ;{\bf v}, {\bf v'}) & = & 0 \ \ {\mathrm for} \ \ X_0 < Y_0
 \nonumber
 \end{eqnarray}
 with
 \begin{equation}
 \hat{C} \ G_{ret}( X, Y ;{\bf v}, {\bf v'}) = \int {d\Omega_{\bf v"}\over 4\pi}
 \ C({\bf v}, {\bf v"}) \ G_{ret} (X, Y; {\bf v"}, {\bf v'})
 \label{RC16}
 \end{equation}
 For the case $ \hat{C} = 0 $, the solution is known \cite{BI3} in terms of the
 parallel transporter along a straight line
 \begin{eqnarray}
 G_{C=0} (X, Y; {\bf v}) &=& -i \int_0^{\infty} d u \ \delta^{(4)}( X-Y-vu) 
 \ U(X,Y) \label{RC17} \\
 U(X, X-vu) &=& P\ \lbrace \exp{-i g u \int_0^1 d s \ v.A(X-vu(1-s))} \rbrace
 \nonumber
 \end{eqnarray}
 where $P$ is the path-ordering operator. An iterative solution to Eq.(\ref{RC15}) may be written as a power series in $\hat{C}$ in terms of
 $G_{C=0}$. As a consequence, the following identity, true for $G_{C=0}
 \delta_{S_2}({\bf v}-{\bf v'})$, also holds for 
 $G_{ret}( X, Y ;{\bf v}, {\bf v'})$
 \begin{equation}
 {\partial G_{ret}( X, Y ;{\bf v}, {\bf v'}) \over{\partial A_c^{\mu}(X_1)}} =
 \int {d\Omega_{\bf v"}\over 4\pi} \ G_{ret}( X, X_1 ;{\bf v}, {\bf v"}) \ g T^c
 {v"}_{\mu} \ G_{ret}( X_1, Y ;{\bf v"}, {\bf v'})
 \label{RC18}
 \end{equation}
 where the time arguments satisfy $X^0 \geq X_1^0 \geq Y^0$ and $T^c$ is in
 the adjoint representation. This identity is of central importance to the
 structure of the amplitudes.
 
 The solution to the transport equation (\ref{RA5}) may be written
 \begin{equation}
 W^a(X, {\bf v})= i \int d^4Y \int{d\Omega_{\bf v'}\over 4\pi} 
 G_{ret}^{ab}( X, Y ;{\bf v}, {\bf v'})\ {\bf v'}.{\bf E}^b(Y)
 \label{RC19}
 \end{equation}
 and from Eq.(\ref{RA2}) the induced current become
 \begin{equation}
 j_{\mu}^{a\  ind}(X) = i m_D^2 \int{d\Omega_{\bf v}\over 4\pi} 
 \int{d\Omega_{\bf v'}\over 4\pi} \int d^4Y \ \ v_{\mu} \ 
  G_{ret}^{ab}( X, Y ;{\bf v}, {\bf v'})\ {\bf v'}.{\bf E}^b(Y)
  \label{RC20}
  \end{equation}
 Writing \cite{BI3}
 \begin{equation}
 {\bf v'}.{\bf E}(Y) = v'^i F_{0i}(Y) = \partial_{Y_O}(v'^{\mu} A_{\mu}) -
 v'.D_Y \ A_0(Y)
 \label{RC20b}
 \end{equation}
 Eq.(\ref{RC20}) may be transformed into a form similar to the case of the HTL
 amplitudes
 \begin{equation}
 j_{\mu}^{a\ ind}(X) = - m_D^2 g_{0\mu} A_0^a(X) + i m_D^2 \int{d\Omega_{\bf v}
 \over 4\pi} \int{d\Omega_{\bf v'}\over 4\pi} \int d^4Y \ v_{\mu} \ 
  G_{ret}^{ab}( X, Y ;{\bf v}, {\bf v'})\  \partial_{Y_0} (v'.A^b(Y))
  \label{RC21}
  \end{equation}
  Indeed, for the second term on the right handside of (\ref{RC20b}) $v'.D_Y$ is
  allowed to act on the left on $G_{ret}( X, Y ;{\bf v}, {\bf v'})$; with the
  use of (\ref{RC15}), it brings the first term on the right handside of 
  (\ref{RC21}), the
  term $G_{ret}( X, Y ;{\bf v}, {\bf v"}) C({\bf v"}, {\bf v'}) $ does not
  contribute as the integration over ${\bf v'}$ makes use of the property
  (\ref{A10}) of $\hat{C}$.
  
  The soft one-particle-irreducible amplitudes are then obtained
  \begin{equation}
  \Pi_{\mu\nu}^{ab}(X, Y) ={\partial j_{\mu}^{a\ ind}(X) 
  \over{\partial A_b^{\nu}(Y)}} \  \vert_{A=0}
  \label{RC22}
 \end{equation}
 and for the $n+1$-gluon amplitude 
 \begin{equation}
 g^n \ V_{\mu \mu_1 \cdots \mu_n}^{a a_1 \cdots a_n}(X ; X_1 \cdots X_n) = 
 {\partial^n \over{\partial A_{a_n}^{\mu_n}(X_n) \cdots
 \partial A_{a_1}^{\mu_1}(X_1)}} \ \ j_{\mu}^a(X) \ \vert_{A=0}
 \label{RC23}
 \end{equation}
 For all these amplitudes, $X^0$ is the largest time. After differentiation, the
 Green function that will enter the amplitudes correspond to the case $A=0$, i.
 e. to
 \begin{equation}
 i (v.\partial_X + \hat{C}) \ G_{ret,A=0}( X, Y ;{\bf v}, {\bf v'}) = 
 \delta^{(4)}(X-Y) \ \delta_{S_2}({\bf v} - {\bf v'})
 \label{RC24}
 \end{equation}
 or in momentum space
 \begin{equation}
 (v.P + i \hat{C})\ G_{ret}(P;{\bf v}, {\bf v'}) ={\cal I} \ \ \ {\mathrm i.e.}\ 
  \ \ G_{ret}(P;{\bf v}, {\bf v'}) = (v.P + i \hat{C})^{-1}
 \label{RC25}
 \end{equation}
 This is the Green function introduced in Sec.\ref{sec2.1} (see Eq.(\ref{A4}))
 and called the $W$ propagator.
 
 Moreover, from Eq.(\ref{RA1}), the induced current obeys the conservation law
 \begin{equation}
 D^{\mu} j_{\mu}^{a\ ind}(X) = 0
 \label{RC26}
 \end{equation}
 Writing $ D^{\mu} = \partial^{\mu}\delta^{ab} + g \ f^{abc} A_c^{\mu} $ and
 formally
 \begin{equation}
 j_{\mu}^{a\ ind} = \Pi_{\mu\nu}^{ab} A_b^{\nu} + {g \over 2!}V_{\mu\nu\rho}^{a
 a_2a_3} A_{a_2}^{\nu} A_{a_3}^{\rho} + {g^2\over 3!} V_{\mu\nu\rho\sigma}^{a
 a_2a_3a_4} A_{a_2}^{\nu} A_{a_3}^{\rho} A_{a_4}^{\sigma} + \cdots
 \label{RC28}
 \end{equation}
 The term in $g^{n-2}$ of (\ref{RC26}) gives a Ward identity for the $X$ variable,
 relating the $n$-point amplitude to $n-1$-point amplitudes.
 
 For $\Pi^{ab}_{\mu\nu}(X)$, the functional derivative (\ref{RC22}) acting on
 the first term in (\ref{RC21}) gives 
 $$-m_D^2 \delta^{ab}g_{0\mu}g_{0\nu} \delta^{(4)}(X-Y)$$
 In the second term, the derivative has to act on $\partial_{Y_0} v'.A^b(Y)$. In
 momentum space, one gets back the expression (\ref{B7}) obtained in 
 Sec.\ref{sec2.2}.
 
 From the properties of the functional derivatives and from the identity
 (\ref{RC18}) for the Green function, follow the consequences:\\
 - the $n+1$-point amplitude is symmetric in the exchange of any pair of the 
 legs $X_1, X_2, \cdots X_n$ \\
 - a functional derivative inserts a vertex along a propagator at an
 intermediate time \\
 - the order in colour space and in $\bf v$ space follows the time order \\
 - a diagram will correspond to a specific time ordering and there will be a sum
 over all chronological orderings of $X_1^0, X_2^0, \cdots X_n^0$. $X^0$ is the
 largest time, a characteristic feature of the $n+1$-Retarded amplitude \\
 - the form has the minimum number of colour $T$ matrices in the adjoint
 representation, none for $\Pi_{\mu\nu}$, one for $V_{\mu\nu\rho}$, two for
 $V_{\mu\nu\rho\sigma}$ \\
 - one goes back to the corresponding HTL amplitudes by replacing $\hat{C}$ by
 $\epsilon{\cal I}$ where ${\cal I}$ is the identity in ${\bf v}$ space.\\
 The Fourier transform of an $n$-point amplitude is defined
 \begin{eqnarray}
 \lefteqn{(2\pi)^4\delta(P_1+P_2+\cdots +P_n) V(P_1, P_2, \cdots P_n) =}
 \nonumber \\ & &
 \int d_4X_1 \cdots
 d_4X_n \ \exp{i(P_1X_1+\cdots+P_nX_n)} \ V(X_1, X_2 \cdots X_n)
 \end{eqnarray}
 all the momenta are incoming the amplitude and the colour indices will 
 be written $a_1=1, \ a_2=2, \cdots$. 
 
 All the properties are consequences of the transport equation for the $W$
 field. The transport equation (\ref{RA5}) has been established in the Coulomb
 gauge, it is expected to be gauge independent \cite{BI1}. At the HTL level,
 there is no induced current for the ghost field, i.e. no effective vertices. It
 is likely to be the same for the soft amplitudes.
 
 \subsection{The 3-gluon vertex}
 \label{sec3.3}
 The result of the functional differentiation (\ref{RC23}) on Eq.(\ref{RC21}) is
 in momentum space
 \begin{eqnarray}
 \lefteqn{V_{\mu\nu\rho}^{123}(P_{1R},P_{2A},P_{3A})
 =im_D^2\lbrace }\label{RC30}\\ & &
f^{123}<v_{\mu}(v.P_1+i\hat{C})^{-1}{v'}_{\nu}(v.(P_1+P_2)+i\hat{C})^{-1}
 {v"}_{\rho} (-p_3^0)>_{vv'v"}\nonumber \\
 & & +f^{132}<v_{\mu}(v.P_1+i\hat{C})^{-1}{v'}_{\rho}(v.(P_1+P_3)+i\hat{C})^{-1}
 {v"}_{\nu}(-p_2^0)>_{vv'v"} \rbrace\nonumber
 \end{eqnarray}
 \begin{eqnarray}
 \lefteqn{V_{\mu\nu\rho}^{123}(P_{1R},P_{2A},P_{3A})
 =im_D^2 f^{123} <v_{\mu}(v.P_1+i\hat{C})^{-1} } \label{RC31} \\ & &
 [\ {v'}_{\nu}(-v.P_3+i\hat{C})^{-1}{v"}_{\rho}(-p_3^0)-
 {v'}_{\rho}(-v.P_2+i\hat{C})^{-1}{v"}_{\nu}(-p_2^0) \ ] >_{vv'v"}
\nonumber
\end{eqnarray}
The first term in (\ref{RC30}) corresponds to the time ordering
$X_1^0>X_2^0>X_3^0$, the second term to the time ordering $X_1^0>X_3^0>X_2^0$.
The amplitude is symmetric in the exchange $2\leftrightarrow3$ of all indices
(momentum, Lorentz, colour), it has no other symmetry. The 3-point amplitude is a
Retarded one, on the left handside of (\ref{RC30}) each momentum has been
given an index R or A according to the rule stated in Sec \ref{sec3.1}. It is
worth noticing that the energy $p_i^0$ that enters the numerator in each term of
(\ref{RC31}) is the one associated with the earliest time, a consequence of the
expression (\ref{RC21}) for the induced current. One may reverse the whole
string of operators in $\bf v$ space. When $\hat{C}$ is replaced 
by $\epsilon\cal{I}$, one recovers Eq.(5.22) of \cite{BI3}.

The Ward identity stemming from the conserved current (\ref{RC26}) writes

\begin{equation}
p_1^{\mu}\ V_{\mu\nu\rho}(P_{1R},P_{2A},P_{3A})=
if^{123}[\Pi_{\nu\rho}((P_1+P_2)_R,P_{3A})-\Pi_{\nu\rho}((P_1+P_3)_R,P_{2A})]
\label{RC32}
\end{equation}
It is obeyed by the expression (\ref{RC31}) in a straightforward way since
from Eq.(\ref{A12}) $$<v.P_1(v.P_1+i\hat{C})^{-1} =  <\cal{I} $$
Moreover there is a Ward identity with respect to $P_2$ (or $P_3$). Indeed,
 in (\ref{RC30}) when $v_{\nu}$ sits in the middle of a string of operators
in $\bf v$ space, one writes
$$ v.P_2 = (v.(P_1+P_2)+i\hat{C}) - (v.P_1+i\hat{C})$$
so that
\begin{equation}
(v.P_1+i\hat{C})^{-1}v.P_2\ (v.(P_1+P_2)+i\hat{C})^{-1}= 
(v.P_1+i\hat{C})^{-1} -(v.(P_1+P_2)+i\hat{C})^{-1}
\label{RC33}
\end{equation}
When $v_{\nu}$ sits at the end of a string one uses
$$ (-v.P_2+i\hat{C})^{-1}v.P_2> = -{\cal I} >$$ and one gets from (\ref{RC30})
\begin{eqnarray}
\lefteqn{p_2^{\nu}\ V^{123}_{\mu\nu\rho}=i f^{123}m_D^2 \lbrace }\label{RC34} \\ & & 
{-<v_{\mu}(v.(P_1+P_2)+i\hat{C})^{-1}v_{\rho}>(-p_3^0) +
<v_{\mu}(v.P_1+i\hat{C})^{-1}v_{\rho}>(-p_3^0-p_2^0)}\rbrace \nonumber
\end{eqnarray}
\begin{equation}
p_2^{\nu}V^{123}_{\mu\nu\rho}=i f^{123} [\Pi_{\mu\rho}(P_{1R},(P_2+P_3)_A)-
\Pi_{\mu\rho}((P_1+P_2)_R, P_{3A})]
\label{RC35}
\end{equation}
To summarize, $V^{123}_{\mu\nu\rho}$ obeys tree-like Ward identities with
respect to all legs, and this property will generalize to the amplitudes $V(X;
X_1, X_2, \cdots X_n)$ although the conservation law for the current enforces it
only for the $X$ leg (see the 4-point vertex in the next Sec.). The reason is,
the expressions are associated with forward-scattering tree diagrams of a
collective excitation, i.e. of the $W$ field. If one thinks about the complex
conjugate amplitude, where $X^0$ is the earliest time, one may interpret 
Eq.(\ref{RC30}) as follows. The incoming soft gluon $P_1$ excites a collective
excitation of momentum $P_1$ which propagates, emits successively the soft
gluons $(-P_2)$ and $(-P_3)$ and disappears at the last emission. This
interpretation is valid at the HTL level as an alternative to the hard-loop
interpretation. When the scale $gT$ is integrated out, the new feature is that
it is the only surviving one. In contrast, in the one-loop interpretation, there
is a symmetry with respect to all legs, so that the leg $P_{1R}$ could sit in
the middle (rather than sitting first), there would be a fork in the
$\epsilon$-flow (an allowed fact) and one can easily check that the Ward
identity (\ref{RC32}) stemming from the conserved current would not be satisfied
when $\epsilon$ is replaced by $\hat{C}$. It is also apparent on the one-loop 
4-point function.

\subsection{Four-gluon Vertices}
\subsubsection{The vertex RAAA}

For the 4-point amplitude, the result of the functional differentiation is
 \begin{eqnarray}
 \lefteqn{V_{\mu\nu\rho\sigma}^{1234}(P_{1R},P_{2A},P_{3A},P_{4A})
 =m_D^2 <v_{\mu}(v.P_1+i\hat{C})^{-1} } \label{RC36} \\ & &
 [f^{12m}f^{34m}{v}_{\nu}(v.(P_1+P_2)+i\hat{C})^{-1}v_{\rho}
 (-v.P_4+i\hat{C})^{-1}{v}_{\sigma}(-p_4^0) + \ \ 5 \ \ {\mathrm terms} \ \ ]
 >_{all\ v} \nonumber
\end{eqnarray}
where $+ 5$ terms \ mean the addition of the permutations $i\leftrightarrow j$
(in all indices) in any pair among  the legs $2, 3, 4 $. The written term corresponds to the
time order $X_1^0 > X_2^0 > X_3^0 > X_4^0 $, the order in colour indices and
Lorentz indices follows the order in time. An interpretation may be given in
terms of tree diagrams, the gluons $(-P_2), (-P_3), (-P_4)$ are emitted by the
collective excitation induced by the gluon $P_1$. This 4-gluon amplitude's form
was written in \cite{Bod3} for the case $\hat{C}=0$,  i.e. for the HTL case in the
Imaginary Time formalism. \\ 
The following tree-like Ward
identities are satisfied
\begin{eqnarray}
 \lefteqn{i p_1^{\mu} \ V_{\mu\nu\rho\sigma}^{1234}(P_{1R},P_{2A},P_{3A},P_{4A})
=} \label{RC37} \\ & &
f^{12m}V_{\nu\rho\sigma}^{m34}((P_1+P_2)_R,P_{3A},P_{4A}) +f^{13m} 
V_{\rho\nu\sigma}^{m24}((P_1+P_3)_R,P_{2A},P_{4A}) \nonumber \\ & &
+f^{14m} V_{\sigma\nu\rho}^{m23}((P_1+P_4)_R,P_{2A},P_{3A}) \nonumber
\end{eqnarray}
a consequence of the conserved current, and for the leg $P_4$ (or $P_3$, or
$P_2$)
\begin{eqnarray}
 \lefteqn{i p_4^{\sigma} \ V_{\mu\nu\rho\sigma}^{1234}(P_{1R},P_{2A},P_{3A},P_{4A})
=} \label{RC38} \\ & &
f^{41m}V_{\mu\nu\rho}^{m23}((P_1+P_4)_R,P_{2A},P_{3A}) +f^{42m} 
V_{\mu\nu\rho}^{1m3}(P_{1R},(P_2+P_4)_A,P_{3A}) \nonumber \\ & &
+f^{43m} V_{\mu\nu\rho}^{12m}(P_{1R},P_{2A},(P_3+P_4)_A) \nonumber
\end{eqnarray}
in a way similar to the case of the 3-point vertex and with the use of the
Jacobi identity
\begin{equation}
f^{12m}f^{34m}\ + \ f^{42m}f^{13m} \ + \ f^{32m}f^{41m} \ = \ 0
\label{RC39}
\end{equation} 
It has to be remembered that the 3-point vertex 
$V_{\mu\nu\rho}^{123}(P_{1R},P_{2A},P_{3A})$ as given by (\ref{RC31}) is
symmetric with respect to the two legs A, and that the R leg sits first in
$\bf v$ space.

It is worth emphasizing the origin of the tree-like structure of the $n$-point
Retarded amplitudes. The existence of the covariant derivative is the only
source of non-linearity in $A(X)$ because the mean gauge field enters linearly in the
induced current in Eq. (\ref{RC21}), just as in the HTL case.
 
 \subsubsection{The vertex RRAA}
 What enters a one-loop self energy diagram is a 4-point function of the type
 RRAA; indeed a loop has to be maid by joining an R leg to an A leg. The 4-point
 function just obtained is of the type RAAA (and its complex conjugate ARRR).
 For the amplitudes of the type RRAA, as explained in Sec.\ref{sec3.1}, a new
 feature appears. The prescription $\epsilon_1>0,\epsilon_2>0,\epsilon_3<0,
 \epsilon_4<0 $ allows for the energy variable $p_1^0 + p_3^0 = -(p_2^0 +
 p_4^0)$, $\epsilon_1 + \epsilon_3$ to be either $>0$ or $<0$, and one has to
 sum over the two possible analytical continuations of the Imaginary Time
 amplitude in that variable with appropriate weights (see Eq.(\ref{RC12})). The
 same for the variable $p_1^0 + p_4^0 = -(p_2^0 + p_3^0)$. A consequence is that
 the Ward identities take forms somewhat different from the ones just written.
 For example, instead of Eq.(\ref{RC37}), one has 
\begin{eqnarray}
\lefteqn{i p_1^{\mu} \ V_{\mu\nu\rho\sigma}^{1234}(P_{1R},P_{2R},P_{3A},P_{4A})
= 
f^{12m}V_{\nu\rho\sigma}^{m34}((P_1+P_2)_R,P_{3A},P_{4A})} \nonumber \\ & &
+ f^{13m} [\  {N(P_3, P_2+P_4) \over{N(P_3,P_4)}} 
V_{\nu\rho\sigma}^{2m4}(P_{2R},(P_1+P_3)_R,P_{4A}) \nonumber \\ & &
+ {N(P_4, P_1+P_3)\over{N(P_3,P_4)}}
V_{\nu\rho\sigma}^{2m4}(P_{2R},(P_1+P_3)_A,P_{4A}) \ ] \nonumber \\ & &  
+ f^{14m}[\ 3\leftrightarrow 4  \ ] \label{RC40}
\end{eqnarray}
where $[ \ 3\leftrightarrow 4 \ ]$  means a term obtained from the factor that
multiplies $f^{13m}$ by the exchange of all indices of $3$ and $4$. $N(P_i,P_j)$
is defined in Eq.(\ref{RC8}), and from (\ref{RC13})
$$ N(P_3,P_2+P_4) + N(P_4, P_1+P_3) = N(P_3, P_4)$$
In the response approach, the amplitudes RRAA have to be extracted from
correlation functions higher than the 2-point ones, i.e. they are beyond the
truncation of the Schwinger-Dyson hierarchy made by Blaizot and Iancu
\cite{BI3}. However the tree-like structure found for the RAA$\ldots$A soft
amplitudes very likely extends to all soft amplitudes. We have been able to
write down a tree-like amplitude which satisfies all the needed requirements,
i.e. \\
- direct causality flow along each tree \\
- symmetry with respect to the two legs of type R, and with respect to the two
legs of type A \\
- Ward identity similar to Eq.(\ref{RC40}) in any leg.

Our procedure follows. At the HTL level, all 4-point amplitudes arise from
different
analytical continuations of the IT amplitude. At the one-loop level, it has been 
shown explicitely  how the R/A formalism, which introduces a weight on the tree-like
vertices RAA, do lead for the RRAA amplitudes to a specific combination of
analytical continuations of the IT amplitude \cite{FG1,Au2}. We have started
from the one-loop HTL 4-point amplitude, written a la Frenkel-Taylor
\cite{FTay}, i.e. with a colour factor Tr$(T_1T_2T_3T_4)$, with appropriate
$\epsilon$-flow and weight $N(P_i,P_j)$ for the case RRAA. Summing over 6
permutations, the colour structure has been reduced to the simplest one, i.e. to
factors such as $f^{12m}f^{34m}$. Then, using partial fraction expansion, it has
been transformed into a direct $\epsilon$-flow along each tree. By construction,
this amplitude satisfies all the required properties listed above. Then a form
has emerged that generalizes to the case when each $\epsilon$ is replaced by
the operator $\hat{C}$ in $\bf v$ space, and satisfies all requirements. It is

\begin{eqnarray}
 \lefteqn{V_{\mu\nu\rho\sigma}^{1234}(P_{1R},P_{2R},P_{3A},P_{4A}) \  N(P_3,P_4)
 m_D^{-2} =
 {1\over 2} f^{12m}f^{34m}N(P_3,P_4) }  \nonumber \\ & & 
 <\  \lbrace v_{\mu}(v.P_1+i\hat{C})^{-1}v_{\nu}
 - v_{\nu}(v.P_2+i\hat{C})^{-1} {v}_{\mu} \rbrace (v.(P_1+P_2)+i\hat{C})^{-1}
 \nonumber \\ & & 
 \lbrace v_{\rho}(-v.P_4+i\hat{C})^{-1} v_{\sigma}(-p_4^0)
 - v_{\sigma}(-v.P_3+i\hat{C})^{-1} v_{\rho} (-p_3^0)\rbrace \ >_{all \ v}
 \nonumber \\ & &  
 + \ f^{13m}f^{24m}(p_1^0+p_3^0) \nonumber \\ & &
 \lbrace N(P_3,P_2+P_4)
  < \ {v}_{\mu}(v.P_1+i\hat{C})^{-1}v_{\rho} (v.(P_1+P_3)+i\hat{C})^{-1}v_{\nu}
 (-v.P_4+i\hat{C})^{-1}v_{\sigma} \ >_{all \ v} \nonumber \\ & &
  - N(P_4,P_1+P_3)
< \ {v}_{\nu}(v.P_2+i\hat{C})^{-1}v_{\sigma} (v.(P_2+P_4)+i\hat{C})^{-1}v_{\mu}
 (-v.P_3+i\hat{C})_{-1}v_{\rho} \ >_{all \ v}\rbrace \nonumber \\ & &
 + \  f^{14m}f^{23m}(p_1^0+p_4^0) \lbrace \ 3\leftrightarrow 4 \ \rbrace  \ \ 
 + \ \ {\cal R} \label{RC41}
\end{eqnarray}
where $\lbrace \ 3\leftrightarrow 4 \ \rbrace$ means a term obtained from the
factor multiplying $f^{13m}f^{24m}$ by the exchange of all indices of $3$ and $4$
, $N(P_i,P_j)$ is defined in (\ref{RC8}) and ${\cal R}$ is another term that
depends on the same propagators that the ones in the factor $f^{12m}f^{34m}$. 
${\cal R}$ is written explicitely in Appendix \ref{secB}, together 
with some details on
the Ward identities. The terms written explicitely in (\ref{RC41}) may be seen as
a naive extension of the RAAA case; in the factor of $f^{12m}f^{34m}$, $X_1^0$
and $X_2^0$ are the later times, $X_3^0$ and $X_4^0$ the earlier ones; in the
factor of $f^{13m}f^{24m}$, the first term corresponds to the time ordering
 $X_1^0>X_3^0>X_2^0>X_4^0$, the second one to $X_2^0>X_4^0>X_1^0>X_3^0$.
 
 What is needed for a self-energy diagram is the forward amplitude with two
 identical colours, for example colour $2$ = colour $3$ and $P_2+P_3 = 0 =
 P_1+P_4$. The terms $f^{14m}f^{23m}$ disappear. Moreover it will be shown later
 on that only the term whose factor is $N(P_3,P_2+P_4)$ contributes to a soft
 loop.

\section{The one-loop self-energy diagrams with effective vertices}
\label{sec4}
\setcounter{equation}{0}
 \subsection{A general expression for the diagram with 3-gluon vertices}

At the HTL level, the Retarded/Advanced formalism is an alternative to 
the Imaginary Time formalism. It avoids the analytic continuations 
that are necessary in the Imaginary Time formalism. 
It has been used for explicit calculations of the photon self-energy 
within the HTL framework at the two-loop level 
by Aurenche and collaborators \cite{Au3}. We adopt their conventions.

The propagators are the retarded and advanced effective propagators, 
i.e. with $\epsilon>0$
\begin{equation}
\Delta (P_R) = \Delta( p_0 + i\epsilon, p ) = 
{i\over{(p_0+i\epsilon)^2 - p^2 -\Pi(p_0+i\epsilon, p)}}
\label{D16}
\end{equation}
\begin{equation}
\Delta(P_A) = \Delta(p_0-i\epsilon, p) = - {\Delta (P_R)}^*   \ \ \ \ \ 
\ \ \ \Delta(P_R) = \Delta((-P)_A)
\end{equation}
from Eqs.(\ref{C4},\ref{C5}) for $\Pi(P)$.  In covariant gauges, the effective 
gluon propagator is
\begin{equation}
D^{\mu \nu}(P) = {\mathcal{P}}_t^{\mu \nu} \Delta^t(P) + 
{\mathcal{P}}_l^{\mu \nu} \Delta^l(P) +i \xi {p^{\mu} p^{\nu} \over {P^4}}
\label{D18}
\end{equation}
\begin{eqnarray}
 {\mathcal{P}}_t^{i j} & = & \delta^{i j} - \hat{p}^i \hat{p}^j \ \ ,\ \   
 {\mathcal{P}}_t^{\mu 0} =0  \\
 {\mathcal{P}}_l^{\mu \nu} & = & g^{\mu \nu} -{ p^{\mu} p^{\nu} \over P^2}  -  
 {\mathcal{P}}_t^{\mu \nu}
\end{eqnarray}
\begin{equation}
\Pi^{\mu \nu} (P) = {\mathcal{P}}_t^{\mu \nu}(P) \Pi^t +  
{\mathcal{P}}_l^{\mu \nu}(P) \Pi^l(P)
\label{D20}
\end{equation}
In a Coulomb-like gauge
\begin{equation}
D^{\mu \nu}(P) = {\mathcal{P}}_t^{\mu \nu} \Delta^t(P) + 
  g_{\mu 0}g_{\nu 0}\ {P^2\over p^2} \Delta^l(P) +
  i \xi {p^{\mu} p^{\nu} \over {p^4}}
\label{D18b}
\end{equation}

The three-gluon vertex has been obtained in Sec.\ref{sec3.3}, i.e.
Eq.(\ref{RC31}). In this section it will be written
\begin{equation}
g\ V_{\mu\rho\sigma}^{123} (Q_R, P_A, S_A)= 
if^{123} g \ V_{\mu\rho\sigma}(Q_R, P_A, S_A)
\label{RD1}
\end{equation}
and the $i$ is dropped to be consistent with the convention for the propagator
(\ref{D16}). As detailed in Sec.\ref{sec3.1}, the 3-point vertices to be used in
the R/A formalism are
 
\begin{eqnarray}
\Gamma^{\mu \rho\sigma}(Q_A, P_R, R_R)  & = &  
V^{ \mu \rho \sigma} (Q_A, P_R, R_R) 
\label{D21} \\
\Gamma^{\mu \rho\sigma}( Q_R, P_A, R_A)  & = & 
 - \  V^{\mu \rho \sigma} ( Q_R, P_A, R_A) (n(p_0) + n(r_0) +1)
\label{D22}
\end{eqnarray}
$ V^{\mu \rho \sigma} ( Q_R, P_A, R_A)$ is antisymmetric in the exchange of the
two legs of type A, it satisfies the general properties Eqs.(\ref{C1},\ref{C2}).
 Two relations will be useful
\begin{eqnarray}
V^{\mu \rho \sigma} ( (-Q)_A, (-P)_R, (-R)_R) &  =  &  
- V^{\mu \rho \sigma} ( Q_R, P_A, R_A) 
\label{D23}\\
 n(p_0)+1/2 & = & -(n(-p_0) +1/2)
\label{D24}
\end{eqnarray}

When one considers the self-energy $\Pi(Q_R)$, the one-loop diagram 
with 3-gluon vertices is, in the R/A formalism, the sum of the three diagrams 
drawn on Fig.\ref{fi1}(a)(b)(c) (a propagator joins an $R$ leg to an $A$ leg, see
Sec.\ref{sec3.1}). 
With the definition of momenta of Fig.\ref{fi1}(d), leaving out a colour 
factor $\delta_{a b}$
\begin{eqnarray}
\lefteqn{ \Pi_{(3g)}^{\mu \nu}(Q_R) = {N g^2\over 2} \int {d_4p \over{i(2\pi)^4}} } \label{D25} \\
& & (n(p_0)+1/2) \lbrack \  D_{\rho \rho'}(P_R)
V^{\rho \mu \sigma}(P_R, Q_R, (-S)_A)  D_{\sigma \sigma'}(S_R)
V^{\rho' \nu \sigma'}((-P)_A, (-Q)_A, S_R) \  \rbrack \nonumber \\
 & & -(n(s_0)+1/2) \lbrack \ D_{\rho \rho'}(P_A)
 V^{\rho \mu \sigma}(P_A, Q_R, (-S)_R)  D_{\sigma \sigma'}(S_A)
 V^{\rho' \nu \sigma'}((-P)_R, (-Q)_A, S_A) \ \rbrack \nonumber \\
 & &  - (n(p_0)- n(s_0)) \lbrack \  D_{\rho \rho'}(P_A)
 V^{\mu \rho \sigma}( Q_R, P_A, (-S)_A)  D_{\sigma \sigma'}(S_R)
 V^{ \nu \rho' \sigma'}( (-Q)_A, (-P)_R, S_R) \  \rbrack \nonumber
\end{eqnarray}
Several comments have to be made \\
- From (\ref{D23}) the two vertices entering each term are identical, 
up to a sign \\
- The prescription $R$ or $A$ on each leg fixes the way one should approach
 the discontinuity in that variable. Then one writes $ S = P + Q $ and 
 one may integrate over $p_0$ either on the  real axis, or in the complex 
 $p_0$ plane. Although $s_0 = p_0 + q_0$ , it make sense to speak of 
 discontinuities in the $p_0$ or $q_0$ or $s_0$ variable, 
 just as at $T=0$ one considers singularities in the $s$ or $t$ or $u$ channel 
 of the 4-point function. \\
 - The properties in the complex $p_0$ plane to be used in the subsequent
 argument are identical whether $(v.P_R)^{-1}$ enters the HTL vertices, or 
 $(v.P +i\hat{C})^{-1}$ enters the soft vertices. Indeed, when $p_0+i\epsilon$
 enters, the meaning is that one should approach the singularities  along the
 real $p_0$ axis from above. The operator $\hat{C}$ has an infinite tower of
 positive eigenvalues $c_l$ and a zero eigenvalue $c_0$, i.e. $(v.P +i\hat{C})$
 vanishes for $p_0={\mathbf v.p}-ic_l$, i.e. those points are in the lower
 complex $p_0$ plane or along the real axis. \\
- The third term of the right handside of Eq.(\ref{D25}) corresponds to diagram(c) 
of Fig.\ref{fi1}. In the first and second term, a term $(n(q_0)+1/2)$ has 
been dropped in the thermal weight. The reason is: all factors in the bracket 
of the first (or second) term have singularities in the $p_0$ complex plane 
only on one side of the real axis; by closing the integration contour on the 
other half plane, they are seen to give a vanishing contribution when they 
are multiplied by $(n(q_0) +1/2)$. By the same token $1/2$  coul be dropped, 
it has been kept because of the parity property (\ref{D24}) \cite{Au1}.
Conversely, $n(p_0)$ has poles at $p_0=\pm i   n 2\pi T$ all along the 
imaginary axis. \\
Thermal weights depending on the external momenta do appear in the R/A 
formalism. They give a vanishing contribution at any loop order because 
they are associated with  terms whose singularities all lie one one side of 
the real axis. It has to be so in order to agree with the Imaginary Time 
formalism, where thermal weights only depend on loop momenta 
(see the coth method \cite{FG2}). This feature will turn out to be an 
useful one when one considers soft momentum exchange $p$. \\
- Eq. (\ref{D25}) is a completely general formula and appears in many cases 
\cite{Au3,Au2}.

\subsection{The case of a soft exchange around the loop}
\label{sec4.2}
We want to use Eq.(\ref{D25}) to compute a soft momentum exchange around the loop,
 i.e. $p_0  << T$. Now comes B\"{o}deker's argument \cite{Bod3} 
 (translated from Imaginary Time to R/A formalism).  For loop momenta 
 $p_0 << T$  one may drop the first and second term in Eq.(\ref{D25}). 
 Indeed, as said above, in the first term, the term in brackets has 
 singularities only on one 
 side of the real $p_0$ axis, and if one closes the integration contour on 
 the other side, the first contribution comes from the pole $p_0 = i 2\pi T$ of
 $n(p_0)$ i.e. a hard energy, and it will lead to terms of order $q_0 / T$. The
 residue of the pole $p_0=0$  of $n(p_0)$ is zero as the two terms in
 (\ref{D25}) whose factor is $n(p_0)$ cancel each other at that point
 ($\Pi(p_0=0,p)=0$, the vertices are identical up to a sign). In what follows,
 only one of these two terms is kept, the contribution to be obtained will come
 from a region away from $p_0=0$. The same argument applies to the variable 
 $s_0$ and the second term. \\
 As a consequence, one is left with the third term, and for $p_0 <<T$  
 and $q_0 << T$, one may approximate the thermal weights i.e. keep only 
 the pole close to the origin and neglect the other ones' contribution. 
 Changing the loop variable from $P$ to $K$
\begin{equation}
P  = K-Q/2   \ \ \ \ , \ \  \   S= K + Q/2
\label{D26} 
\end{equation}
\begin{equation}
n(p_0) - n(s_0) \approx T ({1\over p_0} - {1\over s_0}) = {T q_0 \over{(k_0-q_0/2)(k_0 +q_0/2)}}
\label{D27}
\end{equation}
so that a factor $q_0$ is extracted from the thermal weight. 
Note that in terms of the set of diagrams drawn on Fig.\ref{fi1}, 
this argument amounts to keep only the diagram where the vertex with two legs 
of type $A$ is \textit{not} the one with a $(-Q)_A$ leg. It will be a useful 
feature when one turns to the other case, i.e. one soft loop diagram with a 
4-point vertex. \\

First,  momenta $K\sim gT$ around the loop wil be considered and the result 
 obtained in \cite{Bod3} will be reproduced, then 
 momenta $K\sim g^2T \ln 1/g $ will be studied.  
 The vertex $V$ is the total vertex, bare + effective. However diagrams
 containing only bare vertices (gluon + ghost), or containing one bare and one
 effective vertex have been shown in \cite{Bod3} to give a subleading
 contribution, mostly because a bare vertex contains no factor $1/v.Q$ (or
 $(v.Q+i\hat{C})^{-1}$). At the HTL level, there is no effective ghost vertex,
 and it is likely to be the same for the case $K\sim g^2T\ln1/g$.
  We shall not consider the case $\mu = \nu =0$ in $\Pi^{\mu \nu}$  as our aim
  is the collision operator.

For momenta $K\sim gT$ the effective vertex is the HTL one,
Eqs.(\ref{RD1}),(\ref{RC31}) with 
$ P_1=Q,\  P_2 =-S = -(K+Q/2),\ P_3 = P = K-Q/2 $
and $\hat{C}$ replaced by $\epsilon {\cal I} \ (\epsilon>0)$ where ${\cal
I}=\delta_{S_2}({\mathbf v} -{\mathbf v'})$ is the identity in ${\mathbf v}$
space.
\begin{equation}
V^{\mu \sigma \rho} ( Q_R, (-S)_A, P_A)   =  -  m_D^2<{v_1^{\mu}v_1^{\sigma}v_1^{\rho} \over{v_1.Q_R}}({(k_0+q_0/2)\over{v_1.(K+Q/2)_R}} - {(k_0-q_0/2)\over{v_1.(K-Q/2)_A}} )>_{v_1}
\label{D28b}  
\end{equation}
\begin{equation}
V^{\mu \sigma \rho} ( Q_R, (-S)_A, P_A)   =  - m_D^2  \  {\cal{V}}^{\mu \rho \sigma} ( K, Q)
\label{D28}
\end{equation}
and the resummed propagators (\ref{D18},\ref{D20}) are those with the 
HTL $\Pi^{\mu \nu}$, as in Eq.(\ref{B9}), so that keeping only the third term in Eq.(\ref{D25}). and with (\ref{D26},\ref{D27}) 
\begin{eqnarray} 
\lefteqn{ \Pi_{(3g)}^{\mu \nu}(Q_R) =  q_0 m_D^4 {g^2N T \over 2} 
\int {d_4k\over{i(2\pi)^4}} {1\over{(k_0-q_0/2)(k_0+q_0/2)}} } \nonumber \\
& & D_{\rho \rho'}((K-Q/2)_A) D_{\sigma \sigma'}((K+Q/2)_R) 
{\cal{V}}^{\mu \rho \sigma} ( K, Q) {\cal{V}}^{\nu \rho' \sigma'} ( K, Q)
\label{D29}
\end{eqnarray}
${\cal{V}}^{\nu \rho' \sigma'}$  is obtained 
from ${\cal{V}}^{\mu \rho \sigma} $ with the substitutions 
$\mu \rightarrow \nu , \rho \rightarrow \rho', \sigma \rightarrow \sigma', v_1 
\rightarrow v_2 $

If one now drops $Q\sim g^2 T$ in front of $K\sim gT$ in (\ref{D29},\ref{D28b})
\begin{equation}
k_0({1\over{v_1.K_R}}-{1\over{v_1.K_A}}) = k_0 {\mathrm{disc}}{1\over{v_1.K}} 
= -k_0  2\pi  i \delta(v_1.K)
\end{equation}
and 
\begin{eqnarray} 
\lefteqn{ \Pi_{(3g)}^{\mu \nu}(Q_R) =  q_0 m_D^4 {g^2N T \over 2} 
\int {d_4 k\over{i(2\pi)^4}}  D_{\rho \rho'}(K_A)  
D_{\sigma \sigma'}(K_R) } \nonumber \\
& &  <v_1^{\mu}{1\over{v_1.Q_R}}v_1^{\rho}v_1^{\sigma} {\mathrm{disc}} 
{1\over{v_1.K}}>_{v_1} <v_2^{\nu}{1\over{v_2.Q_R}}v_2^{\rho'}v_2^{\sigma'} 
{\mathrm{disc}} {1\over{v_2.K}}>_{v_2}
\label{D30}
\end{eqnarray}
The result does not depend on the gauge parameter $\xi$ owing 
to the $\delta$ functions. 
$\Pi_{(3g)}^{\mu \nu}$ was found via this perturbative approach 
in \cite{Bod3} in the Imaginary Time formalism.

Comparing  (\ref{D30}) with (\ref{D9})  for the case $\mu=j ,\  \nu=i$, 
one identifies the contribution from this diagram to the collision 
operator $\hat{C}$, or equivalently to $\Phi$ defined in Eq.(\ref{A9})
\begin{equation}
\Phi({\mathbf{v}}_1.{\mathbf{v}}_2) =\int {d_4k\over{(2\pi)^4}}\   
{\vert v_1.D(K).v_2 \vert}^2 \ (2\pi)^2 \delta(v_1.K) \delta(v_2.K)
\label{D31}
\end{equation}
with  $$v_1.D(K).v_2=({\mathbf v_1}.{\mathcal P}_t.{\mathbf v_2})\ \Delta^t(K)
\ +\ {K^2\over k^2}\ \Delta^l(K)  $$
in both covariant and Coulomb gauges.
$\Phi$ was also found by other methods \cite{Bod1,AY2,BI2,Lit}. 
$\Phi({\mathbf{v}}_1.{\mathbf{v}}_2)$ has been interpreted as the collision 
cross section of two fast particles with soft momentum exchange $K$; 
the $\delta$ functions enforce the conservation of energy-momentum at 
each vertex in the near-forward direction \cite{BI1}. The integration range 
for the space momentum $k$ is limited  $\mu_2<k<\mu_1$ (see (\ref{D14}) ). \\

Turning now to the case when the exchange momentum is $K\sim g^2T \ln 1/g$ 
the gluon propagators and the 3-gluon vertices have to be modified. 
The polarization tensor is now as in (\ref{B7}) and its related vertex 
as in (\ref{RD1}),(\ref{RC31}) with $P_1 = Q, P_2 = -(K+Q/2), P_3 = K-Q/2 $. 
The only change in the expression  (\ref{D29}) for the one-loop soft momentum 
exchange $ \Pi_{(3g)}^{\mu \nu} $  is in ${\cal{V}}^{\mu \rho \sigma} $ and 
${\cal{V}}^{\nu \rho' \sigma'} $
\begin{eqnarray}
\lefteqn{ {{\cal{V}}'}^{\mu \rho \sigma} (K, Q) = 
< v_1^{\mu}(v_1.Q+i\hat{C})^{-1}  }  \\
& &  \lbrace(k_0+q_0/2)v_1^{\rho}(v_1.(K+Q/2)+i\hat{C})^{-1} 
v_1^{\sigma} \nonumber \\
& &  - (q_0/2-k_0)v_1^{\sigma}(v_1.(Q/2-K)+i\hat{C})^{-1}
v_1^{\rho} \rbrace>_{v_1,v_1',v_1''}
\nonumber
\end{eqnarray}
and ${{\cal{V}}'}^{\nu \rho' \sigma'} $ is obtained by the 
substitution $\mu \rightarrow \nu, \ \  \rho \rightarrow \rho', \ \ \sigma \rightarrow \sigma' $ and  ${v_1,v_1',v_1''} \rightarrow {v_2,v_2',v_2''}$
\\ The Ward identities (\ref{RC32},\ref{RC35}) imposed on this 3-gluon 
vertex read
\begin{equation}
Q_{\mu}  {{\cal{V}}'}^{\mu \rho \sigma} =m_D^{-2}
(\Pi^{\rho \sigma}((K+Q/2)_R) - \Pi^{\rho \sigma}((Q/2 - K)_R)
\label{D33}
\end{equation}
\begin{equation}
-(K+Q/2)_{\sigma} {{\cal{V}}'}^{\mu \rho \sigma} =m_D^{-2}
(\Pi^{\mu \rho}((Q/2-K)_R) - \Pi^{\mu \rho }(Q_R) )
\end{equation}
$ \Pi_{(3g)}^{\mu \nu} $ is in the desired form (\ref{D15}).  The factors
$v_{\mu}(v.Q+i\hat{C})^{-1} \ldots  \ldots (v.Q+i\hat{C})^{-1}
v_{\nu}$ come from the tree-like structure of the Retarded effective vertices
where $ v_{\mu}(v.Q+i\hat{C})^{-1}$ sits first in $\mathbf v$ space.

 If one now drops $Q\sim g^2T$ in front of $ K \sim g^2 T \ln 1/g$, which is 
 only valid for $\ln 1/g >>1$
\begin{eqnarray}
\lefteqn{ {{\cal{V}}'}^{\mu \rho \sigma} (K, Q) = k_0<v_1^{\mu}(v_1.Q+i\hat{C})^{-1} }  \nonumber \\
& &  \lbrace v_1^{\rho}(v_1.K+i\hat{C})^{-1} v_1^{\sigma} - v_1^{\sigma}(v_1.K-i\hat{C})^{-1}v_1^{\rho} \rbrace>_{v_1,v_1',v_1''}
\label{D35}
\end{eqnarray}
In particular, for $\rho$ and $\sigma $ spacelike
\begin{equation}
{{\cal{V}}'}^{\mu i j}  = k_0<v_1^{\mu}(v_1.Q+i\hat{C})^{-1}  \lbrace v_1^i  
\ W_R^j(K,{\mathbf{v}}_1) -  v_1^j  \ W_A^i(K,{\mathbf{v}}_1)  \rbrace>_{v_1}
\label{D36}
\end{equation}
where the collective field  $W^i_R(K,  \mathbf{v})$ has been defined in 
(\ref{B4}) and $W^i_A(K,  \mathbf{v})$ in a similar way (see (\ref{A7})). 
The resulting contribution to the polarization tensor is
\begin{equation} 
 {\Pi'}_{(3g)}^{\mu \nu}(Q_R) =  q_0 m_D^4 {g^2N T \over 2} \int {d_4k\over{i(2\pi)^4}}  D_{\rho \rho'}(K_A)  D_{\sigma \sigma'}(K_R)  
{1\over k_0^2}{{\cal{V}}'}^{\mu \rho \sigma} (K, Q) 
{{\cal{V}}'}^{\nu \rho' \sigma'} (K, Q)
\label{D37}
\end{equation}
with ${{\cal{V}}'}$ as in (\ref{D35}) or (\ref{D36}). 
A consequence of (\ref{D37}) is, from (\ref{D33})
\begin{eqnarray}
 \lefteqn{ Q_{\nu}{\Pi'}_{(3g)}^{\mu \nu} =  q_0 m_D^2 {g^2N T \over 2} \int {d_4k\over{i(2\pi)^4} } D_{\rho \rho'}(K_A)  D_{\sigma \sigma'}(K_R) } \nonumber \\  
&  &     {1\over k_0^2} {{\cal{V}}'}^{\nu \rho' \sigma'} (K, Q)  
(\Pi^{\rho' \sigma'}(K_R) -\Pi^{\rho' \sigma'}(K_A))
\label{D38}
\end{eqnarray}
In Sec.\ref{sec4.4} the other one-loop diagram with a 4-gluon vertex will be computed and one will obtain 
\begin{equation}
Q_{\nu}( {\Pi'}_{(3g)}^{\mu \nu} + {\Pi'}_{(4g)}^{\mu \nu}) = 0
\end{equation}
\\  Comparing (\ref{D37})( \ref{D35}) with the expression (\ref{D15}) for 
the case $\mu =j \ , \  \nu =i$, one obtains the collision term arising 
from the diagram of Fig.\ref{fi1}(d). Defining
\begin{equation}
C'({\mathbf{v}}_1,{\mathbf{v}}_2) = -m_D^2{ g^2N T \over 2} 
( \lbrack \Phi_{(3g)}'({\mathbf{v}}_1,{\mathbf{v}}_2) 
+\Phi_{(4g)}'({\mathbf{v}}_1,{\mathbf{v}}_2) \rbrack
\label{D40}
\end{equation}
\begin{eqnarray}
\lefteqn { \Phi_{(3g)}'({\mathbf{v}}_1,{\mathbf{v}}_2) = 
  \int {d_4 k\over{(2\pi)^4}}  D_{\rho \rho'}(K_A)  D_{\sigma \sigma'}(K_R) }    \label{D41} \\
& & <v_1^{\rho}(v_1.K+i\hat{C})^{-1}{v'}_1^{\sigma} 
-v_1^{\sigma}(v_1.K-i\hat{C})^{-1}{v'}_1^{\rho}>_{v_1'} \nonumber \\
& &  <v_2^{\rho'}(v_2.K+i\hat{C})^{-1}{v'}_2^{\sigma'} 
-v_2^{\sigma'}(v_2.K-i\hat{C})^{-1}{v'}_2^{\rho'}>_{v_2'} \nonumber
\end{eqnarray}
where the range of integration over the space momentum is 
$\mu_3 < k < \mu_2$ (see(\ref{D14})). The reality of $\Phi_{(3g)}'$ is a
consequence of the Bose symmetry of the two soft gluons in the effective
vertices.      Note that one may go back to 
the expression describing the momentum $K\sim gT$ by replacing the 
operator $\hat{C}$ by $\epsilon {\mathcal I} , \ \epsilon >0$, then the inverse 
operator becomes diagonal in $\mathbf{v}$ space, and one gets 
back $\Phi({\mathbf{v}}_1,{\mathbf{v}}_2)$ as in (\ref{D31}). \\
The constraint
$v_1.K =0 =k_0-{\mathbf v}_1.{\mathbf k} $ in Eq.(\ref{D31}) is essential for
the gauge independence of $\Phi_{(3g)}$ in two ways: \ 
i)  the contraction $k_{\rho}v_1^{\rho}$ vanishes \ 
ii)  $({\mathbf v}.{\mathbf k})^2=k_0^2$ is used to get the same result in
covariant and Coulomb gauges.\\
This is no more true for $\Phi_{(3g)}'$. One has
\begin{eqnarray}
\lefteqn{ k_{\rho} <v_1^{\rho}(v_1.K+i\hat{C})^{-1}{v'}_1^{\sigma} 
-v_1^{\sigma}(v_1.K-i\hat{C})^{-1}{v'}_1^{\rho}>_{v_1'}  } \nonumber \\
 &  & = - < i\hat{C} (v_1.K+i\hat{C})^{-1}{v'}_1^{\sigma} >_{v_1'}
\end{eqnarray}
which is zero when integrated over $v_1$ or contracted with $k_{\sigma}$. For
$k_0,k <<\gamma$, it reduces to $-v_1^{\sigma}$. For a longitudinal gluon
exchange, the result depends on the gauge parameter $\xi$ and on the gauge
(covariant or Coulomb). As the transport equation (\ref{RA5}) has been established
in the strict Coulomb gauge (and expected to be gauge independent), one may wish
to stay in this gauge.

An intuitive picture for the collision term 
$ \Phi_{(3g)}'({\mathbf{v}}_1,{\mathbf{v}}_2)$ is obtained if 
one limits oneself to the dominant exchange of transverse gluons. 
From (\ref{D18})
\begin{equation}
D^{\rho \rho'}(K) \approx \Delta^t(K) {\cal{P}}_t^{\rho \rho'} = 
\Delta^t(K) {\cal{P}}_t^{i i'}
\label{D42}
\end{equation}
then from (\ref{D36}), the collective field $W^i(K,\mathbf{v})$ appears in 
$\Phi_{(3g)}'$, its transverse part is parallel to $v_t^i$ 
\begin{eqnarray}
W ^a(K,{\mathbf{v}}) &=& i\  W^i(K,{\mathbf{v}}) \ E^{i\ a}(K)   \\
W^i(K,{\mathbf{v}}) & = & {\hat{k}}_i\  W^l  + v_t^i \ W^t   \\  
{\cal{P}}_t^{i i'}(k) W^{i'} (K, {\mathbf{v}}) & = & {\cal{P}}_t^{i i'}v^{i'} 
W^t (K, {\mathbf{v}})
\end{eqnarray}
so that
\begin{eqnarray}
\lefteqn{ \Phi_{(3g)}'({\mathbf{v}}_1,{\mathbf{v}}_2) =   
\int {d_4 k\over{(2\pi)^4}} \  
{\vert \Delta^t(K_R) ({\mathbf{v}}_1.{\cal{P}}_t .{\mathbf{v}}_2) \vert }^2  }
\label{D44}\\
  &  &  (-) [ W_R^t(K,{\mathbf{v}}_1)- W_A^t(K, {\mathbf{v}}_1)] \  
  [ W_R^t(K,{\mathbf{v}}_2)- W_A^t(K, {\mathbf{v}}_2)] \nonumber
\end{eqnarray} 
 $\Phi_{(3g)}'(\mathbf{v}_1,\mathbf{v}_2) $ may be interpreted as 
 the near-forward collision cross section of two  collective excitations. 
 The constraint at each 
 vertex is no more a strict particle-like conservation $ 2\pi \delta(v_1.K)$ 
 as in (\ref{D31}), but a smeared one involving the ``width'' of the $W^t$ field. \\
A consequence of (\ref{D41}) is
\begin{eqnarray}
\lefteqn{ \int {d\Omega_{\mathbf{v}_2} \over{4\pi}} 
\Phi_{(3g)}'({\mathbf{v}}_1,{\mathbf{v}}_2)=  
\int {d_4 k\over{(2\pi)^4} } D_{\rho \rho'}(K_A)  D_{\sigma \sigma'}(K_R) 
{1\over k_0m_D^2}  }  \label{D45} \\ 
 &   &  (\Pi^{\rho' \sigma'}(K_R) -\Pi^{\rho' \sigma'}(K_A)) 
 <v_1^{\rho}(v_1.K+i\hat{C})^{-1}{v'}_1^{\sigma} 
 -v_1^{\sigma}(v_1.K-i\hat{C})^{-1}{v'}_1^{\rho}>_{v_1'} \nonumber
\end{eqnarray}
One notices that the contribution to the damping rate of a hard gluon 
 arising from the range $K\sim g^2T \ln 1/g$ is obtained from (\ref{D45}) 
 if one substitutes to the $ \mathbf{v}_1'$ average the quantity 
 $v_1^{\rho}\  v_1^{\sigma}\  (- 2\pi  i )\  \delta(v_1.K)$ .
In Sec. \ref{sec4.4} , one will find that
\begin{equation}
\int {d\Omega_{\mathbf{v}_2} \over{4\pi}}
[ \Phi_{(3g)}'({\mathbf{v}}_1,{\mathbf{v}}_2) 
+\Phi_{(4g)}'({\mathbf{v}}_1,{\mathbf{v}}_2)] =0
\end{equation}
 i.e. a relation similar to (\ref{A10}) for the case of the $\hat{C}$ operator.

\subsection{A sum rule for the collision operators $\hat{C}$ and $\hat{C}'$}
\label{sec4.3}

In order to know the scale of $ \Phi_{(3g)}'$, and of  $\hat{C}'$ defined 
in (\ref{D40}), one may average over  $\mathbf{v}_1$ and $\mathbf{v}_2$
\begin{eqnarray}
\lefteqn{ \bar{\Phi} =  \int {d\Omega_{\mathbf{v}_1} \over{4\pi}} 
{d\Omega_{\mathbf{v}_2} \over{4\pi}} \Phi_{(3g)}'({\mathbf{v}}_1,{\mathbf{v}}_2)   } \nonumber \\
 &  & = \int {d_4 k\over{(2\pi)^4}}  D^*_{\rho \rho'}(K_R)  
 D_{\sigma \sigma'}(K_R)    \nonumber \\   &  & 
 [\Pi^{\rho' \sigma'}(K_R) -\Pi^{\rho' \sigma'}(K_A)]\ 
  [(\Pi^{\rho \sigma}(K_R) -\Pi^{\rho \sigma}(K_A)] ({-1\over{k_0^2m_D^4}})
\label{D47}
\end{eqnarray}
This relation, found in the case $K\sim g^2 T \ln 1/g$ ,  also holds in the 
case $K\sim gT$, i.e. when  $ \Phi_{(3g)}'(\mathbf{v}_1,\mathbf{v}_2)$ is 
replaced by $ \Phi(\mathbf{v}_1,\mathbf{v}_2)$. Indeed, 
for $K\sim gT$ , $\Pi^{\rho \sigma}(K)$  is as in (\ref{B9}) i.e.
\begin{equation}
{i\over{k_0m_D^2}}[\Pi^{\rho \sigma}(K_R) -\Pi^{\rho \sigma}(K_A)] = 
\int{d\Omega_{\mathbf{v}_1} \over{4\pi}}\  v_1^{\rho} \  v_1^{\sigma} \  2\pi \delta(v_1.K)
\label{D48}
\end{equation}
and (\ref{D48}) inserted into (\ref{D47}) leads to $
\Phi(\mathbf{v}_1,\mathbf{v}_2)$ as in (\ref{D31}). $\bar{\Phi}$ is a gauge
independent quantity
\begin{equation}
\bar{\Phi} = \int {d_4k\over{(2\pi)^4}} \ [\ 2 {\vert \Delta^t  \vert }^2  
[ {2{\mathrm Im} \Pi^t\over k_0m_D^2}]^2 \ +\  {\vert \Delta^l  \vert }^2
[ {2{\mathrm Im} \Pi^l\over k_0m_D^2}]^2 \ ]
\label{D49}
\end{equation}

Relation (\ref{D49}) allows an easy comparison between the infrared behaviour 
of the cases $K\sim gT$ and $K\sim g^2T \ln1/g$. One restricts oneself to the 
dominant transverse gluon exchange. For $k<m_D$,  ${\vert \Delta^t (K) \vert
}^2$put a strong weight upon the region $k_0<<k$ and one may write
$${\mathrm Im} \Pi^t(k_0,k) \approx  k_0 \ {\mathrm Im} {\tilde{\Pi}}^t(k_0=0,k) $$

\begin{equation}
 {\vert \Delta^t(K_R)  \vert }^2 = {1\over{(k_0^2-k^2-{\mathrm{Re}} \Pi^t)^2+
 ({\mathrm{Im}} \Pi^t)^2}} \approx {1\over{k^4+k_0^2({\mathrm{Im}} 
 {\tilde{\Pi}}^t)^2}}
\end{equation}
\begin{equation}
 \int{ \frac{dk_0 }{k^4+k_0^2\ ({\mathrm{Im}} {\tilde{\Pi}}^t)^2 } }  = 
{ \frac {\pi}{k^2 \vert{\mathrm{Im}} {\tilde{\Pi}}^t\vert }}
\label{D53}
\end{equation}
so that from (\ref{D49})
\begin{equation}
\bar{\Phi} \approx \int {dk \ k^2\over{4\pi^3}}
{ \frac {\pi}{k^2 \vert{\mathrm{Im}} {\tilde{\Pi}}^t\vert }} \  2 \ 
(\frac{2 {\mathrm{Im}} {\tilde{\Pi}}^t}{m_D^2})^2
\label{D51}
\end{equation}
\begin{equation}
\bar{\Phi}\approx {1\over m_D^2}{2\over \pi^2}\int d k 
{\frac{ \vert{\mathrm{Im}} \Pi^t\vert}{k_0m_D^2}}\vert_{k_0=0}(k)
\label{RD52}
\end{equation}
- For the region $k\sim gT$, from (\ref{D48})
\begin{equation}
{\mathrm{Im}} \Pi^t = -  k_0 m_D^2 {1\over k} {\pi\over 4} (1-{k_0^2\over{k_2}})
 \ \theta(k^2-k_0^2) \approx 
 -m_D^2 {k_0\over k} {\pi\over 4}\ \ =k_0 \ {\mathrm{Im}} {\tilde{\Pi}}^t
\label{D52}
\end{equation}
so that
 \begin{equation}
{\bar{\Phi}}_1 \approx  
 {1 \over m_D^2 }{1\over 2\pi}\int_{\mu_2} {dk \over k}
\end{equation}
the integral has an infrared logarithmic divergence  ($\mu_2, \mu_3$ have been
defined in (\ref{D14})) and  
\begin{equation} 
\gamma = m_D^2 {g^2 NT  \over 2} \ \bar{\Phi} \approx {g^2 NT  \over {4\pi}} 
 \ \ \ln { m_D \over{ \mu_2} } 
\end{equation}
- For the region  $k \sim g^2T \ln 1/g$ , from Eq.(\ref{Ap3}) in Appendix 
\ref{secA}
\begin{equation}
\Pi^t(k_0,k)m= {m_D^2 k_0\over 3} \Sigma_1(k_0, k)
\end{equation}
where $\Sigma_1$ is one matrix element of $(v.K+i\hat{C})^{-1}$
\begin{equation}
{\bar{\Phi}}_2 \approx {1\over m_D^2} {2\over 3\pi^2}\int_{\mu_3}^{\mu_2}
d k  \ {\vert {\mathrm Im}\Sigma_1(k_0=0,k)\vert}
\label{D56}
\end{equation}
From (\ref{Ap2},\ref{Ap1}) one sees that the scale of $\Sigma_1$ is $\gamma$ and
for $k<<\gamma$ 
\begin{equation}
\Sigma_1(k_0=0,k<<\gamma) ={-i\over{\gamma-\delta_1}}  \ \ \ {\mathrm and}\ \ \
 {\mathrm{Im}} {\tilde{\Pi}}^t (0,k) =
-{m_D^2\over3}{1\over{\gamma-\delta_1}}
\label{D54}
\end{equation}
so that $\bar{\Phi}_2$ has no infrared divergence.
In fact it is unlikely that the main contribution to $\bar{\Phi}_2$ 
comes from the momenta $k<\gamma/3$. Indeed, as discussed in  Appendix \ref{secA}, 
there is an imaginary part of $\Pi^t$ in the range $\vert k_0 \vert < k$ 
analogous to (\ref{D52}) i.e. to the Landau-damping term. 
However it only exists for $ k > \gamma/3$.\\
 Eq.(\ref{D56}) sets the scale of the
operator $\hat{C}'$ as $g^2NT$ (see Eq.(\ref{D40})) with a weak dependence 
on both cutoff.

One may look at the contribution from the momenta $k\sim g^2T \ln 1/g$ (i.e. in
Eq.(\ref{D51}), ${\mathrm{Im}} {\tilde{\Pi}}^t$ in the denominator is as in
(\ref{D54})) to two other quantities \\
- the damping rate of a hard point-like gluon. 
Then in $\bar{\Phi}_2$ as in (\ref{D51}), one  ${\mathrm{Im}} {\tilde{\Pi}}^t$ 
in the numerator is as in (\ref{D54}) and 
the other  one  as in (\ref{D52}), and $\bar{\Phi}_2$ has an infrared log  
divergence. \\
- the interaction rate of two point-like hard gluons. 
Then in $\bar{\Phi}_2$ both ${\mathrm{Im}} {\tilde{\Pi}}^t$  in the numerator 
are as in (\ref{D52}), and  $\bar{\Phi}_2$ has a linear divergence, 
as stated in the introduction.

\subsection{The diagram with a 4-gluon effective vertex}
\label{sec4.4}
The contribution to $\Pi_{\mu\nu}(Q)$ of the diagram drawn on Fig.\ref{fi2}(a)
is considered. One treats simultaneously both cases i.e. when the momentum $k$
running around the loop is i) $k\sim gT$ (where 
$\hat{C}$ is replaced by $\epsilon {\mathcal I}$) \  ii) $k\sim g^2T \ln 1/g$. \\
As explained in Sec.\ref{sec3.1}, one has to join the leg $P_{3A}$ to the
leg $P_{2R}$ in the 4-gluon vertex
$$ig^2(-)\ N(P_3,P_4) \ \ 
V_{\mu\sigma\rho\nu}^{1234}(P_{1R},P_{2R},P_{3A},P_{4A}) $$
given in Eq.(\ref{RC41}) with
\begin{equation}
P_3=-P_2 =K \ \ \ , \ \ \ P_1=-P_4=Q
\end{equation}
(Note that the Lorentz indices $\sigma$ and $\nu$ have been exchanged compared
to (\ref{RC41})). With colour $2$ = colour $3$, the term $f^{23m}f^{14m}$
disappears in (\ref{RC41}), all the other colour factors are identical and 
$\sum_{2,m} f^{12m}f^{24m} =-\delta^{14} \ N$. Leaving out the colour factor
$\delta^{14}$
\begin{eqnarray}
\lefteqn{\Pi_{(4g)}^{'\mu\nu}(Q_R)=ig^2Nm_D^2\int{d_4k\over{i(2\pi)^4}} \ \
D_{\rho\sigma}(P_{2R}=(-K)_R) \ \ N(P_3=K, P_4=-Q)} \nonumber \\ & & 
V_{\mu\sigma\rho\nu}^{1234}(P_{1R}=Q_R,P_{2R}=(-K)_R,P_{3A}=K_A,P_{4A}=(-Q)_A)
\label{RF2}
\end{eqnarray}
with
\begin{equation}
N(P_i,P_j)=n(p_i^0)+n(p_j^0)+1=n(p_i^0)-n(-p_j^0)
\end{equation}
One may symmetrize the integrant by adding the term obtained in the change of
variables $K\to -K$ : $ F'(K) =(F(K)+F(-K))/2$.

We now proceed in a way similar to the 3-gluon vertices' case (see
Sec.\ref{sec4.2}) in order to kill most of the terms appearing in
Eqs.(\ref{RC41}),(\ref{Ap5}) : \\
(i) All the terms whose thermal weight depend on the external momentum $Q$
(i.e. $n(p_4^0)=n(-q_0))$ give a vanishing contribution. In the complex $k_0$
plane, they have all their singularities on the same side of the real axis, and
one may close the contour in the other half plane. Indeed the
factor $D_{\rho\sigma}((-K)_R)=D_{\rho\sigma}(-k_0+i\epsilon, -{\bf k})$ is
multiplied for example by
\begin{eqnarray}
\lefteqn{<{v}_{\sigma}(v.P_2+i\hat{C})^{-1}v_{\mu}(v.(P_1+P_2)+i\hat{C})^{-1}v_{\nu}
 (-v.P_3+i\hat{C})^{-1}{v}_{\rho}> =}   \nonumber \\ & &
 <{v}_{\sigma}(-v.K+i\hat{C})^{-1}v_{\mu}(v.(Q-K)+i\hat{C})^{-1}v_{\nu}
 (-v.K+i\hat{C})^{-1}{v}_{\rho}> \nonumber
 \end{eqnarray}
 (ii) For a soft momentum running around the loop $k_0 <<T$, one may drop all
 terms whose factor is $N(P_4,P_i)= n(p_i^0)-n(q_0)$ , ($P_i=P_3=K$ or
 $P_i=P_1+P_3=Q+K$),  because the first contribution will come from the
 pole $p_i^0=2\pi T$ and it will lead to terms of order $q_0/T$. There is no
 singularity when $p_i^0\to 0$ since\\
 - the factor $(p_1^0+p_3^0)N(P_4,P_1+P_3)$ is regular as $p_1^0+p_3^0 \to 0$ \\
 - for $p_3^0=k_0\to 0$, most of the terms whose factor is $n(p_3^0)$ cancel,
 there remain two terms
 $$\ldots{v}_{\rho}[(v.Q-{\bf v.k}+i\hat{C})^{-1}+ (v.Q+{\bf v.k}+i\hat{C})^{-1}]
 v_{\sigma} \ldots \ \  q_0n(k_0)\ D_{\rho\sigma}(k_0=0,-{\bf k}) $$ 
 which cancel with the
 term obtained in the change of variable $K\to -K$. \\
 (iii) The only surviving term in Eqs.(\ref{RC41}),(\ref{Ap5}) has singularities
 on both sides of the real $k_0$ axis, it is
 \begin{eqnarray} 
 \lefteqn{<{v}_{\mu}(v.P_1+i\hat{C})^{-1}v_{\rho}(v.(P_1+P_3)+i\hat{C})^{-1}
 v_{\sigma} (-v.P_4+i\hat{C})^{-1}{v}_{\nu}> \   D_{\rho\sigma}((-K)_R) =}  
 \nonumber \\ & & 
 {v}_{\mu}(v.Q+i\hat{C})^{-1}v_{\rho}(v.(Q+K)+i\hat{C})^{-1}
 v_{\sigma} (v.Q+i\hat{C})^{-1}{v}_{\nu}> D_{\rho\sigma}(-K+i\epsilon)
 \nonumber
 \end{eqnarray}
 with a weight
 \begin{eqnarray}
\lefteqn{N(P_3,P_2+P_4)(p_1^0+p_3^0)=(n(p_3^0)-n(p_1^0+p_3^0))(p_1^0+p_3^0)}
 \nonumber  \\ & &
 \approx ({T\over k_0}-{T\over k_0+q_0})(k_0+q_0) = {Tq_0\over k_0}
 \end{eqnarray}
 This term is associated with the diagram drawn on Fig.\ref{fi2}(b). One adds
 the term obtained in the substitution $K\to -K$, i.e.
  $$ \ldots v_{\rho}(v.(Q-K)+i\hat{C})^{-1}v_{\sigma} \ldots  \ \ 
  D_{\rho\sigma}((K)_R=(-K)_A) \ (-{Tq_0\over k_0}) $$
  and one may complete with irrelevant terms (singularities in $k_0$ on same
  side) to obtain
\begin{eqnarray}
\lefteqn{{\Pi'}_{(4g)}^{\mu \nu}(Q) =  i  q_0 { g^2N T\over2} m_D^2 \int{d_4k\over{i(2\pi)^4}}   {1\over k_0} [ D_{\rho \sigma}((-K)_R)  - D_{\rho \sigma}(K_R) ]  } \nonumber \\   &  &
 <v^{\mu}(v.Q+i\hat{C})^{-1} \lbrack v^{\rho}(v.(K+Q)+i\hat{C})^{-1}v^\sigma   \nonumber \\
 &  & +  v^{\sigma}(v.(Q-K)+i\hat{C})^{-1}v^\rho \rbrack  (v.Q+i\hat{C})^{-1}v^\nu >  
\end{eqnarray}

Neglecting $Q$ in front of $K$, the resulting  ${\Pi'}_{(4g)}^{\mu \nu}$ is
\begin{eqnarray}
\lefteqn{{\Pi'}_{(4g)}^{\mu \nu}(Q) = - i  q_0 { g^2N T\over2} m_D^2 \int{d_4k\over{i(2\pi)^4}}   {1\over k_0} [ D_{\rho \sigma}(K_R)  - D_{\rho \sigma}(K_A) ]  }  \label{F6} \\  &  &
 <v^{\mu}(v.Q+i\hat{C})^{-1} \lbrack v^{\rho}(v.K+i\hat{C})^{-1}v^\sigma -  v^{\sigma}(v.K-i\hat{C})^{-1}v^\rho \rbrack  (v.Q+i\hat{C})^{-1}v^\nu >  \nonumber
\end{eqnarray}
A consequence is
\begin{eqnarray}
\lefteqn{ Q_{\nu} {\Pi'}_{(4g)}^{\mu \nu}(Q) = - i q_0 
{ g^2N T\over2} m_D^2 \int{d_4k\over{i(2\pi)^4}}   {1\over k_0} [ D_{\rho \sigma}(K_R)  -D_{\rho \sigma}(K_A) ]  }  \label{F7} \\  &  &
 <v^{\mu}(v.Q+i\hat{C})^{-1} \lbrack v^{\rho}(v.K+i\hat{C})^{-1}v^\sigma -  v^{\sigma}(v.K-i\hat{C})^{-1}v^\rho \rbrack   >  \nonumber
\end{eqnarray}
where Eq.(\ref{A12}) has been used. One sees  that   the factor that depends 
on $v$ in (\ref{F7}) is ${\cal{V}'}^{\mu \nu \rho}(K,Q) \ /\ k_0$  as 
in Eq.(\ref{D35}). From (\ref{D18},\ref{D20})
\begin{equation}
D_{\rho \sigma} (K_R) - D_{\rho \sigma} (K_A ) =  - i \ D_{\rho \rho'} (K_A) D_{\sigma \sigma'} (K_R) \  ( \Pi^{\rho' \sigma'} (K_R) -  \Pi^{\rho' \sigma'} (K_A)  )
\label{F8}
\end{equation}
so that comparing with (\ref{D38}) one obtains
\begin{equation}
Q_{\mu}{\Pi'}_{(4g)}^{\mu \nu} = - Q_{\mu}{\Pi'}_{(3g)}^{\mu \nu} 
\end{equation}
Moreover, comparing ${\Pi'}_{(4g)}^{\mu \nu}(Q)$ as in (\ref{F6}) with (\ref{D15}) for the case $\mu = j  \ ,\   \nu = i $ one extracts a collision term (see (\ref{D40}))
\begin{eqnarray}
\lefteqn{ {\Phi'}_{(4g)}({\mathbf{v}, \mathbf{v}'}) =  - i  
\int{d_4k\over{(2\pi)^4}}   {1\over k_0 m_D^2} [ D_{\rho \sigma}(K_R)  
-D_{\rho \sigma}(K_A) ]  }  \label{F9} \\  &  &  
  \lbrack v^{\rho}(v.K+i\hat{C})^{-1}_{v,v'}{v'}^{\sigma} -   
  v^{\sigma}(v.K-i\hat{C})^{-1}_{v,v'}{v'}^{\rho} \rbrack    \nonumber  
\end{eqnarray}
${\Phi'}_{(4g)}$ is an operator. When applied  on a function  $>$ that does not depend on $\mathbf{v}'$
\begin{equation}
{\hat{\Phi}'}_{(4g)} > = \int{d\Omega_{\mathbf{v}'_1}\over{4\pi}} {\Phi'}_{(4g)}({\mathbf{v}_1, \mathbf{v}'_1}) = -  \int{ d\Omega_{\mathbf{v}_2} \over{4\pi}} {\Phi'}_{(3g)}({\mathbf{v}_1, \mathbf{v}_2})
\end{equation}
from (\ref{D45}) and (\ref{F8}), i.e. the collision operator has a zero mode
\begin{equation}
\hat{C}' > \ ={\hat{\Phi}'}_{(4g)} >  + {\hat{\Phi}'}_{(3g)} >  =0
\end{equation}
${\Phi'}_{(4g)}$ does not depend on the gauge parameter $\xi$, but the
longitudinal gluon exchange does depend on the gauge (covariant or Coulomb) since
$k_0\neq {\bf v.k}$. 
If one restrict oneself to transverse gluon exchange,  (\ref{F9}) may be written
\begin{eqnarray}
 \lefteqn{ {\Phi'}_{(4g)}({\mathbf{v}, \mathbf{v}'}) = 
  - i  \int{d_4k\over{(2\pi)^4}}  \  {1\over k_0 m_D^2} 
  [\Delta^t(K_R) - \Delta^t(K_A) ]  } \label{F11}\\
&  &  v_t^i \  [ (v.K + i \hat{C})^{-1} - (v.K - i \hat{C})^{-1} ]_{v v'}\ 
 {v'}_t^i \nonumber
 \end{eqnarray}
To go back to the expression that corresponds to momentum exchange
 $K \sim gT$, one replaces $\hat{C}$ by $\epsilon{\mathcal I}$ in (\ref{F9}) and 
 one gets back the local term in $\hat{C}$  (see (\ref{A9})). Indeed, 
 the identity operator in $\mathbf{v}$ space is 
 ${\mathcal{I}} = \delta_{S_2}(\mathbf{v}-\mathbf{v}')$, and 
 one recognizes in (\ref{F8},\ref{F9}) the hard gluon damping rate $\gamma$  
 if $\hat{C}$ is replaced by $\epsilon{\mathcal I}$. ${\Phi'}_{(4g)}$ 
  comes from the self-energy diagram drawn on Fig.\ref{fi2}(b) involving a
  collective-excitation propagator and a soft-gluon propagator, it is
  interpreted as the damping of a collective excitation (at a scale larger than
  $k^{-1}\sim (g^2T\ln 1/g)^{-1}$),  
 its scale is $\bar{\Phi}_2$ discussed in Sec.\ref{sec4.3}.

  \section{Properties of the Collision operator \ $\hat{C}'$}
  \label{sec5}
  \setcounter{equation}{0}
  As a result of Sec.\ref{sec4}, the collision operator $\hat{C}'$ that takes 
  into account the collisions with exchanged gluon $k\sim g^2T \ln 1/g$ is
  \begin{equation}
\hat{C}' = C'({ \mathbf{v}, \mathbf{v}'}) = 
- m_D^2{g^2N T\over2} (  {\Phi'}_{(3g)}({\mathbf{v}, \mathbf{v}'}) 
+  {\Phi'}_{(4g)}({\mathbf{v}, \mathbf{v}'}) )
\label{G01}
\end{equation}
with $ {\Phi'}_{(3g)}(\mathbf{v}, \mathbf{v}')$ as in (\ref{D41}) , 
or (\ref{D44}) for transverse gluon exchange,  
$ {\Phi'}_{(4g)}(\mathbf{v}, \mathbf{v}')$ as in (\ref{F9}) , 
or (\ref{F11}). Both are functions of  $ \mathbf{v}. \mathbf{v}'$ as 
there is no other vector available. \\
In this section one restricts oneself to the dominant transverse gluon 
  exchange 
  \begin{eqnarray}
 \lefteqn{ {\Phi'}_{(3g)}({\mathbf{v}_1, \mathbf{v}_2}) = 
  -   \int{d_4k\over{(2\pi)^4}} \ \vert \Delta^t(K_R) \vert^2 } \label{G03} 
  \\ & &
  < \ v_{1i}^t(v_1.K+i\hat{C})^{-1}{v'}_{1j}^t
-v_{1j}^t(v_1.K-i\hat{C})^{-1}{v'}_{1i}^t\ >_{v_1'} \nonumber \\ & & 
< \ v_{2i}^t(v_2.K+i\hat{C})^{-1}{v'}_{2j}^t
-v_{2j}^t(v_2.K-i\hat{C})^{-1}{v'}_{2i}^t\ >_{v_2'} \nonumber
\end{eqnarray}
   \begin{eqnarray}
 \lefteqn{ {\Phi'}_{(4g)}({\mathbf{v}, \mathbf{v}'}) = 
  - i  \int{d_4k\over{(2\pi)^4}}  \  {1\over k_0 m_D^2} 
  [\Delta^t(K_R) - \Delta^t(K_A) ]  } \label{G04}\\
&  &  v_t^i \  [ (v.K + i \hat{C})^{-1} - (v.K - i \hat{C})^{-1} ] \ 
 {v'}_t^i \nonumber
 \end{eqnarray} 
  where $v^i_t =v^i - {\hat{k}}^i {\bf v.\hat{k}}$

  \noindent The operator $C'({\bf v},{\bf v'})$ shares many features of the operator
   $C({\bf v},{\bf v'})$: \\
  - it commutes with the rotations in ${\bf v}$ space, its eigenvalues are 
  \begin{equation}
  c_l'=< P_l({\bf v.v'}) C'({\bf v},{\bf v'}) >_{v,v'}
  \label{G1}
  \end{equation}
  and its eigenvectors are $Y_l^m({\bf v})$, \\
  - $\hat{C}'$ is written as an integral over the soft momenta $k$ of the
  exchanged gluon in the near-forward collision of two collective excitations of
  direction ${\bf v}$ and ${\bf v'}$, ($k\sim gT$ for $\hat{C}$, $k \sim g^2 T
  \ln 1/g$ for $\hat{C}'$). All the dependance on ${\bf v}$ and ${\bf v'}$  is
  in the effective vertices, i.e. in the (smeared) energy-momentum conservation
  at the vertex \\
  - the soft gluon exchange $k\sim g^2T\ln 1/g$ put a strong weight  upon the
  region $k_0\leq (k^2/T) \ln 1/g$ and one can set $k_0=0$ in the effective
  vertices whose scale is $k_0\sim k\sim g^2T \ln 1/g$ (with one exception). (For
  $\hat{C}$ one also sets $k_0=0$ to leading order in the vertices). The
  resulting scale for $c_l'$ is $g^2NT$, with a weak dependance on both cutoff 
  $\mu_3<k<\mu_2$. \\
  - the collision operator is made of two terms. One term comes from a self-energy
  diagram, in fact it is the damping of the collective excitation at a scale 
  larger  than $1/k$. It dominates the spectrum of the operator, except for 
  the $l=0$ 
  eigenvalue where both terms cancel each other, giving rise to a zero
  eigenvalue. The contribution from the other term vanishes for $l=1, 3, 5
  \ldots$ to leading order. 
  
  The operators $\hat{C}$ and $\hat{C'}$ also show some differences \\
  - the infrared behaviour of the eigenvalues differ. For a transverse gluon
  exchange, the $c_l$ have a logarithmic infrared divergence that shows up in
  both terms of the collision operator. The $c_l'$ are infrared finite with one
  exception: the $l=1$ eigenvalue has a linear infrared divergence (see
  underneath). \\
  - the operator $\hat{C}$ is expected to be gauge independent. As seen in
  Sec.\ref{sec4.2}, $\hat{C}$ takes an identical form in covariant and Coulomb
  gauges. For $\hat{C'}$, the longitudinal soft gluon exchange is very likely
  gauge dependent, apparently a consequence of the finite lifetime of the
  collective excitations at a smaller scale.

  The collision operators $\hat{C}$  and $\hat{C}'$ are so similar that it is
  easy to take into account both types of collisions, i.e. when the exchanged
  gluon has $k\sim gT$ and when $k\sim g^2T \ln 1/g$. One just has to substitute
  the operator $\hat{C}+\hat{C}'$ to the operator $\hat{C}$ in all the relations
  written in Sec.\ref{sec2} and Sec.\ref{sec3}. The  transport
  equation for the $W$ field is now
  $$ (v.D +i\hat{C} +i\hat{C}') \ W \ = \ {\bf v.E} $$
  and from it, one deduces the induced current, then the softer amplitudes. These
  softer amplitudes are tree-like and they obey tree-like Ward identities.

This approach, via the polarization tensor,  proposes an interpretation of    $\hat{C}$  and  
of   $\hat{C}'$ somewhat different from the loss and gain interpretation that results from the transport equation via the Schwinger-Dyson approach \cite{AY2,BI1}. In that alternative approach, the Bose symmetry of the effective vertices with respect to two legs of the same type is the dominant constraint, as a consequence the colour factors differ from the ones that appear in the Schwinger-Dyson case.

  \subsection{The eigenvalues of the operator  $\hat{C}'$}
  \label{sec5.1}
   In Appendix \ref{secC}, matrix elements such as
   \begin{equation}
   M_{(4g)}^{(l)}(K) =<P_l({\bf v.v'}) \  \ v^t_i\ [(v.K+i\hat{C})^{-1} -
   (v.K-i\hat{C})^{-1}] \ {v'}_i^t >_{v,v'}
   \label{G2}
   \end{equation}
   are expressed in terms of matrix elements of $(v.K\pm i\hat{C})^{-1}$, which
   are themselves continued fractions, fonctions of $k_0,k$ and of all the
   eigenvalues $c_l$ of the operator $\hat{C}$. \\
   The case of the eigenvalue $l=0$ of $\hat{C}'$ has been treated in
   Sec.\ref{sec4.3}. Many encountered features are valid for all $c_l'$. For
   $l=0$, the contributions of both terms $\Phi_{(3g)}'$ and  $\Phi_{(4g)}'$ are
   equal and opposite, it has been called $\bar{\Phi}_2$. $\bar{\Phi}_2$ has been 
   expressed in terms of the imaginary part of the $l=1, m=1$ eigenvalue of 
   $(v.K+i\hat{C})^{-1}$ which is finite as $k\to 0$. One consequence is that
   the integral over $k$ in (\ref{G03}) and (\ref{G04}) is infrared finite 
   (see Eq.(\ref{D56})).
   
   One can see that there is no infrared divergence in the contribution of
   $\Phi_{(4g)}'$  to $c_l'$ in the following way. The scale of the matrix
   elements of $(v.K+i\hat{C})$ is the scale of $\hat{C}$, i.e. $\gamma\sim
    g^2T\ln 1/g$. For $k<<\gamma$, the matrix elements of 
    $(v.K+i\hat{C})^{-1} \approx (k_0+i\hat{C})^{-1}$ do not depend on $m={\bf
    l.\hat{k}}$, but only on $l$, they are the inverse of 
    $< P_l({\bf v.v'}) (k_0+i\hat{C}) >_{v,v'} = k_0 +i c_l$. In Eq.(\ref{G2}),
    there is no dependance on $m$ for $k<<\gamma$ and one may write
    \begin{eqnarray}
    {\bf v}_t.{\bf v}_t' \ P_l({\bf v.v'}) &=& 
    {2\over3} \ {\bf v.v'} \ P_l({\bf v.v'}) \nonumber \\
    &=& {2\over3}{1\over 2l+1}[\ (l+1)P_{l+1}({\bf v.v'}) +
     l P_{l-1}({\bf v.v'})\ ] 
     \label{G3}
     \end{eqnarray}
     i.e. for $k<<\gamma$,  Eq. (\ref{G2}) becomes
     \begin{equation}
     M_{(4g)}^{(l)}(k_0)={2\over3}{1\over 2l+1}[ \ {l+1\over{k_0+ic_{l+1}}} + 
     {l\over{k_0+ic_{l-1}}} -\ {\mathrm c.c.}]
     \label{G4}
     \end{equation}     
     where c.c. means complex conjugate. One can set $k_0=0$ in 
     $M_{(4g)}^{(l)}(k_0)$ and one concludes that, just as for the case $l=0$,
     the integral over $k$ has no infrared divergence as $k \to 0$. And the
     region $k<< \gamma$ of the integrant does not depend on $l$ for large $l$
     since $c_l=\gamma-\delta_l$ with $\delta_l\sim \gamma/l$. There is one
     exception: the case $l=1$ for $c_l'$, where one sees from (\ref{G4}) that
     the eigenvalue $c_0=0$ of the operator $\hat{C}$ enters. As fully discussed
     in Appendix \ref{secC3}, one finds a linear divergence as $k\to 0$ in $c_1'$
     (whatever the order of the limits $k_0\to 0$ and $k \to 0$), whose origin 
     is the zero eigenvalue of $\hat{C}$ (see Eqs.(\ref{Ap23},\ref{Ap27})). 

    The contribution from $\Phi_{(3g)}'$ vanishes for $l$ odd, as the entering 
     matrix
     elements vanish for $k_0=0$. For $l$ even, only the sector $|m|=1$ of 
     $(v.K+i\hat{C})^{-1}$ enters, the contribution is infrared finite (see
     Appendix \ref{secC2}).

      \subsection{The colour conductivity}
      \label{sec5.2}
      The properties of this quantity are first summarized for the case when $gT$
      has been integrated out. In Sec.\ref{sec3.2}, the induced current
      $j_{\mu}^{ind}(X)$ has been written in terms of the soft field $E^b(Y)$.
      One may introduce the conductivity tensor
      \begin{equation}
      j_{\mu}^a(X)=\int d_4Y \ {\sigma}_{\mu i}^{a b}(X,Y) \ E_i^b(Y)
      \end{equation}
      The comparison with Eq.(\ref{RC20}) in Sec.\ref{sec3.2} gives
      \begin{equation}
      {\sigma}_{\mu i}^{a b}(X,Y) = i m_D^2\ < v_{\mu} \ 
      G_{ret}^{ab}(X,Y; A\ ;{\bf v},{\bf v'})\ v_i' \ >_{v,v'}
      \end{equation}
      The linearized part is in momentum space
      \begin{eqnarray}
      {\sigma}_{\mu i}^{a b}(K) \vert_{A=0} &=& i m_D^2
      <v_{\mu}\ (v.K+i\hat{C})^{-1} \ v_i' >_{v,v'} \delta^{ab}
      \label{G7} \\
      \Pi^{\mu i}(K)&=&-i k_0\ \sigma^{\mu i}(K)
      \end{eqnarray}
      One introduces the transverse and the longitudinal part of the
      conductivity \cite{BI1}
      \begin{eqnarray*}
      E^i(K) &=& \hat{k}^i E_l +E^i_t \\
      \sigma_t &=& {i\over k_0} \ \Pi^t ={i\over 2 k_0}\ {\cal P}_t^{ij}\  \Pi^{ij} \\
      \sigma_l &=& -i{k_0\over k^2} \ \Pi^l =i{k^i\over k^2} \ \Pi^{0i}
      \end{eqnarray*}
      \begin{eqnarray}
      \sigma_t&=&im_D^2{1\over 2}\ <v_t^i (v.K+i\hat{C})^{-1} v_t^i>
      \label{G8} \\
      \sigma_l&=&im_D^2{1\over k^2}\ < k_0 \ (v.K+i\hat{C})^{-1} {\bf v.k}>
      \end{eqnarray}
      With $v.K =k_0-{\bf v.k}$ and relation (\ref{A12}), the expression for
      $\sigma_l$ may be written in alternative forms
      \begin{eqnarray}
      \sigma_l&=&im_D^2\ {1\over k^2}\ < {\bf v.k} (v.K+i\hat{C})^{-1} {\bf v.k}>
      \label{G9} \\
      \sigma_l&=&im_D^2\ {k_0\over k^2}\ [ \ k_0 \ <(v.K+i\hat{C})^{-1}> - 1] 
      \label{G10}
      \end{eqnarray}
      i.e. in terms respectively of the matrix elements $l=m=1$ and $l=m=0$ of 
      $(v.K+i\hat{C})^{-1}$
      \begin{eqnarray}
      \sigma_t&=&{im_D^2\over3}\ \Sigma_1(k_0.k) \\
      \sigma_l&=&im_D^2\ {k_0\over k^2}\ (k_0\Sigma_0(k_0.k) \ -1)
      \end{eqnarray}
      In Appendix \ref{secC}, Eq.(\ref{Ap46}) gives $k_0(k_0\Sigma_0  -1)/k^2$ as a
      compact continued fraction for an easy comparison with Eq.(\ref{Ap2}) for
      $\Sigma_1$. \\
      For $k<\gamma/3$, the expansion in $k^2$ of the continued fraction 
      converges
      for all $k_0$ (see Appendix \ref{secA}) and one obtains
      \begin{eqnarray*}
      \sigma_t(k_0, k<<\gamma) &=&{im_D^2\over3}\ [{1\over{k_0+ic_1}} +
      {1\over5}\frac{k^2}{(k_0+ic_1)(k_0+ic_2)} + O(k^4)] \\
      \sigma_l(k_0, k^2<<k_0\gamma) &=&{im_D^2\over3}\ [{1\over{k_0+ic_1}} +
      {4\over15}\frac{k^2}{(k_0+ic_1)(k_0+ic_2)} + O(k^4)]
      \end{eqnarray*}
      i.e. $\sigma_t$ and $\sigma_l$ differ by terms of order 
      $k^2/(k_0+ic_1)(k_0+ic_2)$. For $\sigma_l$, a different limit is (see
      (\ref{Ap46}))
      $$\sigma_l(k_0\gamma<<k^2) = -im_D^2 \ {k_0\over k^2} $$ 
      
      We now examine how the collisions at the scale $g^2T \ln 1/g$ affect the
      conductivities. As said at the beginning of Sec.\ref{sec5}, the solution
      that includes both types of collisions is
      \begin{equation}
      \Pi^{ji}(K)= k_0 m_D^2<v^j\ (v.K+i\hat{C}+i\hat{C}')^{-1} \ v^i>
      \label{G11}
      \end{equation}
      As a consequence, in the expressions just written for $\sigma_t,\sigma_l$,
      i.e. Eqs.(\ref{G8}, \ref{G9}, \ref{G10}), one just have to substitute 
      $\hat{C}+\hat{C}'$ to $ \hat{C}$, i.e. substitute $c_l+c_l'$ to $c_l$ in
      every continued fraction, in particular
      \begin{equation}
      \sigma_t(k_0,k<<\gamma)=\sigma_l(k_0, k^2<<k_0\gamma) = 
      {m_D^2\over3}\ {i\over{k_0+i(c_1+c_1')}}
      \label{G12}
      \end{equation}
       so that
       \begin{equation}
       \sigma_t=\sigma_l=\sigma(k_0<<\gamma, k<k_0) ={m_D^2\over3}\ {1\over
       c_1+c_1'}
       \end{equation}

       \noindent{\it  A Comparison with related work}\\
       We now examine how these results are related to the pioneers' work of
       Arnold and Yaffe \cite{AY1}. Their interest is in the next-to-leading-log
       contribution to the colour conductivity.\\
       Their approach makes use of effective theories. They are lead to compute
       the same pair of one-loop diagrams contributing to a self-energy in an
       effective static theory, in the Coulomb gauge. Their operator
       $\hat{O}({\bf 0})$ is identical to the collision operator $i\hat{C}'$, if
       in  $\hat{C}'$ one makes the static approximation $k_0=0$ in the effective
       vertices, and one performs the integral over $k_0$ with only the weight
       of the soft transverse propagator; it is an approximation which is done
       at the very end in this work, they do it at the start. Their $W$ field
       propagator
       $$\hat{G}_0({\bf k}) = i  (v.K+i\hat{C})^{-1} \vert_{k_0=0} $$
       is a real quantity (see the matrix elements $i \Sigma_m(k_0=0,k)$ in
       Eq.(\ref{Ap2})). They are lead to consider the same quantity 
       $$<v_l \  \hat{O}({\bf 0}) \ v_l > = i \ c_1'  $$
       and they find that the relevant matrix element is
       $$<v_lv_i\  \hat{G}_0({\bf k})\ v_j v_l> {\cal P}_t^{ij}(k) $$
       in agreement with $M_{(4g)}^{(l=1)}(k_0=0,k)$ in Eq.(\ref{G2}). The
       expression of this matrix element in terms of those of 
       $\hat{G}_0({\bf k})$ agrees with Eq.(\ref{Ap45}) in Appendix \ref{secC3}.
       They encounter the infrared linear divergence of $ c_1'$ which
       disappears upon dimensional regularization. At the soft level ($gT$
       integrated out), their conductivity tensor agrees with Eq.(\ref{G9}) for
       $\sigma_l$. However, in their approach, once the contribution of the
       momenta $k\sim g^2T\ln 1/g$ is included, the effective conductivity is
       not merely obtained by the substitution of $c_1+c_1' =c_1(1 +c_1'/c_1)$
       to $c_1$ in the expression for $\sigma_l$, another term is added arising
       from the small momentum expansion of their self-energy (see a detailed
       comparison after Eq.(\ref{Ap35}) in App.\ref{secC3}).
    
  \section{Conclusion}
  To the near-forward collision of two collective colour excitations of the
  thermal gluons $p \sim T$ is associated a collision operator $\hat{C}$ when
  the gluon exchanged during the collision has $k\sim gT$. This
  operator has an infinite number of eigenvalues $c_l$ that may be interpreted
  as the multipole moments of a rate. The
  $l=0$ eigenvalue is zero, a cancellation between a damping term and another
  term, the other $c_l$ are dominated by the damping term, whose scale is
  $g^2T\ln 1/g$. 
  When inserted into the transport equation that describes the evolution at
  some space-time scale $x$ of the collective excitations, the collision term
  accounts for the effective damping of the excitations due to the integration
  over smaller scales. \\
  This work has presented two results. The first result is the explicit form of
  the effective $n$-gluon amplitudes when the scale $T$ and $gT$ are integrated
  out. These amplitudes exhibit remarkable properties. \\
  The picture that emerges follows. The central role is played by the collective
  colour excitations. The soft gluons of momentum $k << gT$ are emitted by the
  collective excitations that occur at the scale $1/k$. These excitations are 
  associated with
  the transport equation, not with the structures seen in the resummed gluon
  propagator (such as quasiparticle poles).  The effective amplitudes are
  tree-like and they obey tree-like Ward identities. The propagator along the
  tree, the one of the collective excitation, propagates an infinite tower of
  damping rates  and one undamped mode. All the rates show
  an infrared log divergence. The presence of the undamped mode (the
  eigenvalue $c_0=0$) turns out to be essential i) for the Ward identities to be
  satisfied, ii) for the strong similitude with the HTL amplitudes. \\
  The second result is, a new collision operator $\hat{C'}$ that allows to take
  into account the collisions with soft gluon exchange $k\sim g^2T\ln 1/g$ has
  been found by means of a perturbative approach, i.e. by computing the
  one-soft-loop diagrams (with loop momentum $k\sim g^2T\ln 1/g$) which enter
  the polarization tensor. The operator $\hat{C'}$ shares many features of the
  operator $\hat{C}$ : a zero mode, an
  infinite number of eigenvalues $c_l'$ which are expressed in terms of the
  $c_l$.\\
  For a transverse gluon exchange, the landscape in the infrared has changed.
  The logarithmic divergence that was entering the damping rates $c_l$ has
  disappeared. The eigenvalues $c_l'$ are finite and of order $g^2T$  (with
  some dependence on $\ln 1/g$). There is one exception, the $l=1$ eigenvalue
  exhibits a linear infrared divergence, which is linked to the zero mode of
  the operator $\hat{C}$ (as a result of angular momentum combination).\\
  
  The properties of $\hat{C}$ and $\hat{C'}$ are so similar that it is easy to
  take into account the collisions with $k\sim g^2T\ln 1/g$. One just has to
  substitute $\hat{C} + \hat{C'}$ to $\hat{C}$ in the transport equation, i.e.
  $c_l+c_l'$ to $c_l$ in its solution. All what  has  just been said for the
  $n$-gluon amplitudes where $\hat{C}$ enters ($T$ and $gT$ integrated out) are
  valid for the softer amplitudes where $\hat{C} + \hat{C'}$ enters; they are
  tree-like and satisfy tree-like Ward identities. The quantities $c_l+c_l'$ are
  infrared finite except for the $l=1$ case. This $l=1$ eigenvalue dominates the
  polarization tensor at momentum $q << g^2T\ln 1/g$. One possible
  interpretation of this linear infrared divergence is that it is the signal of
  new physics occurring at the scale $(g^2T)^{-1}$. \\
   The collision operator  $\hat{C}$ is gauge
  independent, in contrast the longitudinal gluon exchange in the operator
  $\hat{C'}$ is very likely gauge dependent. \\
  It remains to be seen whether integrating out the scale $g^2T\ln 1/g$ only
  amounts to take into account the collisions through $\hat{C'}$.
  
  Also, this work sheads a new light on what kind of process is being summed by
  means of the operator $\hat{C}$ or $\hat{C'}$ in the polarization tensor 
  \cite{AY1,BI1}.
   The soft gluon polarization tensor may be written as an infinite
  series in powers of  $\hat{C}$ (at scale $q << gT$) or of $\hat{C'}$ (at
  scale $q<< g^2T\ln 1/g$). $\hat{C}$ and $\hat{C'}$ are obtained from a truncation
  of part of the effective vertices that enter the one-soft-loop diagrams. The
  truncation amounts to cut-off one initial propagator of the collective
  excitation. One term entering the collision operator has been shown to
  come from a self-energy diagram involving a collective excitation and a soft
  gluon (in fact a damping rate); the terms in the series in powers of that term
   have alternating
  factors, one collective excitation's propagator, one self-energy factor, one 
  propagator \ldots, i.e. the series is just the resummation of the self-energy 
  of the
  collective excitation. For the collision's other term, the alternating factors
  are one collective excitation's propagator, two soft gluons, one propagator
  \ldots. Since the exchanged gluons have $k_0<k$, this is a $t$ channel
  picture; in the crossed channel, one collective excitation scatters and dies
  out, it is replaced by another collective excitation at a different space-time
  point, this transition rate involves a two-gluon exchange.  
This picture is scale invariant. When one goes from $\hat{C}$ to $\hat{C'}$ the change is in the collective excitation's propagator, a straight-line propagation for $\hat{C}$, fluctuations in  the direction $\bf v$ for $\hat{C'}$ as a result of the processes included in $\hat{C}$.

\subsection*{Acknowledgments}

 The author wishes to thank E. Iancu for a  useful discussion at INT and 
 for his constant interest, and E. Petitgirard for an e-correspondence.\\
Part of this work was done while the author stayed at the INT session 
"Non-equilibrium Dynamics  in Quantum Field Theory"(INT-99-3) and the author 
wishes to thank the INT and L. Yaffe for the extended hospitality and for 
generating many friendly, open discussions. The hospitality of the LAPTH-Annecy
is acknowledged.

\appendix
\setcounter{equation}{0}
\section{The analytic properties of  $\Pi$ in the complex $k_0$ plane}
\label{secA}
\subsection{An explicit expression for $\Pi^t$ and $\Pi^l$}
 Arnold and Yaffe \cite{AY1}  have written down the characteristics of 
 the operator $ [k_0-{\mathbf{v}.\mathbf{k}} + i C({\mathbf{v}, \mathbf{v}'}) ]
  ^{-1}$ for the case $k_0 =0$. The extension to the case $k_0\neq 0$ is 
  straightforward. $C(\mathbf{v},\mathbf{v}')$ commutes with the rotations 
  in $\mathbf{v}$ space, its eigenvectors are the spherical harmonics
  $\sqrt{4\pi}\ Y_l^m({\mathbf{v}}) = \vert l m >$ with the measure 
  ${d\Omega_{\mathbf v}/4\pi}$
\begin{equation}
\hat{C} \  \vert lm > = c_l \vert l m > = (\gamma - \delta_l) \vert l m >
\label{Ap1}
\end{equation}
\begin{equation}
c_l = < C({\mathbf{v}, \mathbf{v}'}) P_l({\mathbf{v}.\mathbf{v}'}) >_{v v'} \ \ \ \ c_0 =0 \  , \  \ \ c_l > 0\ \ \ {\mathrm{for}}\ \   l>0
\end{equation}
$C({\mathbf{v},\mathbf{v}'})$ is given in (\ref{A9}).
\\ If one chooses $\mathbf{\hat{k}}$ as the $z$ axis
\begin{equation}
{\mathbf{v}.\mathbf{k}} \  \vert l\  m > = k\ \  ( \  b_l^{(m)} \vert l+1\  m > +
 b_{l-1}^{(m)} \vert l-1 \ m > \ \ )
\end{equation}
\begin{equation}
b_l^{(m)} = \sqrt{{(l+1)^2-m^2}\over{4(l+1)^2 - 1}}
\label{RAp1}
\end{equation}
For fixed $m$, the matrix $ [k_0-{\mathbf{v}.\mathbf{k}} + i C({\mathbf{v}, \mathbf{v}'}) ] $ is a tri-diagonal matrix whose inverse is known. In particular, the element 
$$ \Sigma_m = < l=m \ \   m \vert  ( k_0-{\mathbf{v}.\mathbf{k}} + i C({\mathbf{v}, \mathbf{v}' })) ^{-1} \vert l=m \ \   m > $$
is the continued fraction
\begin{equation}
\Sigma_m(k_0, k) = {1\over{k_0 + i c_m}}\  { \frac{1}{1- b_m^{(m)2} {\displaystyle{\frac{\rho_m^2}{1-b_{m+1}^{(m) 2}{\displaystyle{{\frac{ \rho^2_{m+1}}{1 -b_{m+2}^{(m)2}{\displaystyle{\frac{\rho^2_{m+2}}{1- \  \ . \ . \ .}}}}}}}}}}}}
\label{Ap2}
\end{equation}
where 
\begin{equation}
\rho_m^2 = {k^2\over{(k_0+ic_m)(k_0+ic_{m+1})}}
\end{equation}
the transverse part of the self-energy is
\begin{equation}
\Pi_R^t(k_0, k) = {m_D^2\over 3} k_0 \ \Sigma_1(k_0, k)
\label{Ap3}
\end{equation}
since from (\ref{B7}),(\ref{D20})
\begin{equation}
\Pi^t_R(k_0,k)=
m_D^2 k_0 < v^i (v.K + i \hat{C})^{-1} {v'}^j >_{v v'} (\delta_{i j } - \hat{k}_i\hat{k}_j){1\over 2}
\end{equation}
 and
\begin{equation}
{1\over{4\pi}}{\mathbf{v}_t . \mathbf{v}'_t} = {1\over3}[ Y^1_1({\mathbf{v}})Y^{1*}_1({\mathbf{v}'}) + Y^{-1}_1({\mathbf{v}})Y^{-1*}_1({\mathbf{v}'}) ]
\end{equation}
Similarly, from (\ref{B7})
\begin{equation}
\Pi_R^l = - \Pi_{00} = -m_D^2 [ k_0 \Sigma_0(k_0 , k) - 1 ]
\label{Ap4}
\end{equation}
If one restricts oneself to transverse gluon exchange, to leading order,
 the  eigenvalues of the collision operator, defined in (\ref{Ap1}),   have the properties
\begin{equation}
\delta_{2l+1} =0 \ \ \  , \ \ \ 0 < \delta_{2l} \leq \delta_2 ={5\over8}\gamma  \ \ \ \ \ \delta_{2l} \rightarrow {2\gamma\over l} \ \ { \mathrm{as}}\ \  l \rightarrow \infty
\end{equation} 
so that the upper bound  $\bar{\delta}$ for $\delta_l$ is $\delta_2$. In the following, the discussion is restricted to $\Pi^t$. The extension to any $\Sigma_m$ is immediate.

\subsection{ The singularity-free region in the complex $k_0$ plane}

i)The Legendre Functions

If one sets $\gamma\neq 0$ and $\delta_l =0$ ,  the continued fraction (\ref{Ap2}) is a function of $\rho^2 = ( k / (k_0 + i \gamma) )^2$, its value is well known from the case $\gamma=0$
\begin{equation}
\Pi^t_R = { m_D^2\over 3} {k_0\over k} [ Q_0({{k_0 + i \gamma }\over k}) -Q_2({{k_0 + i \gamma }\over k})  ]
\end{equation}
The analytic properties of $Q_l(z)$ in the complex $z$ plane are seen from
$$ Q_l(z) = {1\over 2} \int_{-1}^{1} dx {P_l(x) \over{z-x}} $$
$Q_l(z)$ has a logarithmic singularity at $ z+1$ and $z=-1$, and for $\parallel z \parallel >1 $, $Q_l(z)$ has a series expansion 
$$Q_l(z) = \sum_{n \geq l} a_{l n} z^{-n}   \ \ \ \ \ \ {\mathrm{with}}\ \ \  a_{l n}  \geq 0 $$
For $ \parallel z \parallel  >  1 $ this series is absolutely convergent i.e. $\sum_n a_{l n} \parallel z \parallel^{-n} $ converges. \\
ii) The case $\delta_l\not= 0$

The continued fraction  (\ref{Ap2})  can be expanded in powers of $k^2$ and the domain of convergence of this expansion delimited. In this expansion there appear products of 
$${k^2\over{(k_0 + i (\gamma -\delta_m))(k_0 + i (\gamma -\delta_{m+1}))} }$$  
Since all numerical coefficients of the expansion in $k^2$ are positive, the series has as an upperbound the series of the modulus of its terms. Then one just needs a lower bound for
$${\parallel k_0 + i (\gamma -\delta_m) \parallel}^2 = ({\mathrm{Re}}k_0)^2 + ( {\mathrm{Im}}k_0 + \gamma -\delta_m)^2 $$ 
If $\bar{\delta}$ is the upper bound of the $\delta_m \ \ (\bar{\delta}< 2\gamma /3)$, the lower bounds are
\begin{eqnarray}
{\mathrm{Im}}k_0 + \gamma - \bar{\delta} >0  & \rightarrow &  \parallel k_0 + i(\gamma - \delta_m) \parallel >  \parallel k_0 + i(\gamma - \bar{\delta}) \parallel  \nonumber \\
{\mathrm{Im}}k_0 + \gamma < 0  & \rightarrow &  \parallel k_0 + i(\gamma - \delta_m) \parallel  > \parallel k_0 + i \gamma  \parallel  \nonumber \\
-\gamma <{ \mathrm{Im}} k_0 < -(\gamma - \bar{\delta})  & \rightarrow & \parallel k_0 + i(\gamma - \delta_m) \parallel  > \vert {\mathrm{Re}} k_0 \vert \nonumber
\end{eqnarray}
For the region $ {\mathrm{Im}}k_0 + \gamma - \bar{\delta} >0$, the expansion in $k^2$ of the continued fraction is certainly convergent in the domain  
$\parallel  k /(k_0 + i ( \gamma - \bar{\delta})) \parallel  < 1 $ as it is the domain of absolute convergence of the expansion in $k^2$ of $ Q_l(( k_0 + i ( \gamma - \bar{\delta}))/k)$, i.e., in the complex $k_0$ plane, the expansion is convergent out of a half disk of radius $k$ centered at $ k_0 = - i ( \gamma - \bar{\delta})$. Similarly, for the region ${\mathrm{Im}}k_0 < - \gamma $, it is certainly convergent out of a half disk centered at $k_0 =- i \gamma $. And for $-\gamma < {\mathrm{Im}} k_0 < -(\gamma - \bar{\delta})$, it is convergent for  $\vert {\mathrm{Re}} k_0 \vert  > k$.
\\
To summarize, the expansion of $\Pi^t_R$ in powers of $k^2$ is certainly 
convergent in the complex $k_0$ plane out of a domain made of two half disks 
and a rectangle (see Fig.\ref{fi3}). In particular, for $ {\mathrm{Im}} k_0 \geq 0 $ the expansion is certainly convergent for
\begin{equation}
({\mathrm{Re}}k_0)^2 + ( {\mathrm{Im}}k_0 + \gamma -\bar{\delta})^2   > k^2      \ \ \ \ {\mathrm{with}} \ \ \ \  \bar{\delta} = {\mathrm{sup}}_m \delta_m  < 2 \gamma /3
\label{Ap5b}
\end{equation}

\subsection{The Imaginary parts of $\Pi^t_R$}
$\Pi^t_R$ has two types of imaginary parts along the real $k_0$ axis. \\
i) all along the axis, $\Pi^t_R$ gets an imaginary part from the eigenvalues $c_m$ of the collision operator, i.e. from the damping of the color excitations arising from collisions at the scale $(gT)^{-1}$. \\
ii) out of the domain of convergence of the expansion in $k^2$, $\Pi^t_R$ gets another imaginary part similar to the $i\pi$ term of $Q_{2l}(z)$.  For example, for $Q_0(z)$ in the complex $z$ plane,
\begin{eqnarray}
\parallel z \parallel   > 1  &  &  \ \ Q_0(z) = {1\over 2} \ln ({ {z+1}\over{z-1}})  \nonumber \\
\parallel z \parallel   <  1  &   & \ \    Q_0(z) =  - i {\pi\over2}  + {1\over 2} \ln ({ { 1+ z }\over{1-z}})   \ \ \ {\mathrm{above}  \ \mathrm{the}  \ \mathrm{cut}}  \ \  z=-1 \   {\mathrm{to}} \ z=1  \nonumber
\end{eqnarray}
From the shape  of the domain of convergence of the expansion 
in $k^2$ (see Fig. \ref{fi3}), this imaginary part may only appear in the region  $\vert {\mathrm{Re}} k_0 \vert <  k $. It may be interpreted as a Landau-type effect for the $W$ field, i.e. the propagating fluctuating $W$ field absorbs (emits) a soft gluon $k \sim g^2 T \ln 1/g$ from the plasma.

From the inequality (\ref{Ap5b}) one sees that for 
$ k < (\gamma - \bar{\delta})$, the expansion in powers of $k^2$ is convergent 
all along the $k_0$ axis, i.e. the half disk on Fig. \ref{fi3} dont intersect the real axis. As a consequence, $\Pi^t$ has no Landau-type imaginary part for $k <  (\gamma - \bar{\delta}) \sim \gamma/3 $

This appendix discusses the properties of the retarded amplitude 
$\Pi^t_R(k_0, k)$. A general property of a retarded propagator is that 
it is analytic in the upper $k_0$ plane. So it is for $\Pi^t_R$ 
(see Fig.\ref{fi3}). The advanced amplitude   $\Pi^t_A(k_0, k)$ is just the mirror picture, it is analytic in the lower $k_0$ plane, its singularities are in the upper plane
$$   \Pi^t_A(k_0, k) =   [ \Pi^t_R(k_0, k) ]^*  $$
Note that   $\Pi^t_R(k_0, k)$ and  $\Pi^t_A(k_0, k)$ have no common boundary 
in the complex $k_0$ plane. ($\Pi^t_R$ and $\Pi^t_A$ are on different Riemann
sheets of the Landau cut that arises at the scale $k \sim gT$). When one considers 
along the real axis   
$\Pi^t_R(k_0, k)  -  \Pi^t_A(k_0, k)$ one is substracting two different functions.

\subsection{Locations of the singularities}

 We study the tail of the continued fraction $\Sigma_m$  (see (\ref{Ap2})) .\\  As $ l \rightarrow \infty \ \ , \ \ b_l^{(m) 2} \rightarrow 1/4 $ \\
i)  case $\gamma \not= 0 \ \ \ , \ \ \ \delta_l =0$
\\ The tail of the continued fraction $\Sigma_m$ obeys 
$$ X = 1- { \rho^2 \over{4 X} } \ \ \ \  \mathrm{i.e.}  \ \ \ \ X ={{1\pm \sqrt{1-\rho^2}}\over 2} $$
Hence one recovers the fact that the continued fraction has a singularity at $ \rho^2 = 1 = (k / (k_0 + i \gamma))^2$
\\ ii) case  $\gamma \not= 0 \ \ \ , \ \ \ \delta_l \not=0$ \\
The same argument gives
\begin{equation}
\rho^2_l = 1 = { k^2 \over{ [k_0 +i(\gamma - \delta_l) ] [k_0 +i(\gamma - \delta_{l+1})]}}
\end{equation}
i.e.
\begin{equation}
k_0 = -i (\gamma -{ {\delta_l + \delta_{l+1}}\over 2}) \pm \sqrt{  k^2 - ({ {\delta_l - \delta_{l+1}}\over 2})^2}
\end{equation}
Since $( {\delta_l + \delta_{l+1}) / 2} < \bar{\delta}$ and 
$( \delta_l - \delta_{l+1})^2 < {\bar{\delta}^2 } $  ($\bar{\delta}$ is 
the upper bound of $\delta_l$), for $ k>\bar{\delta}/2 \sim \gamma/3$ all 
the singularities are inside the two regions in the complex $k_0$ plane drawn 
on Fig.\ref{fi3}
$$ \sqrt{k^2- \bar{\delta}^2/4} < \vert{ \mathrm{Re}}k_0 \vert \leq k \ \ \ , \ \ \ \bar{\delta}> {\mathrm{Im}}k_0 + \gamma > 0  $$
 Very likely, a cut links these two regions.
As $ l\rightarrow \infty$ the singularities tend towards $k_0 = \pm k - i \gamma$ since $ \delta_{2 l} \rightarrow \gamma /( 2 l) $ for  $ l \rightarrow \infty$

\section{The 4-gluon vertex}
\label{secB}
 The 4-gluon vertex 
 $V_{\mu\nu\rho\sigma}^{1234}(P_{1R},P_{2R},P_{3A},P_{4A})$ is given by Eq.
(\ref{RC41}) where $\cal R$ is the sum of two terms
\begin{eqnarray}
\lefteqn{{\cal R}=-{1\over 2}N(P_3,P_4)(f^{14m}f^{23m}+f^{13m}f^{24m})} \nonumber
\\ & &
<\ [ v_{\mu}(v.P_1+i\hat{C})^{-1}v_{\nu}
 + v_{\nu}(v.P_2+i\hat{C})^{-1} {v}_{\mu} ] (v.(P_1+P_2)+i\hat{C})^{-1}
 \nonumber \\ & & 
 [ v_{\rho}(-v.P_4+i\hat{C})^{-1} v_{\sigma}(-p_4^0)
 + v_{\sigma}(-v.P_3+i\hat{C})^{-1} v_{\rho} (-p_3^0) ] \ >_{all \ v}
\nonumber \\ & &  +
<\ [ v_{\mu}(v.P_1+i\hat{C})^{-1}v_{\nu}
 + v_{\nu}(v.P_2+i\hat{C})^{-1} {v}_{\mu} ] (v.(P_1+P_2)+i\hat{C})^{-1}
 \nonumber \\ & & 
 \lbrace f^{13m}f^{24m}(p_1^0+p_3^0)\ [\ N(P_3,P_2+P_4) v_{\rho}
 (-v.P_4+i\hat{C})^{-1} v_{\sigma} \nonumber \\ & &
 -N(P_4,P_1+P_3) v_{\sigma}
 (-v.P_3+i\hat{C})^{-1} v_{\rho} >\  \ ]  \nonumber \\ & &
 + \ f^{14m}f^{23m} \ [ \ 3\leftrightarrow 4 \ ] \ \  \rbrace 
 \label{Ap5}
 \end{eqnarray}
 where $[ \ 3\leftrightarrow 4 \ ]$ means a term obtained from the factor
 multiplying $f^{13m}f^{24m}$ by the exchange of all indices of $3$ and $4$. The
 first term in (\ref{Ap5}) is the analogue of the first term in (\ref{RC41}) with
 a colour factor symmetric in $1$ and $2$ rather than antisymmetric. In both
 terms of ${\cal R}$, $X_1^0,X_2^0$ are later times, $X_3^0,X_4^0$ earlier ones, both are
 essential for the Ward identities to be satisfied. The symmetries 
 $1\leftrightarrow 2$ and $3\leftrightarrow 4$ are explicit in (\ref{RC41}) and
 in (\ref{Ap5}).
 
 The Ward identity relating $p_1^{\mu} \ V_{\mu\nu\rho\sigma}^{1234}$ to 3-point
 vertices is written in Eq.(\ref{RC40}) where a vertex  of type RRA is related
 to the one of type ARR (as given in (\ref{RC31})) as follows
 \begin{equation}
 V_{\nu\rho\sigma}^{2m4}(P_{2R},P_{3R},P_{4A}) =
 [V_{\sigma\nu\rho}^{42m}(P_{4R},P_{2A},P_{3A})]^{+}
 \label{Ap6}
 \end{equation}
 where  the $+$ operation reverses the string of operators, 
  changes $i\hat{C}$ into $- i\hat{C}$ and reverses the colour order. This is
  consistent with Eqs.(\ref{C1},\ref{C2}) since $i f^{42m}$ is unchanged in the
  $+$ operation. The Ward identity for a leg of type A is
\begin{eqnarray}
\lefteqn{i p_4^{\sigma} \ 
V_{\mu\nu\rho\sigma}^{1234}(P_{1R},P_{2R},P_{3A},P_{4A}) = 
f^{43m}V_{\mu\nu\rho}^{12m}(P_{1R},P_{2R},(P_3+P_4)_A)} \nonumber \\ & &
+ f^{42m} [ \ {N(P_3, P_2+P_4) \over{N(P_3,P_4)}} 
V_{\mu\nu\rho}^{1m3}(P_{1R},(P_2+P_4)_A,P_{3A}) \nonumber \\ & &
+ {N(P_4, P_1+P_3)\over{N(P_3,P_4)}}
V_{\mu\nu\rho}^{1m3}(P_{1R},(P_2+P_4)_R,P_{3A}) \ ] \nonumber \\ & &  
+ f^{41m}[\ 1\leftrightarrow 2  \ ] \label{Ap7}
\end{eqnarray}
where the 3-point vertices are those of Eq.({\ref{RC31}) and Eq.(\ref{Ap6}).

\section{The eigenvalues of the collision operator $\hat{C}'$}
\label{secC}
\subsection{The eigenvalues as an integral}
\label{secC1}
$\hat{C}'({\bf v},{\bf v'})$ commutes with the rotations in ${\bf v }$ space,
its eigenvalues ${c'}_l$ depend on the only available quantity ${\bf v.v'}$
\begin{equation}
c_l' = <\ \hat{C}'({\bf v},{\bf v'}) \ P_l({\bf v.v'}) \ >_{v,v'}
\end{equation}
where $P_l$ is the Legendre polynomial. In this appendix an analytic expression
of $c_l'$ is given for the case of the dominant transverse gluon exchange in
terms of the matrix elements of the operator $(v.K+i\hat{C})^{-1}$
\begin{equation}
c_l' =m_D^2{g^2NT \over 2}\int {d_4k\over(2\pi)^4} \ {\vert\Delta^t(K_R)\vert}^2
 \ [ \ N_{(3g)}^{(l)}(K) \ + \ N_{(4g)}^{(l)}(K) \ ]
\label{Ap11}
\end{equation} 
from Eqs (\ref{D40},\ref{D41},\ref{D42}) and (\ref{F6},\ref{F8}), with
\begin{eqnarray}
\lefteqn{N_{(3g)}^{(l)}(K) = < P_l({\bf v}_1.{\bf v}_2) \
< \ v_{1i}^t(v_1.K+i\hat{C})^{-1}{v'}_{1j}^t
-v_{1j}^t(v_1.K-i\hat{C})^{-1}{v'}_{1i}^t\ >_{v_1'}} \nonumber \\ & & 
< \ v_{2i}^t(v_2.K+i\hat{C})^{-1}{v'}_{2j}^t
-v_{2j}^t(v_2.K-i\hat{C})^{-1}{v'}_{2i}^t\ >_{v_2'} \ \ >_{v_1,v_2}
\label{Ap12}
\end{eqnarray}
\begin{equation}
N_{(4g)}^{(l)}(K) = -({2i{\mathrm Im}\Pi^t \over{k_0m_D^2}})
< \  P_l({\bf v.v'}) \ v_i^t \ [(v.K+i\hat{C})^{-1}-(v.K-i\hat{C})^{-1}] \  {v'}_i^t
 \ >_{v,v'}
 \label{Ap13}
 \end{equation}
 where $v_i^t =v_i - {\hat{k}}_i {\bf v.\hat{k}}$, and the integration range on $k$ is
 limited to $k\sim g^2NT\ln 1/g$, i.e. $\mu_3<k<\mu_2$ where $\mu_2$ and $\mu_3$
 have been defined in Eq.(\ref{D14}). In this range
 \begin{equation}
 \Pi^t(k_0,k) ={k_0m_D^2\over 3} \ \Sigma_1(k_0,k)
 \end{equation}
 where $\Sigma_1(k_0,k)$ is a matrix element of $(v.K+i\hat{C})^{-1}$ given in
 Eq.(\ref{Ap2}) whose scale is $\gamma\sim g^2NT\ln 1/g$ for $k_0$ and $k$.
 Then
 \begin{equation}
  {\vert\Delta^t(K_R)\vert}^2 = {1\over{(k_0^2-k^2+{\mathrm Re}\Pi^t)^2
  +(\ {\mathrm Im}\Pi^t)^2}\ } \approx 
  {1 \over{k^4+k_0^2m_D^4({{\mathrm Im}\Sigma_1 /3})^2}}
  \label{RAp13}
  \end{equation}
  In the space $(k_0, k)$, ${\vert\Delta^t\vert}^2$ put a strong weight upon the
  region
  \begin{equation}
  k_0{m_D^2\over3} {\mathrm Im}\Sigma_1(k_0,k)  \leq \ k^2
  \label{Ap14}
  \end{equation}
  As $m_D^2\Sigma_1 \sim m_D^2 /\gamma \ \sim T/ \ln 1/g $ the weight is on the
  domain
  \begin{equation}
  k_0 \leq {k^2\over T} \ln 1/g
  \label{Ap15}
  \end{equation}
  The matrix elements of  $(v.K+i\hat{C})^{-1}$ all have the same scale $\gamma$
  for both $k_0$ and $k$ (see Eq.(\ref{Ap2}) and see Sec.\ref{secC4} of this
  Appendix) except for a  few that depend on the zero eigenvalue of $\hat{C}$.
  For those, one may substitute $N^{(l)}(k_0=0,k)$ to $N^{(l)}(k_0,k)$ in the
  integrant of (\ref{Ap11}) and perform the integration over $k_0$ (see Eq.
  (\ref{D53})) with the result
  \begin{equation}
  c_l' \approx {g^2NT\over2} \int_{\mu_3}^{\mu_2} {d k \over 4 \pi^2} 
  {3\over{\vert{\mathrm Im}\Sigma_1(k_0=0,k)\vert}} 
  [ \ N_{(3g)}^{(l)}(k_0=0,k) \ + \ N_{(4g)}^{(l)}(k_0=0,k) \ ]
  \label{Ap16}
 \end{equation}
 The integral will have no infrared divergence if $N^{(l)}(k_0=0,k)$ is finite
 as $k\to 0$ since ${\mathrm Im}\Sigma_1(k_0=0,k) \to -1/\gamma$
 as $k\to 0$. The exceptional cases will be discussed later on.
 
 In the following, $N_{(3g)}^{(l)}(k_0,k)$ and $N_{(4g)}^{(l)}(k_0,k)$ are
 expressed in terms of matrix elements of $(v.K+i\hat{C})^{-1}$ whose explicit
 expressions are given in Sec.\ref{secC4}. Useful properties are:
 \\ - The matrix elements of
  $(v.K+i\hat{C})^{-1}$ are divided into subspaces where 
  $l_z=m={\bf l.\hat{k}}$ is fixed and $l=m,m+1,m+2,\ldots$.
  (see Appendix \ref{secA}) \\
  - Matrix elements for $l_z=m$ and $l_z=-m$ are equal. \\
  - For $k_0=0$ and all $k$, all the relevant matrix elements are imaginary
  numbers. \\
  One will write
  \begin{equation}
  G \ = \  (v.K+i\hat{C})^{-1} \ \ {\mathrm and} \ \ \ G^+ \ = \ (v.K-i\hat{C})^{-1}
  \label{RAp17}
  \end{equation}
  ${\bf \hat{k}}$ is taken as the $z$ axis, and in $N^{(l)}(k_0,k)$ one writes
  \begin{equation}
  P_l({\bf v.v'}) = {4\pi\over{2l+1}} \sum_{m=-l}^{m=l} \ Y_l^{m*}({\bf v}) \
  Y_l^m({\bf v'})
  \label{Ap17}
  \end{equation}

  \subsection{The 3-gluon vertices diagram's contribution}
  \label{secC2}
  
  $N_{(3g)}^{(l)}(K)$ is given by Eq.(\ref{Ap12}) and (\ref{Ap17}) is used. Only
  the subspace $ |m| =1$ of $(v.K+i\hat{C})^{-1}$ enters. Indeed,
  ${v'}^t_{1i}$ is an element with $ |m|\  =1$, $(v.K+i\hat{C})^{-1}$ acts
  in this subspace, $v_{1i}^t$ and $Y_l^m({\bf v}_1)$ are combined with the use
  of Clebsh-Gordan coefficients and the ${\bf v}_1$ integration project them
  into the subspace.
  Writing for compactness
  \begin{equation}
  \langle \ l'\ m=1 \ \vert \ G\ -\ G^+\ \vert \ 1\ m=1\rangle =
  \langle\ l' \  \vert \  \Delta G \ \vert \ 1\ \rangle
  \label{Ap18}
  \end{equation}
  the result is
  \begin{eqnarray}
  \lefteqn{N_{(3g)}^{(l)}={1\over{3(2l+1)^2}}\lbrace\ [\ \left({l(l-1)\over{2l-1}}\right)^{1/2}
  \langle\ l-1 \  \vert \  \Delta G \ \vert \ 1\ \rangle 
  -\left({(l+1)(l+2)\over{2l+3}}\right)^{1/2}
  \langle\ l+1 \  \vert \  \Delta G \ \vert \ 1\ \rangle \ ]^2} \nonumber \\ & &
   + [\ \left({(l+1)(l+2)\over{2l-1}}\right)^{1/2}
  \langle\ l-1 \  \vert \  \Delta G \ \vert \ 1\ \rangle
  -\left({l(l-1)\over{2l+3}}\right)^{1/2}
  \langle\ l+1 \  \vert \  \Delta G \ \vert \ 1\ \rangle \ ]^2
  (1-\delta_{l0})(1-\delta_{l1})  \ \ \rbrace \nonumber \\ & & 
  \label{Ap19}
  \end{eqnarray}
  The first term comes from the $m=0$ term in the sum (\ref{Ap17}), the second
  from the $|m|=2$ term. \\
  As the matrix elements have the property (see Eq.(\ref{Ap42}))
  \begin{equation}
  \langle\ 1+n \  \vert \  G^+(k_0,k) \ \vert \ 1\ \rangle =
  (-1)^{n-1} \langle\ 1+n \  \vert \  G(-k_0,k) \ \vert \ 1\ \rangle
  \label{Ap20}
  \end{equation}
  $\langle\ l\pm1 \  \vert \  \Delta G \ \vert \ 1\ \rangle$ 
  is an even function of $k_0$ for $l$ even, an odd function of
  $k_0$ for $l$ odd. $N_{(3g)}^{(l)}(k_0=0,k)=0$ for $l$ odd, i.e. with the
  approximation of (\ref{Ap16}) 
  the contribution of $N_{(3g)}^{(l)}$ to $c_l'$ vanishes for $l$ odd.
   
   The matrix elements $\langle\ l\pm1 \  \vert \   G \ \vert \ 1\ \rangle$ may
   be expressed in terms of $\langle\  1 \  \vert \  G \ \vert \ 1\ \rangle \
   =\Sigma_1 \ $ (See Eq.(\ref{Ap36})). Explicitely
   \begin{equation}
   N_{(3g)}^{(l=0)}(k_0, k) = -{2\over9}[2\ {\mathrm Im}\Sigma_1\ ]^2
   \end{equation}
   \begin{equation}
    N_{(3g)}^{(l=1)}(k_0, k) = {1\over 5^2}{2\over9}\ k^2\ [\ 
    {\Sigma_1\over{(k_0+ic_2)X_2}} -{\mathrm c.c.} ]^2
   \end{equation}
   \begin{eqnarray}
   \lefteqn{N_{(3g)}^{(l=2)}(k_0,k) = {1\over5^2}{2\over9}
   \lbrace \ [\ \Sigma_1(1-{3.4\over5.7}
   {\frac{k^2}{(k_0+ic_2)(k_0+ic_3)}}{1\over X_2X_3}) 
   - {\mathrm c.c.}]^2} \nonumber \\ & & 
   + 6 [\Sigma_1(1-{2\over5.7}{\frac{k^2}{(k_0+ic_2)(k_0+ic_3)}}{1\over X_2X_3})
    - {\mathrm c.c.}]^2 \rbrace
   \label{Ap21b}
   \end{eqnarray}
   where c.c. means complex conjugate, $\Sigma_1(k_0,k)$ is given in (\ref{Ap2}),
   $X_2, X_3$ are defined in term of $\Sigma_1$ in (\ref{Ap34},\ref{Ap35}), and
   $c_l$ is the $l$ eigenvalue of the operator $\hat{C}$. \\ 
   For $k <<\gamma$, $\Sigma_1\approx (k_0+ic_1)^{-1}$ and from
   Eq.(\ref{Ap36}) $\langle\ l+1 \  \vert \  G \ \vert \ 1\ \rangle$ behaves as
   $k^l$, $\langle\ l-1 \  \vert \  G \ \vert \ 1\ \rangle$ as $k^{l-2}$. One
   concludes that the contribution of $N_{(3g)}^{(l)}$ to the integral
   (\ref{Ap16}) is infrared finite for all $l$.
   
   \subsection{The 4-gluon vertex diagram's contribution}
   \label{secC3}
   
   $N_{(4g)}^{(l)}(K)$ is written Eq.(\ref{Ap13}) and (\ref{Ap17}) is used.
   $v_i^t\ Y_l^m({\bf v})$ has components in the $m+1$ and $m-1$ subspaces of 
   $G=(v.K+i\hat{C})^{-1}$. The result is written as a sum over the different
   subspaces $M$ of $G$, the two first terms correspond to $M=l+1$ and $M=l$, 
   $\Delta G = G-G^+$.
   \begin{eqnarray}
   \lefteqn{N_{(4g)}^{(l)}=-i {2\over3}\ {\mathrm Im} \Sigma_1 \ {2\over {2l+1}}
    \lbrace}
   \label{Ap20b} \\ & &
   {l+1\over{2l+3}}
   \langle\ l+1 \ l+1 \vert \ \Delta G \ \vert \ l+1\ l+1\rangle
   + {l\over{2l+3}}
   \langle\ l+1 \ l \vert \ \Delta G \ \vert \ l+1\ l\rangle \nonumber \\ & & 
   +\sum_{M=0}^{M=l-1}(1-{1\over2}\delta_{M0})  [\ {l(l+1)+M^2\over{2l+1}}
    (\ {1\over{2l+3}} 
   \langle\ l+1 \ M \vert \ \Delta G \ \vert \ l+1\ M\rangle \nonumber \\ & & 
   + {1\over{2l-1}} 
   \langle\ l-1 \ M \vert \ \Delta G \ \vert \ l-1\ M\rangle\  )
   -b_l^{(M)}b_{l-1}^{(M)}2 
   \langle\ l+1 \ M \vert \ \Delta G \ \vert \ l-1\ M\rangle\ (1-\delta_{l0}) 
   \  ] \  \rbrace
   \nonumber
   \end{eqnarray}
   with $b_l^{(M)}$ given in (\ref{RAp1}). With the property of G (see
   Eq.(\ref{Ap42}))
   \begin{equation}
   \langle\ l'+n \ M \vert \ G^+(k_0,k) \ \vert \ l'\ M\rangle =
   (-1)^{n+1} \langle\ l'+n \ M \vert \ G(-k_0,k) \ \vert \ l'\ M\rangle
   \label{Ap21}
   \end{equation}
   each matrix element of (\ref{Ap20b}) is an even function of $k_0$. The matrix
   elements in a given subspace $M$ may be written in terms of the lowest one
   \begin{equation}
   \Sigma_M = \langle\ l=M \ M \vert \ G \ \vert \ l=M \ M\rangle
   \label{Ap22}
   \end{equation}
   For example
   \begin{equation}
   N_{(4g)}^{(l=0)} = {2\over9}[ 2\ {\mathrm Im}\Sigma_1 ]^2
   \end{equation}
   the contribution of $ N_{(4g)}^{(l=0)}$ and $ N_{(3g)}^{(l=0)}$ cancel each
   other and $c_0'=0$ as it was shown in Sec.\ref{sec4.4} by another method.
   \begin{eqnarray}
   \lefteqn{N_{(4g)}^{(l=1)}=-i {2\over3}\ {\mathrm Im} \Sigma_1 \ {2\over3} 
   \lbrace \ 
   {2\over5} \langle\ 2 \ 2 \  \vert \ \Delta G \ \vert \ 2 \ 2\ \rangle
   + {1\over5} \langle\ 2 \ 1 \  \vert \ \Delta G \ \vert \ 2 \ 1\ \rangle}
     \label{Ap23} \\ & & 
   +{1\over3}[\  \langle\ 0 \ 0 \  \vert \ \Delta G \ \vert \ 0 \ 0\ \rangle
   +{1\over5} \langle\ 2 \ 0 \  \vert \ \Delta G \ \vert \ 2 \ 0\ \rangle
   -{2\over{\sqrt{5}}} \langle\ 2 \ 0 \ \vert \ \Delta G \ \vert \ 0 \ 0\ \rangle
   \ ]\ \rbrace
   \nonumber
   \end{eqnarray}
   With the use of the relations between the matrix elements of $G$, it may be
   written in terms of $\Sigma_2, \Sigma_1, \Sigma_0$ associated respectively
   with the subspaces $M=2,1,0$. For example
   \begin{equation}
    \langle\ 2 \ 1\  \vert \ \Delta G \ \vert \ 2 \ 1\ \rangle
    =5{d_1\over k^2}(d_1\ \Sigma_1 -1) = {d_1\over d_2}\  {\Sigma_1\over X_2^{(1)}}
    \end{equation}
    where $d_l=k_0+ic_l$ and the $X_l^{(m)}$ are expressed in terms of $\Sigma_m$
    in (\ref{Ap35},\ref{Ap34}).
    \begin{eqnarray}
    \lefteqn{N_{(4g)}^{(l=1)}(k_0,k)=-i {2\over3}\ {\mathrm Im} \Sigma_1 \ {2\over3}
    \lbrace \ {2\over5}(\Sigma_2 -{\mathrm c.c.})
    +{1\over5}[{k_0+ic_1\over{k_0+ic_2}}{1\over X_2^{(1)}}\Sigma_1 -{\mathrm
    c.c.}] } \label{Ap24} \\ & &
    +{1\over3}[\ \Sigma_0\ (1-{1\over3}{\frac{k^2}{(k_0+ic_1)(k_0+ic_2)}}
    {1\over{X_1^{(0)}X_2^{(0)}}} +{1\over5}{\frac{k_0}{k_0+ic_2}}
    {1\over{X_1^{(0)}X_2^{(0)}}} \ ) - {\mathrm c.c.}]
    \nonumber
    \end{eqnarray}
    where the $c_l$ are the eigenvalues of $\hat{C}$, and the $X_l^{(m)}$ are
    expressed in terms of $\Sigma_m$ in (\ref{Ap34},\ref{Ap35}).\\

    \noindent{\it The infrared sector} \\
    For $k<<\gamma$, the diagonal matrix elements of $G$ do not depend on the
    subspace $M$
    \begin{equation}
    \langle\ l' \ M\  \vert \  G \ \vert \ l \ M\ \rangle =
    \langle\ l' \ M\  \vert \ (v.K+i\hat{C})^{-1}  \ \vert \ l \ M\ \rangle
    \approx (k_0+ic_l)^{-1}\ \delta_{ll'}
    \end{equation}
    With $\sum_1^{l-1}M^2=l(l-1)(l-1/2)/3$, one can check that the full
    expression (\ref{Ap20b}) reduces to
    \begin{equation}
    N_{(4g)}^{(l)}(k_0, k<<\gamma)=-i {2\over3}\ {\mathrm Im} \Sigma_1 \ 
    {2\over 2l+1}{1\over3}[\ {l+1\over{k_0+ic_{l+1}}} + {l\over{k_0+ic_{l-1}}}
    - \ {\mathrm c.c.}\ ]
    \label{Ap26}
    \end{equation}
    a form immediately obtained from the initial expression of $N_{(4g)}^{(l)}$,
    Eq.(\ref{Ap13}), if one writes
    \begin{equation}
    {\bf v}_t.{\bf v'}_t \ P_l({\bf v.v'}) = {2\over3}{\bf v.v'} \ P_l({\bf v.v'})
     ={2\over3}{1\over{2l+1}}[\ (l+1)P_{l+1} + l P_{l-1}]
     \end{equation}
    Because of the eigenvalue $c_0=0$, care has to be given to the subspace
    $M=0$. As shown in Sec.\ref{secC4}  of this appendix, the limits $k\to 0$ and 
    $k_0\to 0$ commute for all 
    $\langle\ l' \ 0\  \vert \  G \ \vert \ l \ 0\ \rangle $ with $l$ and
    $l'\geq 2$. One concludes that the contribution of $N_{(4g)}^{(l)}$ to the
    eigenvalue $c_l'$ is infrared finite for $l>2$. For the case $l=2$, the
    elements $\langle\ 1 \ 0\  \vert \  G \ \vert \ l' \ 0\ \rangle $ enter in
    (\ref{Ap20b}), the limits  $k\to 0$ and $k_0\to 0$ do not commute, these
    elements vanish as $k_0\to 0$, $k$ fixed (see Eq.(\ref{Ap43})). There is no
    infrared divergence in $N_{(4g)}^{(l=2)}$.

    \noindent{\it The case of $c_1'$} \\
    As the relation (\ref{Ap26}) suggests, one is left with the case $l=1$ whose
    explicit expression is written in (\ref{Ap23}) or (\ref{Ap24}). In
    (\ref{Ap23}), the matrix elements
    $\langle\ 2 \ M\  \vert \  \Delta G \ \vert \ 2 \ M\ \rangle  \to 2/i\ c_2$ as
    $k\to 0$ irrespective of the order of the limits $k_0$ and $k$, they give a
    finite contribution to $c_1'$. The other terms are
    \begin{eqnarray}
    \lefteqn{{1\over3}[\ \langle\ 0 \ 0\  \vert \  \Delta G \ \vert \ 0 \ 0\ 
    \rangle  -{2\over{\sqrt 5}}
    \langle\ 0 \ 0\  \vert \  \Delta G \ \vert \ 2 \ 0\ \rangle \ ] =}
    \nonumber \\ & & 
    {1\over3}\ \Sigma_0 \ 
     (1-{k^2\over{(k_0+ic_1)(k_0+ic_2)}}{1\over{X_1^{(0)}X_2^{(0)}}}{4\over15})
     - {\mathrm c.c.} 
    \label{Ap27}
    \end{eqnarray}
    if one uses (\ref{Ap36}) to relate the second matrix element to the first 
    one. With
    \begin{equation}
    \Sigma_0(k_0=0,k)=-{3ic_1\over k^2} \ X_1^{(0)}
    \label{Ap28}
    \end{equation}
    where $X_l^{(0)}(k_0=0,k)$ real $>1$ from Eqs.(\ref{Ap34},\ref{Ap35}), one 
    sees that the integral  (\ref{Ap16}) has a linear infrared divergence which 
    is solely due to the existence of the matrix element 
    $\langle\ 0 \ 0\  \vert \  \Delta G \ \vert \ 0 \ 0\ \rangle = 
    2i \ {\mathrm Im}\Sigma_0$ in (\ref{Ap27}).

    Instead of considering $\Sigma(k_0=0,k)$ as it enters (\ref{Ap16}), one may 
    wish to go back one step ahead and study the relevant domain $(k_0,k)$ in 
    $\int k^2 dk \ dk_0$ of Eq.(\ref{Ap11}). Here
    \begin{equation}
    \int k^2 dk \ dk_0 {1\over{k^4+k_0^2a^2}} \ {k^2\over{k^4+k_0^2b^2}}=
    \pi\int{dk\over k^2} {1\over a+b}
    \label{Ap29}
    \end{equation}
    From $|\Delta^t|^2$ in (\ref{RAp13}) one has $a\sim m_D^2/c_1$, and for 
    ${\mathrm Im} \Sigma_0$ one has $b\sim c_1$ since from (\ref{Ap34}), 
    for an estimate
    \begin{equation}
    {\mathrm Im} \Sigma_0 \approx 
    {\mathrm Im} 
   {\displaystyle{\frac{1}{k_0-{k^2\over{3(k_0+ic_1)}}}}} 
   =-{1\over3}{\displaystyle{\frac{c_1k^2}{(k_0^2-{k^2\over3})^2+c_1^2k_0^2}}} 
   \approx -3{\frac{c_1k^2}{k^4+9c_1^2k_0^2}}
   \end{equation}
   With $m_D^2=g^2NT^2/3 \ >> c_1^2\sim(g^2NT \ln 1/g)^2$, one has $a>>b$ in
   (\ref{Ap29}) and 
   one concludes that the linear divergence is indeed given by 
   $\Sigma_0(k_0=0,k)$. \\

   If one sets $k_0=0$ and  one writes $\Sigma_l(k_0=0,k) 
   =-i \tilde{\Sigma}_l(k)$, one gets from (\ref{Ap23})
   \begin{equation}
    N_{(4g)}^{(l)}(k_0=0, k) = {\tilde{\Sigma}_1} \ {4\over9} [\
    {4\over5}\tilde{\Sigma}_2 + 2{c_1\over k^2}(1-c_1\tilde{\Sigma}_1)
    +{2\over3}(\tilde{\Sigma}_0-3{c_1\over k^2})(1+{5\over4}) + \ 
    2{c_1\over k^2}\ \ ]
    \label{Ap45}
    \end{equation}
    As $k\to 0$, all the terms are finite, except for the term $2c_1/k^2$. In the
    factor $( 1+ 5/4)=9/4$, $1$ comes from the infrared singular element 
    $\langle\ 0 \ 0\  \vert \  \Delta G \ \vert \ 0 \ 0\ \rangle$ and $5/4$ from
    the elements $\langle\ 2 \ 0\  \vert \  \Delta G \ \vert \ 2 \ 0\ \rangle$
    and $\langle\ 0 \ 0\  \vert \  \Delta G \ \vert \ 2 \ 0\ \rangle$. The form
    (\ref{Ap45}) allows an easy comparison with the result of Arnold and Yaffe
    \cite{AY1} at the end of their subsec.7 of Sec.II.C. As it has been
    discussed in Sec.\ref{sec5.2}  , comparing their expressions at the end of their 
    subsecs.7 and 5, one has to substract out the $\rho$-dependent part which 
    does not
    come fron $<v_lv_i G_0(p)v_jv_l> {\cal P}_t^{ij}$ but from 
    $${2\over 3p^2}(4-3{\sigma_p\over \sigma_0}) = {8\over3p^2} -
    {2\Sigma_1\over p^2}$$
    i.e. for the term in $(-\Sigma_1)$, writing in their subsec.7's form 
    $8/3=2+2/3$, it remains
    $$ {1\over2}(\Sigma_0 -{3\over \rho^2}) + {2\over{3\rho^2}}(1-\Sigma_1)
    +{4\over 15}\Sigma_2 $$
    which agrees with the convergent term in the bracket in Eq.(\ref{Ap45}) when
    one factors out $1/3$.

    \subsection{Matrix elements of $G=(v.K+i\hat{C})^{-1}$}
    \label{secC4}
    The eigenvectors of $\hat{C}({\bf v.v'})$ are 
    $\vert \ l\ m\ \rangle = \sqrt{4\pi} \ Y_l^m({\bf v})$ with the measure
    ${d\Omega_{\bf v}/ 4\pi}$, its eigenvalues are $c_l$. It is convenient
    to choose $ {\bf\hat{k}}$ as the z axis, then the operator $(k_0+i\hat{C}
    -v_zk)$ changes $l$ but does not change $l_z=m$, so does its inverse
    \begin{equation}
    \langle\ l' \ m\  \vert \ (k_0+i\hat{C} -v_zk)^{-1}  
    \ \vert \ l \ m\ \rangle =\delta_{m m'}
    \langle\ l' \ m\  \vert \ (k_0+i\hat{C} -v_zk)^{-1}  
    \ \vert \ l \ m\ \rangle
    \label{Ap30}
    \end{equation}
    In a given subspace $m$, all matrix elements of $G$ may be written in terms
    of the lowest one \cite{AY1}
    \begin{equation}
    \langle\ l=m \ m\  \vert \   G \ \vert \ l=m \ m\ \rangle =\Sigma_m
    \label{Ap31}
    \end{equation} 
    because of the recursion relations
    \begin{equation}
    \langle\ l' \ m\  \vert \   G\ G^{-1} \ \vert \ l \ m\ \rangle =
    \delta_{l l'}
    \end{equation}
    \begin{eqnarray}
    \delta_{l l'} &=&-kb_{l'} \langle\ l'+1 \  \vert \   G \ \vert \ l \ \rangle
     -kb_{l'-1} \langle\ l'-1 \  \vert \   G \ \vert \ l \ \rangle 
     +d_{l'}  \langle\ l' \  \vert \   G \ \vert \ l \ \rangle  
	\nonumber \\
    \delta_{l l'} &=&-kb_l \langle\ l' \  \vert \   G \ \vert \ l +1 \ \rangle
     -kb_{l-1} \langle\ l'-1 \  \vert \   G \ \vert \ l \ \rangle 
     +d_{l}  \langle\ l' \  \vert \   G \ \vert \ l \ \rangle
     \label{Ap32}
     \end{eqnarray}
     with the definition
     \begin{equation}
     \langle\ l \  \vert \   k_0+i\hat{C} \ \vert \ l \ \rangle =
     k_0+ic_l =d_l
     \label{Ap33}
     \end{equation}
     The $b_l=b_l^{(m)}$ are the matrix elements of $v_z$ defined in
     (\ref{RAp1}), and  the reference to the subspace $m$  has been
     dropped. Note that a consequence of (\ref{Ap32}) is, \\
     as $k \to 0$,  $\langle\ l' \  \vert \   G \ \vert \ l \ \rangle \to
     \delta_{l l'} /\ d_l$ independent of $m$.
     
     It is convenient to exhibit the threshold factors arising from angular
     momentum conservation. $\Sigma_m$ is the continued fraction explicited in 
     Eq.(\ref{Ap2}), it may be written
     \begin{equation}
     \Sigma_m(k_o,k) = {1\over d_m} \ {\displaystyle{\frac{1}{1-{\displaystyle{\frac{k^2
     b_m^2}{d_md_{m+1}}}\ {1\over X_{m+1}}}}}}
     \label{Ap34}
     \end{equation}
     \begin{equation}
     X_{m+1}= 1-{\frac{k^2 b_{m+1}^2}{d_{m+1}d_{m+2}}} \ {1\over X_{m+2}}
     \label{Ap35}
     \end{equation}
     with $d_m=k_0+ic_m$ (see (\ref{Ap33})), $b_{m+n}=b_{m+n}^{(m)}$ is the
     quantity that depends on the subspace $m$ (see (\ref{RAp1}))
     \begin{equation}
     {b_0^{(0)}}^2 = {1\over3} \ \ , \  \  {b_1^{(0)}}^2 = {4\over15} \ \ , \  \ 
     {b_1^{(1)}}^2 = {1\over5}
     \end{equation}
     So do the $X_m$. Then the solution to the recursion relations {\it in the
     subspace} $m$ may be written for any $ l\geq m$, $n>0$
     \begin{eqnarray}
     \lefteqn{\langle\ l+n \  \vert \   G \ \vert \ l \ \rangle =
     \langle\ l \  \vert \   G \ \vert \ l +n \ \rangle =} \nonumber \\ & & 
     {\frac{b_lb_{l+1}\ldots b_{l+n-1}}{X_{l+1}X_{l+2}\ldots X_{l+n}}} \ 
     {\frac{k^n}{d_{l+1}d_{l+2}\ldots d_{l+n}}} 
     \langle\ l \  \vert \   G \ \vert \ l \ \rangle
     \label{Ap36}
     \end{eqnarray}
     and $\langle\ l \  \vert \   G \ \vert \ l \ \rangle$ is related as follows
     to the lowest $l=m$
     \begin{equation}
     \langle\ l=m \  \vert \   G \ \vert \ l=m \ \rangle = \Sigma_m
     \label{Ap37}
     \end{equation}
     \begin{equation}
     \langle\ l=m+1 \  \vert \   G \ \vert \ l=m+1 \ \rangle =
     {\frac{d_m}{d_{m+1}X_{m+1}}} \ \Sigma_m
     \label{Ap38}
     \end{equation}
     \begin{equation}
     \langle\ l=m+2 \  \vert \   G \ \vert \ l=m+2 \ \rangle =
     {d_m\over d_{m+2}}{1\over{X_{m+1}X_{m+2}}}(1-{\frac{k^2b_m^2}{d_md_{m+1}}}) \ 
     \Sigma_m
     \label{Ap39}
     \end{equation}
     \begin{equation}
     \langle\ l=m+3 \  \vert \   G \ \vert \ l=m+3 \ \rangle =
     {d_m\over{d_{m+3}}}{1\over X_{m+1}X_{m+2}X_{m+3}}(1-{\frac{k^2b_m^2}{d_md_{m+1}}}
     -{\frac{k^2b_{m+1}^2}{d_{m+1}d_{m+2}}}) \ 
     \Sigma_m
     \end{equation}
     \\
     
     \noindent {\it Limits as} $k_0$ {\it or} $k \ \to 0$ \\
     i) \ For any subspace $m \neq 0$ \\
     - as $k\to 0$, all $X_m\to 1$, $\Sigma_m \to 1/d_m$, one finds 
     $\langle\ l \  \vert \   G \ \vert \ l \ \rangle \to 1/\ d_l$ independent
     of the subspace $m$ as expected, and one obtains the correct threshold
     factor for the non-diagonal matrix elements. \\
     - as $k_0 \to 0$, $d_m =i \ c_m$, $X_m>1$, all
     $\langle\ l \  \vert \   G \ \vert \ l \ \rangle$ are imaginary, and the
     non-diagonal matrix elements are alternatively real and imaginary. \\
     ii) \ For the subspace $m=0$, $d_0=k_0$ as $c_0=0$, and care is needed when
     taking the limits \\
     - $k_0\neq 0$ the limit $k\to 0$ is as in the other subspaces $m$ \\
     - $k$ fixed, $k_0 \to 0$, $d_0\to 0$, $d_m\to i\ c_m$, $X_m>1$. From 
     (\ref{Ap34})
     \begin{equation}
     \langle\ 0 \ 0 \  \vert \   G \ \vert \ 0 \ 0 \  \rangle =\Sigma_0 =
     -{d_1X_1\over{k^2b_0^2}} =-i \ {c_1X_1\over{k^2b_0^2}}
     \label{Ap40}
     \end{equation}
     For $l\geq 2$, the elements 
     $\langle\ l \ 0 \  \vert \   G \ \vert \ l \ 0 \  \rangle$ have factors
     $${1\over d_l}(\ d_0(1-{k^2b_0^2\over{d_0d_1}}) +\ldots)\ \Sigma_0 \ 
     {{\longrightarrow}_{k_0\to 0}} \ \ {1\over d_l}(-{k^2b_0^2\over d_1} + O(k^4)) \ \Sigma_0 $$
     so that the limit $k\to 0$ again gives $1/d_l$. As a consequence, the
     limits $k\to 0$ and $k_0\to 0$ commute in all matrix elements 
     $\langle\ l \ 0 \  \vert \   G \ \vert \ l' \ 0 \  \rangle$ with $l$ and
     $l'\ \geq2$. \\
     The exceptions are seen from Eqs.(\ref{Ap36}),(\ref{Ap38}),(\ref{Ap40}). For
     $k_0\to 0$, $k$ fixed,
     \begin{equation}
     \langle\ 1 \ 0 \  \vert \   G \ \vert \ l \ 0 \  \rangle \ \to 0 \ \ 
     {\mathrm for}  \ \ l  \geq 1
     \label{Ap43}
     \end{equation}
     \begin{equation}
     \langle\ 0 \ 0 \  \vert \   G \ \vert \ 1 \ 0 \  \rangle =
     {b_0 k\Sigma_0 \over {d_1 X_1}} \ \to -{1\over k \ b_0}
     \label{Ap41}
     \end{equation}
     and $\langle\ 0 \ 0 \  \vert \   G \ \vert \ l \ 0 \  \rangle$ have the
     anomalous threshold factor $k^{l-2}$. \\
     Another quantity to be used is
     \begin{equation}
     {k_0\over k^2}(k_0\Sigma_0 -1) = {1\over3} \ 
     {\displaystyle{\frac{k_0}{k_0d_1(1-{\displaystyle{\frac{k^2\ {b_1^{(0)}}^2}{d_1d_2 \ X_2^{(0)}}})
     -{k^2\over3}}}}}
     \label{Ap46}
     \end{equation}
     where $d_1=k_0+ic_1$.
      There are two limits: \\
     $k^2 << k_0|k_0+ic_1| $, the limit is $1/3 d_1 $ \\
     $k_0|k_0+ic_1| << k^2$, the limit is $-k_0/k^2$  \\
     
     \noindent {\it Complex conjugation} \\
     The matrix elements of $G=(v.K+i\hat{C})^{-1}$ and of 
     $G^+=(v.K-i\hat{C})^{-1}$ may be related. 
     Writing $d_m^*(k_0) =k_0-ic_m=-d_m(-k_0)$, one has from Eqs.(\ref{Ap34}) to
     (\ref{Ap37})
     \begin{eqnarray}
     X_m^*(k_0,k) &=&  X_m(-k_0,k) \nonumber \\
     \Sigma_m^*(k_0,k) &=& - \ \Sigma_m(- k_0,k) \nonumber \\
     \langle\ l \  \vert \   G^+(k_0,k) \ \vert \ l+n \  \rangle &=&
     (-1)^{n+1} \ \langle\ l \  \vert \   G(-k_0,k) \ \vert \ l+n \  \rangle
     \label{Ap42}
     \end{eqnarray} \\
     
     \noindent {\it Limit when $\hat{C}$ is replaced by $\epsilon$}
     \begin{eqnarray}
     \lefteqn{\langle\ l' \ m \  \vert \ (v.K+i\epsilon)^{-1} \ \vert \ l \ m \  \rangle
     =}  \\ & &
     (-1)^m\sum_{L=l-l'}^{L=l+l'} {1\over k}Q_L({k_0+i\epsilon\over k}) \ 
     [(2l+1)(2l'+1)]^{1/2} \langle\ l \ l';0\ 0 | L \ 0   \rangle 
     \langle\ l \ l';m -m | L \ 0   \rangle \nonumber
     \end{eqnarray}

\begin{figure}[p]
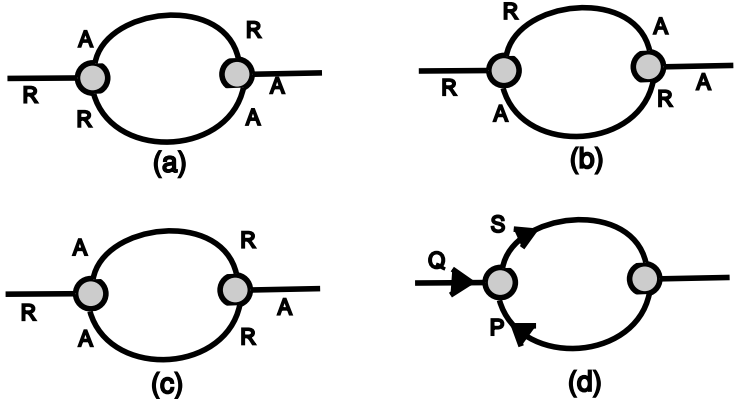

\caption{\label{fi1} (a) (b) (c) are the three possible ways of joining two 
3-point vertices in the R/A formalism.}
\end{figure}

\begin{figure}
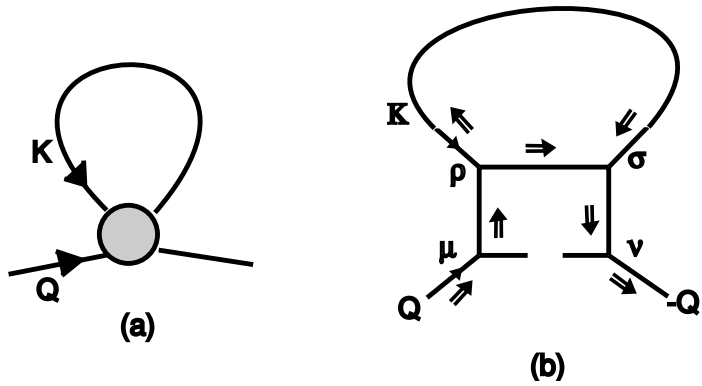

\caption{\label{fi2} (a) Contribution to the self-energy  of the 4-gluon 
vertex' diagram. (b) The surviving diagram for a soft loop; the double arrow
indicates the $\epsilon$ flow.}
\end{figure} 
 
\begin{figure}
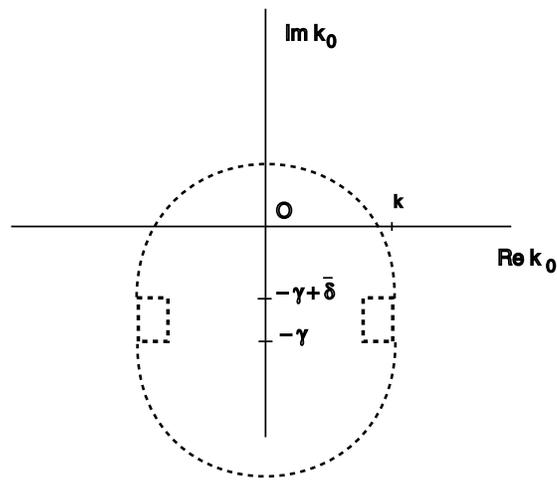

\caption{\label{fi3} Analytic properties of $\Pi_R^t(k_0, k)$. 
The singularities are inside the rectangles; the expansion in $k^2$ is 
convergent out of the domain limited by the dashed lines.}
\end{figure}


\begin{references}   

\bibitem{Pi1} R. Pisarski,  Phys. Rev. Letters {\bf 63}, 1129 (1989); E. Braaten and R. Pisarski, Nucl. Phys. B {\bf 337}, 569 (1990); {\it ibid.} {\bf 339}, 310 (1990).

\bibitem{FTay} J. Frenkel and J. C. Taylor, Nucl. Phys. B {\bf 334}, 199 (1990).

\bibitem{MLB} M. Le Bellac, {\it Thermal Field Theory} (Cambridge University Press, Cambridge, 1996).

\bibitem{Bod1} D. B\"{o}deker, Phys. Lett. B {\bf 426}, 351 (1998).

\bibitem{Bod2} D. B\"{o}deker, {\it Proceedings of the 5th International Workshop in thermal field theories and their applications} ed. U. Heinz,  hep-ph/9811469 ; Nucl. Phys. B {\bf 559}, 502 (1999)

\bibitem{AY2} P. Arnold, D. Son and L.G. Yaffe, Phys. Rev. D {\bf 59}, 105020 (1999); {\it ibid.} {\bf 60}, 025007 (1999).

\bibitem{BI2} J. P. Blaizot and E. Iancu, Nucl. Phys. B {\bf 557}, 183 (1999)

\bibitem{Lit} D. Litim and C. Manuel, Phys. Rev. Lett. {\bf 82}, 4981 (1999);  Nucl. Phys. B {\bf 562}, 237 (1999)

\bibitem{BI3} J. P. Blaizot and E. Iancu, Nucl. Phys. B {\bf 417}, 608 (1994); 
{\it ibid.} {\bf 434}, 662 (1995)

\bibitem{BI1} J. P. Blaizot and E. Iancu, Nucl. Phys. B {\bf 570}, 326 (2000), hep-ph/9906485. 

\bibitem{AY1} P. Arnold and L. G. Yaffe,  hep-ph/9912306.

\bibitem{Bod3} D. B\"{o}deker, Nucl. Phys. B {\bf 566}, 402 (2000), hep-ph/9903478.

\bibitem{Ev} T. S. Evans, Nucl. Phys. B {\bf 374}, 340 (1992).

\bibitem{Au1} P. Aurenche and T. Becherrawy, Nucl. Phys. B {\bf 379}, 259 (1992).

\bibitem{Au2} P. Aurenche, T. Becherrawy, and E. Petitgirard, hep-ph/9403320 (1993) (unpublished)

\bibitem{vW} C. Van Eijck and Ch. G.  Van Weert, Phys. Lett. B {\bf 278}, 305 (1992)

\bibitem{FG1} F. Guerin, Nucl. Phys. B {\bf 432}, 281 (1994)

\bibitem{Au3} P. Aurenche, F. Gelis, R. Kobes and E. Petitgirard, Phys. Rev. D {\bf 54}, 5274 (1996). ; Z. Phys. C. {\bf 75}, 315 (1997).

\bibitem{FG2} F. Guerin, Phys. Rev. D {\bf 49}, 4182 (1994). 

\end{references}
  \end{document}